\documentclass[aps,prl,reprint,groupedaddress]{revtex4-1}

\usepackage{graphicx}
\usepackage{dcolumn}
\usepackage{amsmath}
\usepackage{amssymb}
\usepackage{wasysym}
\usepackage{color}

\usepackage{mathtools}

\begin{document}

%\preprint{}

\title{Magic gap ratio for optimally robust fermionic condensation and its implications for high-$T_{\rm c}$ superconductivity
%Resonance condensation in the high-$T_{\rm c}$ cuprates
}


\author{N.~Harrison and M.~K.~Chan}
%\thanks{}
\affiliation{National High Magnetic Field Laboratory, Los Alamos National Laboratory, Los Alamos, NM 87545, USA}


\date{\today}

\begin{abstract}
Bardeen-Schrieffer-Cooper (BCS) and Bose-Einstein condensation (BEC) occur at opposite limits of a continuum of pairing interaction strength between fermions.  A crossover between these limits is readily observed in a cold atomic Fermi gas. Whether it occurs in other systems such as the high temperature superconducting cuprates has remained an open question. 
We uncover here unambiguous evidence for a BCS-BEC crossover in the cuprates by identifying a universal magic gap ratio $2\Delta/k_{\rm B}T_{\rm c}\approx$ 6.5 (where $\Delta$ is the pairing gap and $T_{\rm c}$ is the transition temperature) at which paired fermion condensates become optimally robust. 
At this gap ratio, corresponding to the unitary point in a cold atomic Fermi gas, the condensate fraction $N_0$ and the height of the jump $\delta\gamma(T_{\rm c})$ in the coefficient $\gamma$ of the fermionic specific heat at $T_{\rm c}$ are strongly peaked. 
In the cuprates, $\delta\gamma(T_{\rm c})$ is peaked at this gap ratio when $\Delta$ corresponds to the antinodal spectroscopic gap, thus reinforcing its interpretation as the pairing gap. 
We find the peak in $\delta\gamma(T_{\rm c})$ also to coincide with a normal state maximum in $\gamma$, which is indicative of a pairing fluctuation pseudogap above $T_{\rm c}$. 
%\\\\
%
\end{abstract}

%\keywords{}

\maketitle
A crossover in the pairing interactions between the weak coupling Bardeen-Schrieffer-Cooper (BCS)~\cite{bardeen1957} and the strong coupling Bose-Einstein Condensation (BEC)~\cite{jochim2003,zwierlein2003} limits was proposed in the high transition temperature $T_{\rm c}$ superconducting cuprates soon after their discovery~\cite{sademelo1993,randeria1989,drechsler1992,friedberg1989,micnas1990}.  
On the BCS side, pairing takes place at the Fermi surface below $T_{\rm c}$ as in a conventional superconductor, whereas on the BEC side, fermions pair up to produce bosons whose subsequent condensation at $T_{\rm c}$ is determined by the phase stiffness of the superfluid.  
Whereas the cuprates provided the motivation for much of the early theoretical work on the BCS-BEC crossover, today it is in a cold atomic Fermi gas~\cite{bloch2008,giorgini2008} where this phenomenon is well established. 
The relative simplicity of a cold atomic Fermi gas, consisting of pairing interactions tuned via a Feshbach resonance in an otherwise weakly interacting Fermi gas, has made it the ideal paradigm for cementing~\cite{ku2012,regal2004,zwielein2004} our theoretical understanding of condensation in the crossover region~\cite{sademelo1993,haussmann2007}. 
Yet the question of whether such a crossover occurs in other paired fermion systems such as the cuprates has remained. The other proposed BCS-BEC crossover candidates include nuclear matter, quark-gluon plasmas, iron-based superconductors and twisted graphene~\cite{strinati2018,randeria2014,chen2005,hazra2019,park2021,kasahara2014,rinott2017}.

%%%%%%%%%%%%%%%

While various experiments are suggestive of a non BCS pairing scenario in the cuprates~\cite{uemura1989,emery1995,dubroka2011,hu2014,kaiser2014,li2010,zhou2019}, uncertainty has surrounded the question of whether $T_{\rm c}$ is a sufficiently large fraction of the Fermi temperature $T_{\rm F}$ for a BCS-BEC crossover to be viable~\cite{hazra2019}.
For example, electronic band theory predicts a ratio $T_{\rm c}/T_{\rm F}\sim10^{-2}$ that is clearly too small for a BCS-BEC crossover to occur~\cite{giorgini2008}. However, thermodynamic measurements, including magnetic quantum oscillations, have revealed strongly renormalized quasiparticle effective masses~\cite{supplemental}. 
It can be argued on the basis of such measurements that the ratio is close to that $T_{\rm c}/T_{\rm F}=1/8$  required to be in the BCS-BEC crossover regime of a two-dimensional superconductor~\cite{hazra2019,supplemental}. Yet, given the increased effective mass renormalizations at low temperatures~\cite{ramshaw2015,michon2019} and various poorly understood phenomena such as the Fermi surface reconstruction~\cite{prevailingview,sebastian2012} and `Fermi arcs~\cite{keimer2015,supplemental},' 
it is unclear whether the parabolic band approximation upon which $T_{\rm F}$ estimates are based~\cite{hazra2019} is valid in the cuprates.

Studies aiming to address the question of whether a BCS-BEC crossover occurs in the cuprates~\cite{chen2014,chen2005,randeria2014} have instead focused on the pseudogap~\cite{timusk1999}, which is a partial gap in the fermionic density of states above $T_{\rm c}$. 
In a cold atomic Fermi gas, a pseudogap is reported to develop in the BEC-BCS crossover region~\cite{gaebler2010,magierski2011,perali2011,chin2004}, and is unambiguously the result of normal state pairing correlations~\cite{chen2014,chen2005,randeria2014,tsuchiya2009,jensen2020,richie2020,perali2011}. 
In the cuprates, the pseudogap is maximal in the antinodal region of momentum-space where the $d$-wave pairing gap is maximal~\cite{timusk1999}. But while pairing has been proposed as the origin of the pseudogap in the cuprates~\cite{timusk1999,chen2005,chen2014,hufner2008}, antiferromagnetic correlations and unconventional broken symmetry phases have also been proposed to produce a pseudogap~\cite{rice2012,chakravarty2001,varma2006,nie2014,lee2006,schmalian1999}.

\begin{figure}
\begin{center}
\includegraphics[width=0.9\linewidth]{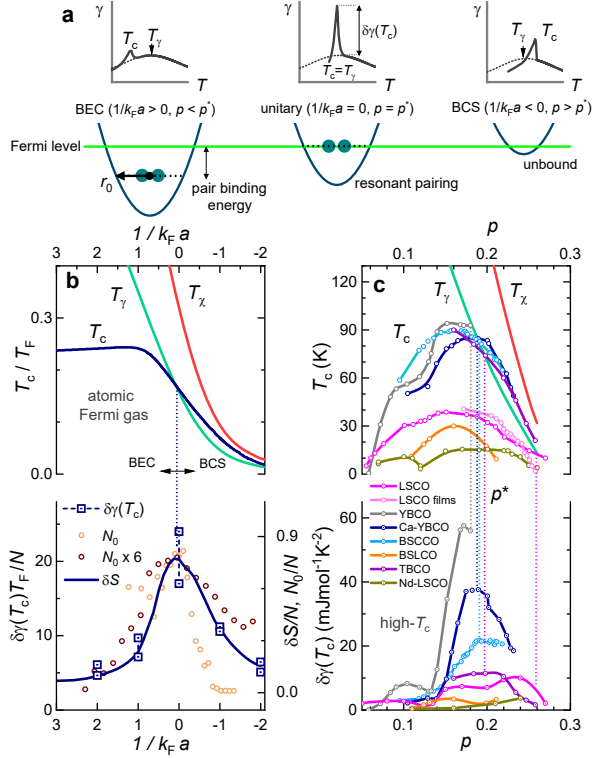}
\textsf{\caption{%{\bf Phase diagrams and evidence for a peak in $\delta\gamma(T_{\rm c})$}. 
{\bf a}, Schematic $\gamma$($T$) with (solid lines) and without (dotted lines) a phase transition. Because of the transition, the normal state maximum is visible only when $T_\gamma>T_{\rm c}$ (or $1/k_{\rm F}a>0$ or $p<p^\ast$). Also shown is a schematic of resonant pairing, occurring when the bound state energy coincides with the Fermi level~\cite{supplemental}, producing a sharp peak in $\delta\gamma(T_{\rm c})$. 
{\bf b}, Unitary regime of a Fermi gas~\cite{sademelo1993,haussmann2007}. Upper panel: $T_{\rm c}$ from Ref.~\cite{haussmann2007} and $T_\gamma=2\Delta/6.5k_{\rm B}$ and $T_\chi=2\Delta/3k_{\rm B}$ (using $\Delta$ at the lowest $T$ from Ref.~\cite{haussmann2007}). Lower panel: $\delta\gamma(T_{\rm c})$ (lower and upper bound estimates extracted~\cite{supplemental} from $S(T)$ in Fig.~5 of Ref.~\cite{haussmann2007}), $\delta S$ (from Fig.~6 of Ref.~\cite{haussmann2007}; this closely follows $\delta\gamma(T_{\rm c})$, providing a guide to the eye), 
and $N_0$ (brown~\cite{zwielein2004} and yellow~\cite{regal2004} circles). {\bf c}, Cuprates. Upper panel: $T_{\rm c}(p)$~\cite{michon2019,loram1998,wade1994,loram2001,loram1993,wen2009,mirmelstein1995,compositions,bozovic2016} and $T_\gamma$ and $T_\chi$ from Fig.~2d. Lower panel: $\delta\gamma(T_{\rm c})$;  
Spline fits connect points. In ({\bf b}) and ({\bf c}), dotted lines indicate $p=p^\ast$ (for each cuprate family~\cite{compositions}) and $1/k_{\rm F}a=0$, at which $T_\gamma=T_{\rm c}$, coinciding approximately with peaks in $\delta\gamma(T_{\rm c})$. 
}
\label{resonance}}
\end{center}
\vspace{-0.7cm}
\end{figure}

In this paper, we show that the key to establishing a universal thermodynamic signature of the BCS-BEC crossover, %
is the identification of a magic gap ratio~\cite{carbotte1990,inosov2011} $2\Delta/k_{\rm B}T_{\rm c}\approx$ 6.5 at which paired fermion condensates become optimally robust~\cite{randeria2014}; throughout, we use $\Delta$ to refer to the magnitude of the pairing gap at low $T$~\cite{schirotzek2008,chin2004,haussmann2007}. 
At this gap ratio, corresponding to the unitary point in a cold atomic Fermi gas, experimental indicators of a robust condensate exhibit a sharp peak.  These include the condensate fraction $N_0$ and the height of the jump $\delta\gamma(T_{\rm c})$ in the fermionic (or electronic) contribution $C=\gamma T$ to the specific heat at $T_{\rm c}$ (see schematic in Fig.~\ref{resonance}a). 
In the cuprates, we find $\delta\gamma(T_{\rm c})$ to be peaked at the magic gap ratio when $\Delta$ corresponds to the antinodal gap~\cite{cuprategap}. 
Reinforcing its interpretation as the pairing gap~\cite{timusk1999,chen2005,chen2014,hufner2008}, we find (i) nearly identical asymmetric line shapes of $\delta\gamma(T_{\rm c})$ versus $2\Delta/k_{\rm B}T_{\rm c}$ in the cuprates as for the unitary regime of a Fermi gas and (ii)
coincidence of the peak in $\delta\gamma(T_{\rm c})$ with a normal state maximum in $\gamma$. The latter, along with an accompanying maximum in the spin susceptibility $\chi$, can be understood as a signature of normal state pair amplitude fluctuations. 

In the unitary regime of a Fermi gas, corresponding to $1\gtrsim1/k_{\rm F}a\gtrsim-1$ in Figs.~\ref{resonance}a and b, continuous tuning of the pairing interactions through the crossover occurs by way of the dimensionless parameter $1/k_{\rm F}a~$\cite{leggett1980}, where $a$ is the pair scattering length and $k_{\rm F}$ is the Fermi radius. 
The BCS side~\cite{bardeen1957} corresponds to $k_{\rm F}a<0$, while the BEC side corresponds to $k_{\rm F}a>0$. 
The divergence in the elastic scattering cross section at $1/k_{\rm F}a=0$, which defines the location of the unitary point, causes the condensate to become optimally robust. This leads to peaks in $\delta\gamma(T_{\rm c})$ and in the entropy change $\delta S$ accompanying condensation at $T_{\rm c}$~\cite{haussmann2007,supplemental} (see Fig.~\ref{resonance}b). 
An optimally robust condensate is confirmed experimentally by the observation of a peak in $N_0$ as a function of $1/k_{\rm F}a$ (see Fig.~\ref{resonance}b)~\cite{regal2004,zwielein2004,supplemental} and a maximally large $\delta\gamma(T_{\rm c})$~\cite{ku2012,zwerger2016}, which also occurs at the value of $T_{\rm c}$ predicted by theory~\cite{haussmann2007}.

Turning to the cuprates in Fig.~\ref{resonance}, the measured $\delta\gamma(T_{\rm c})$ changes by as much as a factor of $\sim$~30 in YBCO~\cite{michon2019,loram1998,wade1994,loram2001,loram1993,wen2009,mirmelstein1995,compositions}. 
This change is far larger than the variations in $\delta\gamma(T_{\rm c})$ that are ordinarily explained by Eliashberg theory in regular BCS systems~\cite{carbotte1990,inosov2011,supplemental}, or have been predicted in various strongly coupled pairing models of the cuprates~\cite{curty2003,banerjee2011,noat2021}. The $\delta\gamma(T_{\rm c})$ curves do, however, exhibit maxima as a function of $p$ resembling the behavior as a function of $1/k_{\rm F}a$ in the unitary regime of a Fermi gas in Fig.~\ref{resonance}b.

\begin{figure}
\begin{center}
\includegraphics[width=0.9\linewidth]{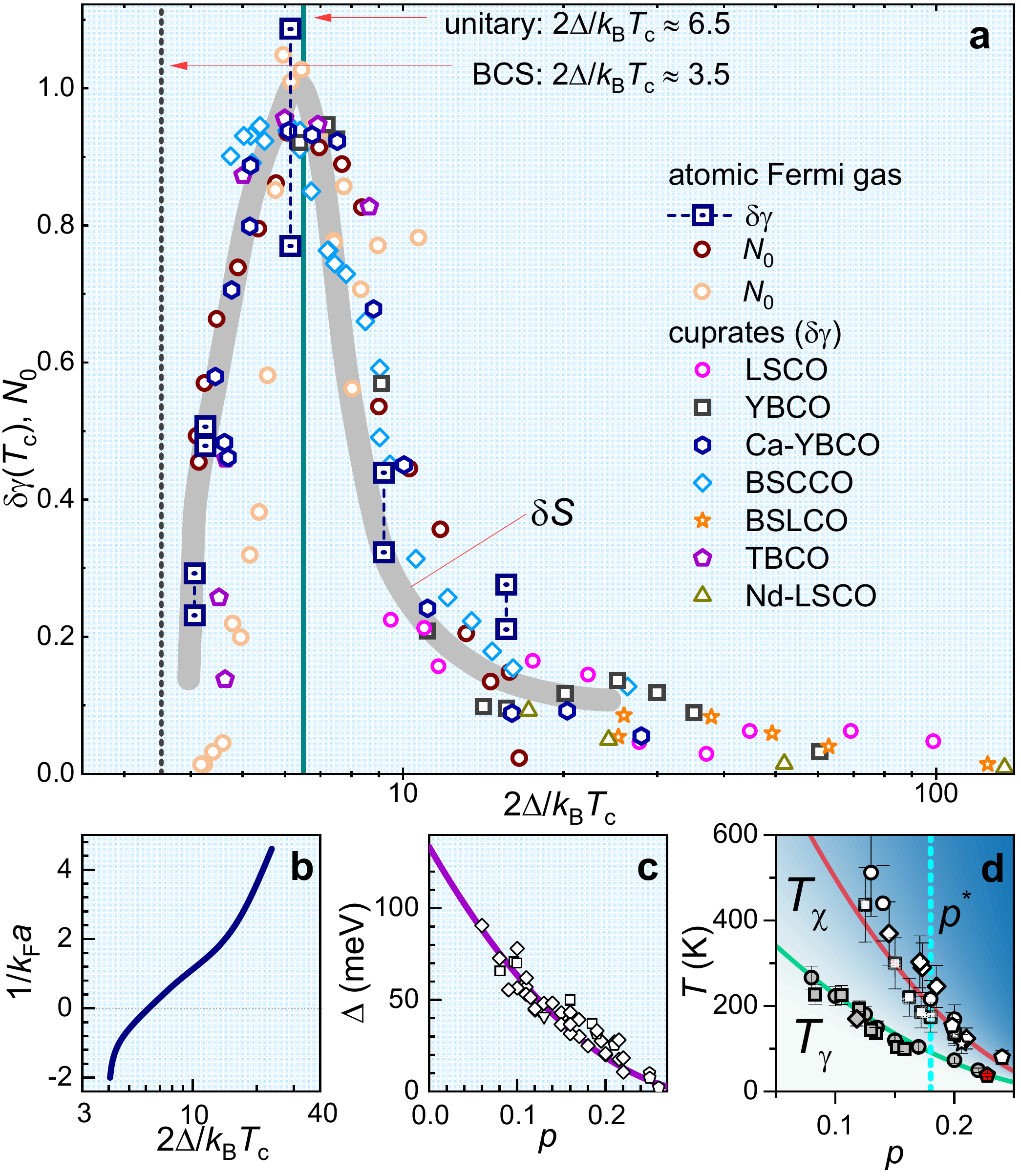}
\textsf{\caption{%{\bf Evidence for resonance condensation.}
{\bf a}, $\delta\gamma(T_{\rm c})$, $N_0$ and $\delta S$ (rescaled to unity from Figs.~\ref{resonance}b and c) versus $2\Delta/k_{\rm B}T_{\rm c}$. 
{\bf b}, $1/k_{\rm F}a$ versus $2\Delta/k_{\rm B}T_{\rm c}$~\cite{haussmann2007}. 
{\bf c}, Spectroscopic and thermal antinodal gap measurements~\cite{sutherland2003,hufner2008,mukhopadhyay2019}.  %
{\bf d}, Maxima in $\gamma$ (grey symbols) and $\chi$ (white symbols) from the raw data~\cite{supplemental}. $T_\gamma$ (green line) is a polynomial fit to the grey symbols~\cite{polynomial}, from which we obtain $\Delta=6.5k_{\rm B}T_\gamma/2$ (i.e. the purple line in ({\bf d})) and $T_\chi=2\Delta/3k_{\rm B}$ (red line). Symbol shapes identify the cuprate family in ({\bf a}), ({\bf c}) and ({\bf d}). The down triangle in ({\bf c}) refers to HBCO~\cite{compositions}.
}
\label{gap}}
\end{center}
\vspace{-0.7cm}
\end{figure}

The similar behavior of the cuprates to the unitary regime of a Fermi gas becomes clear 
once the data from Figs.~\ref{resonance}b and~c are replotted on the same $2\Delta/k_{\rm B}T_{\rm c}$ axis in Fig.~\ref{gap}a. While $2\Delta/k_{\rm B}T_{\rm c}$ is not a tuning parameter, it has the advantage in that it can be determined in both systems. 
In a cold atomic Fermi gas, there exists a direct correspondence between $1/k_{\rm F}a$ and $2\Delta/k_{\rm B}T_{\rm c}$~\cite{haussmann2007,moshe2019} (see Fig.~\ref{gap}b). 
Studies of the unitary regime differ on the precise values of $\Delta$ and $T_{\rm c}$ at the unitary point~\cite{chen2005,randeria2014,chen2014,sademelo1993,randeria1989,bloch2008,giorgini2008,drechsler1992,pisani2018}. However, they are found to be consistent with respect to the ratio $2\Delta/k_{\rm B}T_{\rm c}=$~6.5~$\pm$~0.2~\cite{supplemental} (see for example Fig.~9 of Ref.~\cite{moshe2019} and Table 1 of Ref.~\cite{randeria2014}), indicating this magic gap ratio to be a robust property of such a point.

Various non cuprate superconductors, including classic BCS~\cite{carbotte1990} and iron-based systems~\cite{inosov2011}, while spanning comparatively limited ranges in $2\Delta/k_{\rm B}T_{\rm c}$, are found to exhibit trends in $\delta\gamma(T_{\rm c})/\bar{\gamma}$ versus $2\Delta/k_{\rm B}T_{\rm c}$ consistent with Fig.~\ref{gap}a~\cite{supplemental}. In these systems, dividing by an assumed constant Sommerfeld coefficient $\bar{\gamma}$ enables universal trends in $\delta\gamma(T_{\rm c})$ to be established~\cite{supplemental} for materials with different electronic structures. 
The iron-based superconductors with the highest $\delta\gamma(T_{\rm c})/\bar{\gamma}$ values are found to have gap ratios consistent with the magic value. Of these, iron selenide has also recently been reported to exhibit a BCS-BEC crossover~\cite{kasahara2014,rinott2017} --- albeit without accompanying measurements of $\delta\gamma(T_{\rm c})$. 
A similar gap ratio at unitarity is further reported in gated layered superconductors~\cite{nakagawa2021}.

The asymmetric line shape in Fig.~\ref{gap}a can be understood to result from the fact that the gap ratio has a hard cutoff on the left-hand side at a value similar to that $\approx$~3.5 of an ideal BCS superconductor~\cite{bardeen1957}, while there is no cutoff on the BEC side~\cite{chen2005,randeria2014,chen2014,sademelo1993,randeria1989,bloch2008,giorgini2008,drechsler1992}.  
On the BCS side, $\delta\gamma(T_{\rm c})$ increases with $2\Delta/k_{\rm B}T_{\rm c}$ similarly to that in Eliashberg theory~\cite{carbotte1990,supplemental}, while on the BEC side, $T_{\rm c}$ and consequently $\delta\gamma(T_{\rm c})$ are limited by the phase stiffness of the condensate~\cite{emery1995}. 
We find precisely this line shape in the cuprates when $\Delta$ (purple line in Fig.~\ref{gap}c)~\cite{polynomial,supplemental} corresponds to the antinodal gap dominating spectroscopic and thermodynamic measurements (symbols in Fig.~\ref{gap}c)~\cite{hufner2008,sutherland2003,mukhopadhyay2019}. 
The same asymmetric behavior is displayed for multiple cuprate families~\cite{michon2019,loram1998,wade1994,loram2001,loram1993,wen2009,mirmelstein1995}.
On averaging the values of $2\Delta/k_{\rm B}T_{\rm c}$ in Fig.~\ref{gap}a at which $\delta\gamma(T_{\rm c})$ is peaked (near $p\approx$~0.2 in Fig.~\ref{resonance}c) for the higher $T_{\rm c}$ cuprates (YBCO, Ca-YBCO, BSCCO and TBCO~\cite{compositions}), we obtain $2\Delta/k_{\rm B}T_{\rm c}=$~6.4~$\pm$~0.3, which is the same within experimental uncertainty as for a unitary Fermi gas. 
Validity of the universal magic gap ratio is therefore strongly suggested in the cuprates.

\begin{figure}
\begin{center}
\includegraphics[width=0.9\linewidth]{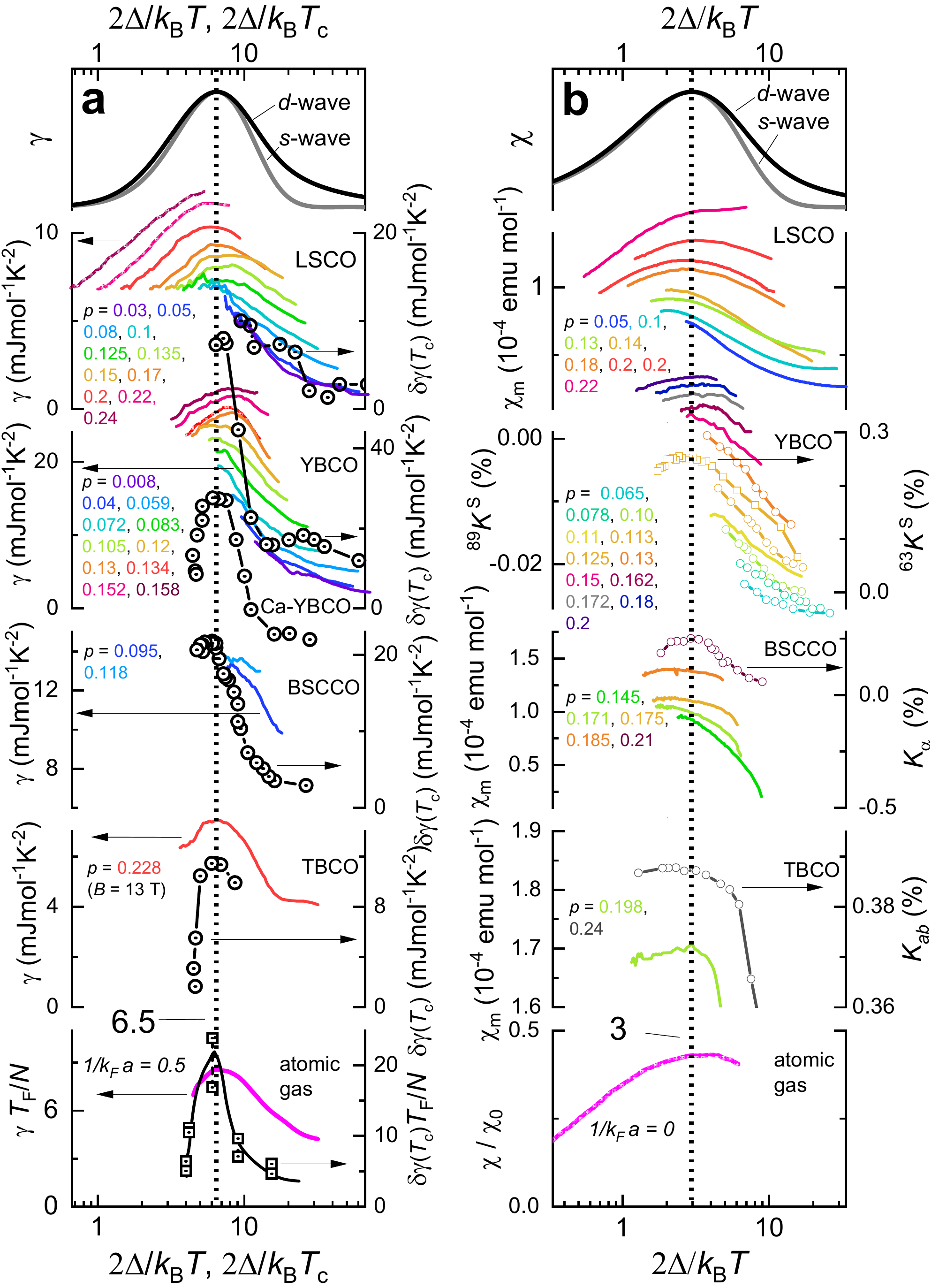}
\textsf{\caption{%s{\bf Electronic heat capacity and susceptibility for various cuprate compositions.} 
{\bf a}, Left-hand axes: $\gamma$ for $T>T_{\rm c}$~\cite{loram1993,loram2001,radcliffe1996,loram1994,chen2014,supplemental} (colored lines; including 2\% Zn substitution for the highest 2 YBCO dopings~\cite{loram1994})
versus $2\Delta/k_{\rm B}T$; $p$ and $1/k_{\rm F}a$ values are indicated throughout. Right-hand axes: $\delta\gamma(T_{\rm c})$ at $T_{\rm c}$ versus $2\Delta/k_{\rm B}T_{\rm c}$ for the cuprates~\cite{loram1998,michon2019,loram2001,loram1993,wade1994,mirmelstein1995} (center-dot circles; shifted by $\pm$~10mJmol$^{-1}K^{-2}$ for YBCO and Ca-YBCO) and a Fermi gas from Fig.~\ref{resonance} (center-dot squares).  
{\bf b}, $\chi_{\rm m}$~\cite{nakano1994,johnston1989,loram2001,kubo1991}, $K$~\cite{curro1997,alloul1989,alloul2016,crocker2011,vyaselev1994} and $\chi$~\cite{enss2012i} versus $2\Delta/k_{\rm B}T$~\cite{susceptibilitynote}. 
Included are $\gamma$ and $\chi$ for model $d$- (black) and $s$-wave (grey) gaps of magnitude $\Delta$~\cite{supplemental}. Some YBCO $K$ curves are shifted vertically for clarity and spline fits connect coarsely spaced points. Dotted lines indicate $2\Delta/k_{\rm B}T_{\rm c}=6.5$ and $2\Delta/k_{\rm B}T=3$.   
}
\label{data}}
\end{center}
\vspace{-0.7cm}
\end{figure}

The association of $\Delta$ with the antinodal gap in the cuprates is reinforced by thermodynamic evidence for pairing correlations in the normal state. 
In the cuprates, normal state pair amplitude fluctuations associated with the pseudogap have been proposed to account for maxima in $\gamma$ and $\chi$ as a function of $T$~\cite{curty2003,banerjee2011,noat2021,engelbrecht1998}. 
Pair amplitude fluctuations in the unitary and BEC regimes of a Fermi gas also produce normal state maxima in $\gamma$ (or $C=\gamma T$)~\cite{wyk2016,supplemental} and $\chi$~\cite{tajima2014,jensen2020,enss2012i,richie2020}. 
Figure~\ref{data} shows that on plotting $\gamma$~\cite{loram1994,loram2001} and $\chi$~\cite{nakano1994,johnston1989,alloul2016,loram2001,alloul1989,curro1997,kubo1991,crocker2011,vyaselev1994} versus $2\Delta/k_{\rm B}T$, maxima in $\gamma$ and $\chi$ emerge as ubiquitous properties of the normal state 
(in the cuprates, the shape of $\chi$ versus $T$ is provided by magnetic susceptibility $\chi_{\rm m}$ and nuclear magnetic resonance Knight shift $K$ measurements~\cite{susceptibilitynote}). 
The model $\gamma$ and $\chi$ curves (black and grey in Fig.~\ref{data}) produced by an excitation gap of width $\Delta$~\cite{supplemental,cuprategap} exhibit maxima at $T_\gamma\approx2\Delta/6.5k_{\rm B}$ and $T_\chi\approx2\Delta/3k_{\rm B}$. 
A pairing pseudogap~\cite{gapnote} is strongly suggested in the cuprates by the consistency of the observed maxima in Fig.~\ref{data} with $T_\gamma$ and $T_\chi$. 
In fact, we find overall consistency between each of the $\Delta(p)$, $T_\gamma(p)$ and $T_\chi(p)$ curves and the experimental data points for the antinodal gap and maxima in $\gamma$ and $\chi$~\cite{supplemental} in Figs.~\ref{gap}c and d. Thermodynamic and spectroscopic measurements can therefore both be understood in terms of a $\Delta(p)$ that is approximately the same for all cuprate families, regardless of their optimal $T_{\rm c}$.

A direct association of the normal state maxima with pairing amplitude fluctuations is strongly suggested by the alignment of the maxima in $\gamma$ with the peaks in $\delta\gamma(T_{\rm c})$ when $\gamma$ and $\delta\gamma(T_{\rm c})$ are respectively plotted versus $2\Delta/k_{\rm B}T$ and $2\Delta/k_{\rm B}T_{\rm c}$ in Fig.~\ref{data}a. 
Since $2\Delta/k_{\rm B}T$ and $2\Delta/k_{\rm B}T_{\rm c}$ are both scaled by $\Delta$, the alignments of $\gamma$ and $\delta\gamma(T_{\rm c})$ are independent of any experimental uncertainties in the functional form of $\Delta(p)$~\cite{polynomial}. 
We find the alignments to originate from $\delta\gamma(T_{\rm c})$ being peaked close to the points of intersection of $T_{\rm c}$ with $T_\gamma$ (see Figs.~\ref{resonance}b and c), corresponding to $1/k_{\rm F}a=0$ (i.e. the unitary point) in a unitary Fermi gas and a characteristic doping $p=p^\ast$ in the cuprates.

In the unitary regime of a Fermi gas, $\delta\gamma(T_{\rm c})$ exhibiting a strong peak at $T_{\rm c}=T_\gamma$ can be understood as a consequence of the heavily broadened pseudogap transitioning into a regular pairing gap~\cite{gapnote} as long range phase coherence is established below $T_{\rm c}$~\cite{chen2000,chen2001}. The entropy change contributing to $\delta\gamma(T_{\rm c})$ is naturally largest when $T_{\rm c}$ coincides with the maximum in $\gamma$ resulting from excitations across $\Delta$. This is therefore suggested also to occur in the cuprates at $T_{\rm c}=T_\gamma$~\cite{supplemental}. 
$\delta\gamma(T_{\rm c})$ exhibiting a strong peak at $T_{\rm c}=T_\gamma$ can also be understood as a consequence of the normal state entropy $S_{\rm n}$ at $T_{\rm c}$ (in addition to $\delta S$) exhibiting a maximum (as a function of $1/k_{\rm F}a$) close to this point, owing to this region of the normal state consisting of a maximally disordered mixture of a bosonic and fermionic degrees of freedom~\cite{wyk2016,haussmann2007}.
At $T>T_{\rm c}$, a peak in $S_{\rm n}(1/k_{\rm F}a)$ is also seen to extend vertically in $T$ at the unitary point~\cite{haussmann2007,bloch2008}, with the loss of fermion degrees of freedom at $1/k_{\rm F}a>0$ leading to a sharp drop $S_{\rm n}$ on the BEC side of the phase diagram. An examination of $S_{\rm n}$ in several cuprates~\cite{cooper2000} reveals that this too exhibits a sharp peak that extends vertically in $T$ near $p^\ast$, accompanied by a drop in $S_{\rm n}$ at $p<p^\ast$.

One consequence of $\delta\gamma(T_{\rm c})$ being peaked close to the point of intersection of $T_\gamma$ and $T_{\rm c}$ is that $p^\ast$ is distinct from the hole doping $p\approx$ 0.16 at which $T_{\rm c}$ is optimal. In fact, $p^\ast$ moves towards the upper end of the superconducting dome as the optimal $T_{\rm c}$ is reduced, and appears to be accompanied by a strong suppression of the overall peak height of $\delta\gamma(T_{\rm c})$. %In 
In LSCO, for example, an extrapolation of $T_\gamma$ in Fig.~\ref{resonance}c suggests that $p^\ast\approx$~0.26~$\pm$~0.03, which is consistent with the higher value of $p=$~0.23~$\pm$~0.1 (compared e.g. to YBCO) at which $\delta\gamma(T_{\rm c})$ is peaked~\cite{supplemental} and the higher value of $p=$~0.24~$\pm$~0.01 (compared to Ca-YBCO) at which $S_{\rm n}$ is peaked~\cite{cooper2000}. In LSCO films, by contrast, $T_{\rm c}$ lies significantly below $T_\gamma$ in Fig.~\ref{resonance}c, suggesting that they do not exhibit a crossover into the BCS regime, as has also been suggested on the basis of the superfluid density measurements~\cite{bozovic2016,supplemental}. The tiny $p$-dependent $\delta\gamma(T_{\rm c})$ in Nd-LSCO, meanwhile, suggests that its peak value occurs at higher dopings than have been accessed experimentally~\cite{michon2019,ma2021}.

Given the prior reports of quantum criticality in the cuprates at similar hole dopings to $p^\ast$~\cite{giraldogallo2018,legros2019,cooper2009}, one intriguing possibility is that the BCS-BEC crossover and quantum criticality share a common origin. 
Indeed, some of the reported phenomenology of quantum criticality in the cuprates bears similarities to that of the unitary regime of a Fermi gas~\cite{zaanen2019}. This includes Plankian dissipation and scale invariance~\cite{zaanen2019,cao2010,enss2012ii,giraldogallo2018,legros2019,cooper2009}, and a minimum in the pair coherence length~\cite{engelbrecht1997,strinati2018,pistolesi1996} inferred from the maximum in the superconducting upper critical magnetic field~\cite{supplemental,grissonnanche2014,ramshaw2015,chan2020}. 
It should be noted, however, that thermodynamic evidence for quantum criticality in the form of a sharply increasing $\gamma$ or an upturn in the effective mass, has thus far only been reported at low temperatures ($T\ll$~10~K)~\cite{michon2019,ramshaw2015}, and has yet to be accompanied by evidence for a divergence in the correlation length of a broken symmetry phase~\cite{supplemental}.

\begin{acknowledgments}
Supported by the Department of Energy (DoE) BES project `Science of 100 tesla.' The National High Magnetic Field Laboratory is funded by NSF Cooperative Agreements DMR-1157490 and 1164477, the State of Florida and DoE.
We thank John Cooper, Mohit Randeria, Patrick Lee, Arkady Shekter, John Singleton, Greg Boebinger and Giancarlo Strinati for helpful discussions.
\end{acknowledgments}

\bibliographystyle{naturemag}

\bibliography{basename of .bib file}

\begin{thebibliography}{99}

%
%

\bibitem{bardeen1957} Bardeen, J., Cooper, L.~N., Schrieffer, J.~R. Theory of superconductivity. {\it Phys. Rev.} {\bf 108}, 1175-1204 (1957).

%

\bibitem{jochim2003} Jochim, S.,Bartenstein, M., Altmeyer, A ., Hendl, G., Riedl, S., Chin, C.., Denschlag, J.H., Grimm, R., Bose-Einstein condensation of molecules, {\it Science} {\bf 302}, 2101-2103 (2003). 

\bibitem{zwierlein2003} Zwierlein, M.W., Stan, C.A., Schunck, C.H., Raupach, S.M.F., Gupta, S., Hadzibabic, Z., Ketterle, W.,  Observation of Bose-Einstein condensation of molecules, {\it Phys. Rev. Lett.} {\bf 91}, 250401 (2003).

%

\bibitem{randeria1989} Randeria, M., Duan, J.-M., Shieh, L.-Y., Bound states, Cooper pairing, and Bose condensation in two dimensions, {\it Phys. Rev. Lett.} {\bf 62}, 981-984 (1989). 

\bibitem{friedberg1989} Friedberg, R., Lee, T. D., Gap energy and long-range order in the boson-fermion model of superconductivity. {\it Phys. Rev. B} {\bf 40}, 6745-6762 (1989). 

\bibitem{micnas1990} Micnas, R., Ranninger, J., Robaszkiewicz, S., Superconductivity in narrow-band systems with local nonretarded attractive interactions. {\it Rev. Mod. Phys.} {\bf 62}, 113-171 (1990).

\bibitem{drechsler1992} Drechsler, M., Zwerger, W., Crossover from BCS-superconductivity to Bose-condensation. {\it Ann. Physik} {\bf 1}, 15-23 (1992). 

\bibitem{sademelo1993} S\'{a} de Melo, C. A. R., Randeria, M., Engelbrecht, J. R., Crossover from BCS to Bose superconductivity: transition temperature and time-dependent Ginzburg-Landau theory. {\it Phys. Rev. Lett.} {\bf 71}, 3202-3205 (1993).

%

\bibitem{bloch2008} Bloch, I., Dalibard, J., Zwerger, W., Many-body physics with ultracold gases. {\it Rev. Mod. Phys.} {\bf 80}, 885-964 (2008).

\bibitem{giorgini2008} Giorgini, S., Pitaevskii, L. P., Stringari, S., Theory of ultracold atomic Fermi gases. {\it Rev. Mod. Phys.} {\bf 80}, 1215-1274 (2008).

%

\bibitem{ku2012} Ku, M. J. H., Sommer, A. T., Cheuk, L. W., Zwierlein, M. W.,  Revealing the superfluid Lambda transition in the universal thermodynamics of a unitary Fermi gas. {\it Science} {\bf 335}, 563-567 (2012).

\bibitem{regal2004} Regal, C. A., Greiner, M., Jin, D. S., Observation of resonance condensation of fermionic atom pairs. {\it Phys. Rev. Lett.} {\bf 92}, 040403 (2004).

\bibitem{zwielein2004} Zwierlein, M. W., Stan, C. A., Schunck, C. H., Raupach, S. M. F., Kerman, A. J., Ketterle, W., Condensation of pairs of Fermionic atoms near a Feshbach resonance. {\it Phys. Rev. Lett.} {\bf 92}, 120403 (2004). 

%

\bibitem{haussmann2007} Haussmann, R., Rantner, W., Cerrito, S., Zwerger, W., Thermodynamics of the BCS-BEC crossover. {\it Phys. Rev. A} {\bf 75}, 023610 (2007).

%

\bibitem{chen2005} Chen, Q., Stajic, J., Tan, S., Levin, K. BCS–BEC crossover: From high temperature superconductors to ultracold superfluids. {\it Phys. Reports} {\bf 412}, 1-88 (2005).

\bibitem{randeria2014} Randeria, M., Taylor, E., Crossover from Bardeen-Cooper-Schrieffer to Bose-Einstein condensation and the unitary Fermi gas. {\it Annu. Rev. Condens. Matter Phys.} {\bf 5}, 209-232 (2014). 

\bibitem{strinati2018} Strinati, G. C., Pieri, P. P., R\"{o}pke, G., Schuck, P., Urban, M. The BCS–BEC crossover: From ultra-cold Fermi gases to nuclear systems. {\it Phys. Rep.} {\bf 738}, 1-76 (2018).

\bibitem{hazra2019} Hazra, T., Verma, N., Randeria, M., Bounds on the superconducting transition temperature: applications to twisted bilayer graphene and cold atoms. {\it Physical Review X} {\bf 9}, 031049 (2019).

\bibitem{park2021} Park, J.~M., Cao, Y., Watanabe, K., Taniguchi, T., Jarillo-Herrero, P., Tunable strongly coupled superconductivity in magic-angle twisted trilayer graphene. {\it Nature} {\bf 590}, 249-255 (2021).

\bibitem{kasahara2014} Kasahara, S., Watashige, T., Hanaguri, T., Kohsaka, Y., Yamashita, T., Shimoyama, Y., Mizukami, Y., Endo, R., Ikeda, H., Aoyama, K., Terashima, T., Uji, S., Wolf, T., von Lohneysen, H., Shibauchi, T., Matsuda, Y. Field-induced superconducting phase of FeSe in the BCS-BEC cross-over. {\it Proc. Nat. Acad. Sci. USA} {\bf 111}, 116309-16313 (2014).

\bibitem{rinott2017} Rinott, S., Chashka, K. B., Ribak, A., Rienks, E. D. L., Taleb-Ibrahimi, A., Le Fevre, P., Bertran, F., Randeria, M., Kanigel, A., Tuning across the BCS-BEC crossover in the multiband superconductor Fe$_{1+y}$Se$_x$Te$_{1-x}$: An angle-resolved photoemission study. {\it Sci. Adv.} {\bf 3}, 1602372 (2017).

%%%%%%%%%%%%%%%%%%%%%

\bibitem{uemura1989} Uemura, Y. J., Luke, G. M., Sternlieb, Brewer, J. H., Carolan, J. F., Hardy, W. N., Kadono, R., Kempton, J. R., Kiefl, R. F., Kreitzman, S. R., Mulhern, P., Riseman, T. M., Williams, D. L., Yang, B. X., Uchida, S., Takagi, H., Gopalakrishnan, J., Sleight, A. W., Subamanian, M. A., Chien, C. L., Cieolak, M. Z., Xiao, G., Lee, V. Y., Statt, N. W., Stronach, C. E., Kossler, W. J., Yu, X. H., Universal correlations between $T_{\rm c}$ and $n_{\rm s}/m^\ast$ (carrier density over effective mass) in high-$T_{\rm c}$ cuprate superconductors. {\it Phys. Rev. Lett.} {\bf 62}, 2317-2320 (1989).

\bibitem{emery1995} Emery, V.J., Kivelson, S.A., Importance of phase fluctuations in superconductors with small superfluid density. {\it Nature} {\bf 374}, 434-437 (1995).

\bibitem{li2010} Li, L., Wang, Y., Komiya,. S., Ono, S., Ando, Y., Gu, G.D., Ong, N.P. Diamagnetism and Cooper pairing above $T_{\rm c}$ in cuprates. {\it Phys. Rev. B} {\bf 81}, 054510 (2010).

\bibitem{dubroka2011} Dubroka, A., R\"{o}ssle, M., Kim, K.W., Malik, V.K., Munzar, D., Basov, D.N., Scafgans, A.A., Moon, S.J., Lin, C.T., Haug, D., Hinkov, V., Keimer, B., Wolf, Th., Storey, J.G., Tallon, J.L., Bernhard, C., Evidence of a precursor superconducting phase at temperatures as high as 180~K in $R$Ba$_2$Cu$_3$O$_{7-\delta}$ ($R=$~Y, Gd, Eu) superconducting crystals from infrared spectroscopy. {\it Phys. Rev. Lett.} {\bf 106}, 047006 (2011).

\bibitem{hu2014} Hu, W., Kaiser, S., Nicoletti, D., Hunt, C.R., Gierz, I., Homann, M.C., Le Tacon, M., Loew, T., Keimer, B., Cavalleri, A., Optically enhanced coherent transport in YBa$_2$Cu$_3$O$_{6.5}$ by ultrafast redistribution of interlayer coupling. {\it Nature Materials} {\bf 13}, 705-711 (2014).

\bibitem{kaiser2014} Kaiser, S., Hunt, C.R., Nicoletti, D., Hu, W., Gierz, I., Liu, H.Y., Le Tacon, M., Loew, T., Haug, D., Keimer, B., Cavalleri, A., Optically induced coherent transport far above $T_c$ in underdoped YBa$_2$Cu$_3$O$_{6+\delta}$. {\it Phys. Rev. B} {\bf 89}, 184516 (2014).

\bibitem{zhou2019} Zhou, P., Chen, L., Liu, Y., Sochnikov, I., Bollinger, A. T., Han, M.-G., Zhu, Y., He, X., Bo\u{z}ovi\'{c}, I., Natelson, D., Electron pairing in the pseudogap state revealed by shot noise in copper oxide junctions. {\it Nature} {\bf 572}, 493-496 (2019).

%

\bibitem{supplemental} See Supplementary Information,
which discusses data pertaining to $\delta\gamma(T_{\rm c})$ in conventional BCS and Fe- and Ni based superconductors; information about cuprate hole dopings $p$ used in constructing graphs; raw $\gamma$ and $\chi$ data including the locations of $T_\gamma$ and $T_\chi$; Knight shift data on BLSCO; a discussion pertaining to prior modeling of $\delta\gamma(T_{\rm c})$ versus $p$ by Chen {\it et al.}; details concerning cold atomic gas data used; estimates of the gap ratio at the unitary point of a cold atomic gas; a discussion of the thermodynamics of maxima in $\gamma$ and $\chi$; coherence length estimates in various cuprates based on $H_{\rm c2}$; estimates of the Fermi energy in the cuprates; a discussion of the higher values of $p^\ast$ in cuprates with lower $T_{\rm c}$'s; the origin of `Fermi arcs;' the $T$-dependence of the antinodal gap; lifetime effects on $N_0$; problems with a smaller gap option for $\Delta$ in the cuprates;  a hidden maximum in the heat capacity of a cold atomic Fermi gas; reports of peaks in $\delta\gamma(T_{\rm c})$, $\gamma$ and $m^\ast$ associated with quantum criticality, and includes Refs.~\cite{radcliffe1996,liang2006,zhou1996,tallon1995,meingast2009,kawasaki2010,kittel2004,nolthing2009,suzuki1991,rourke2010,leboeuf2007,rullieralbenque2008,fukuzumi1996,nachumi1996,takagi1989,carrington1994,chubukov2007,demello2012,fradkin2015,agterberg2020,ding1996,renner1998,norman2014,park2008,leblanc2009}.






%supplemetary references
%%%%%%%%%%%%%%%%%%%%%%
%%%%%%%%%%%%%%%%%%%%%%
%%%%%%%%%%%%%%%%%%%%%%
%%%%%%%%%%%%%%%%%%%%%%
%%%%%%%%%%%%%%%%%%%%%%
%%%%%%%%%%%%%%%%%%%%%%
%%%%%%%%%%%%%%%%%%%%%%
%%%%%%%%%%%%%%%%%%%%%%
%%%%%%%%%%%%%%%%%%%%%%
%%%%%%%%%%%%%%%%%%%%%%

\bibitem{radcliffe1996} Radcliffe, J.W., Loram, J.W., Wade, J.M., Wltschek, G., Tallon, J.W., Electronic specific heat of overdoped TI$_2$Ba$_2$CuO$_{6+\delta}$ in a magnetic field. {\it J. Low. Temp. Phys.} {\bf 105}, 903-908 (1996).

%
\bibitem{takagi1989} Takagi, H., Ido, T., Ishibashi, S., Uota, M., Uchida, S., Tokura, Y., Superconductor-to-nonsuperconductor transition in (La$_{1-x}$Sr$_x$)$_2$CuO$_4$ as investigated by transport and magnetic measurements. {\it Phys. Rev. B} {\bf 40}, 2254-2261 (1989).

%

\bibitem{liang2006} Liang, R., Bonn, D. A., Hardy, W. N. Evaluation of CuO$_2$ plane hole doping in YBa$_2$Cu$_3$O$_{6+x}$ single crystals. {\it Phys. Rev. B} {\bf 73}, 180505 (2006).

%

\bibitem{zhou1996} Zhou, J.-S., Goodenough, J.B., Dabrowski, B., Rogacki, K., 
Transport properties of a YBa$_2$Cu$_4$O$_8$ crystal under high pressure. {\it Phys. Rev. Lett.} {\bf 77}, 4253-4256 (1996)

%

\bibitem{tallon1995} Tallon, J.L., Bernhard, C., Shaked, H., Hitterman, R.L., Jorgensen, J.D., Generic superconducting phase-behavior in high-$T_{\rm c}$ variation with hole concentration in YBa$_2$Cu$_3$O$_{7-\delta}$, {\it Phys. Rev. B} {\bf 51}, 12911-12914 (1995). 

%

\bibitem{meingast2009} Meingast, C., Inaba, A., Heid, R., Pankoke, V., Bohnen, K.-P., Reichardt, W., Wolf, T., Specific-heat of YBa$_2$Cu$_3$O$_x$ up to 400~K: high-resolution adiabatic measurements and {\it Ab-initio} LDA phonon calculations. {\it J. Phys. Soc. Japan} {\it 78}, 074706 (2009).

%

\bibitem{kawasaki2010} Kawasaki, S., Lin, C., Kuhns, P.L., Reyes, A.P., Zheng, G.-Q., Carrier-concentration dependence of the pseudogap ground state of superconducting Bi$_2$Sr$_{2-x}$La$_x$CuO$_{6+\delta}$ revealed by $^{63,65}$Cu-nuclear magnetic resonance in very high magnetic fields. {\it Phys. Rev. Lett.} {\bf 105}, 137002 (2010).

%

\bibitem{kittel2004} Kittel, C.,  Introduction to Solid State Physics, eighth ed. (Wiley, New York, 2004).

\bibitem{nolthing2009} Nolthing, W., Ramakanth, A., Quantum Theory of Magnetism (Springer, New York, 2009).

%

\bibitem{suzuki1991} Suzuki M., Hikita, M., Resistive transition, magnetoresistance, and anisotropy in La$_{2-x}$Sr$_x$CuO$_4$ single-crystal thin films. {\it Phys. Rev. B} {\bf 44}, 249-261 (1991).

%

\bibitem{rourke2010} Rourke, P. M. C., Bangura, A. F., Benseman, T. M., Matusiak, M., Cooper, J. R., Carrington, A., Hussey, N. E., A detailed de Haas–van Alphen effect study of the overdoped cuprate Tl$_2$Ba$_2$CuO$_{6+\delta}$, {\it New J. Phys.} {\bf 12}, 105009 (2010).

%

\bibitem{leboeuf2007} LeBoeuf, D., Doiron-Leyraud, N., Levallois, J., Daou, R., Bonnemaison, J. B., Hussey, N. E., Balicas, L., Ramshaw, B. J., Liang, R. X., Bonn, D. A., Hardy, W. N., Adachi, S., Proust, C., Taillefer, L. Electron pockets in the Fermi surface of hole-doped high-$T_{\rm c}$ superconductors. {\it Nature} {\bf 450}, 533-536 (2007).

%

\bibitem{rullieralbenque2008} Rullier-Albenque, F., Alloul, H., Balakirev, F., Proust, C., Disorder, metal-insulator crossover and phase diagram in high-$T_{\rm c}$ cuprates. {\it EPL} {\bf 81}, 37008 (2008).

\bibitem{fukuzumi1996} Fukuzumi, Y., Mizuhashi, K., Takenaka, K., Uchida, S., Universal superconductor-insulator transition and $T_{\rm c}$ depression in Zn-substituted high-$T_{\rm c}$ cuprates in the underdoped regime. {\it Phys. Rev. Lett.} {\bf 76}, 684-687 (1996).

\bibitem{nachumi1996} Nachumi, B., Keren, A., Kojima, K., Larkin, M., Luke, G. M., Merrin, J., Tchernysh\"{o}v, O., Uemura, Y. J., Ichikawa, N., Goto, M., Uchida, S., Muon spin relaxation studies of Zn-substitution effects in high-$T_{\rm c}$ cuprate superconductors. {\it Phys. Rev. Lett.} {\bf 77}, 5421-5424 (1996).

\bibitem{carrington1994} Carrington, A., Mackenzie, A. P., Sinclair, D. C., Cooper, J. R., Field dependence of the resistive transition in Tl$_2$Ba$_2$CuO$_{6+\delta}$. {\it Phys. Rev. B} {\bf 49}, 13243-13246 (1994).

\bibitem{chubukov2007} Chubukov, A.V., Norman, M.R., Millis, A.J., Abrahams, E., Gapless pairing and the Fermi arc in the cuprates. {\it Phys. Rev. B} {\bf 76}, 180501 (2007).

\bibitem{demello2012} de~Mello, E.V.L., Disordered-based theory of pseudogap, superconducting gap, and Fermi arc of cuprates. {\it Euro. Phys. Lett.} {\bf 99}, 37003 (2012).


\bibitem{fradkin2015} Fradkin, E., Kivelson, S.A., Tranquada, J.M., Colloquium: Theory of intertwined orders in high temperature superconductors. {\it Rev. Mod. Phys.} {\bf 87}, 457-482 (2015).

\bibitem{agterberg2020} Agterberg, D.F., Davis, J.C.S., Edkins, S.D., Fradkin, E.,Van~Harlingen, D.J., Kivelson, S.A., Lee, P.A., Radzihovsky, L., The physics of pair-density waves: cuprate superconductors and beyond. {\it Annu. Rev. Condens. Matter Phys.} {\bf 11}, 231–270 (2020).

\bibitem{ding1996} Ding, H., Campuzano, J.C., Takahashi, T., Randeria, M., Norman, M.R., Mochikull, T., Kadowaki, K., Giapintzaki, J., Spectroscopic evidence for a pseudogap in the normal state of underdoped high-$T_{\rm c}$ superconductors. {\it Nature} {\bf 382}, 51-54 (1996).

\bibitem{renner1998} Renner, Ch., Revaz, B., Genoud, J.-Y., Kadowaki, K., Fischer, \O., Pseudogap precursor of the superconducting gap in under- and overdoped Bi$_2$Sr$_2$CaCu$_2$O$_{8+\delta}$. {\it Phys. Rev. Lett.} {\bf 80}, 149-152 (1998).

\bibitem{norman2014} Mishra, V., Chatterjee, U., Campuzano, J. C., Norman, M. R., Effect of the pseudogap on the transition temperature in the cuprates and implications for its origin. {\it Nature Phys.} {\bf 10}, 357-360 (2014).

\bibitem{park2008} Park, T., Graf, M. J., Boulaevskii, L., Sarrao, J. L., Thompson, J. D. Electronic duality in strongly correlated matter. {\it Proc. Nat. Acad. Sci. USA} {\bf 105}, 6825-6828 (2008).

\bibitem{leblanc2009} LeBlanc, J.P.F., Nicol, E.J., Carbotte, J.P., Specific heat of underdoped cuprates: Resonating valence bond description versus Fermi arcs. {\it Phys. Rev. B} {\bf 80}, 060505 (2009).

%%%%%%%%%%%%%%%%%%%%%%
%%%%%%%%%%%%%%%%%%%%%%
%%%%%%%%%%%%%%%%%%%%%%
%%%%%%%%%%%%%%%%%%%%%%
%%%%%%%%%%%%%%%%%%%%%%
%%%%%%%%%%%%%%%%%%%%%%
%%%%%%%%%%%%%%%%%%%%%%
%%%%%%%%%%%%%%%%%%%%%%
%%%%%%%%%%%%%%%%%%%%%%
%%%%%%%%%%%%%%%%%%%%%%
%%%%%%%%%%%%%%%%%%%%%%

%

\bibitem{ramshaw2015} Ramshaw, B. J., Sebastian, S. E., McDonald, R. D.,  Day, J., Tan, B. S., Zhu, Z., Betts, J.B., Liang, R.-X., Bonn, D.A., Hardy, W.N., Harrison, N., Quasiparticle mass enhancement approaching optimal doping in a high-$T_{\rm c}$ superconductor. {\it Science} {\bf 348}, 317-320 (2015).

\bibitem{michon2019} Michon, B., Girod, C., Badoux, S., Ka\u{c}mar\u{c}\'{i}k, J., Ma, Q., Dragomir, M., Dabkowska, H. A., Gaulin, B.D., Zhou, J.-S., Pyon, S., Takayama, T., Takagi, H., Verret, S., Doiron-Leyraud, N., Marcenat, C., Taillefer, L., Klein, T., Thermodynamic signatures of quantum criticality in cuprate superconductors. {\it Nature} {\bf 567}, 218-222 (2019). 

%

\bibitem{sebastian2012} Sebastian, S.E., Harrison, N., Lonzarich, G.G. Towards resolution of the Fermi surface in underdoped high-$T_{\rm c}$ superconductors. {\it Rep. Prog. Phys.} {\bf 75}, 102501 (2012).

\bibitem{prevailingview} The prevailing view~\cite{keimer2015} is that the Fermi surface reconstruction producing the pockets, for instance by a charge density wave~\cite{hucker2014,blancocanosa2014}, cannot by itself account for the pseudogap. 

%

\bibitem{keimer2015} Keimer, B.,  Kivelson, S. A., Norman, M. R., Uchida, S. Zaanen, J. From quantum matter to high-temperature superconductivity in copper oxides. {\it Nature} {\bf 518}, 179-186 (2015).



\bibitem{hucker2014} H\"{u}cker, Christensen, N.B., Holmes, A.T., Blackburn, E., Forgan, E.M., Liang, R., Bonn, D.A., Hardy, W.N., Gutowski, O., Zimmermann, M.v., Hayden, S.M., Chang, J., Competing charge, spin, and superconducting orders in underdoped YBa$_2$Cu$_3$O$_y$.  {\it Phys. Rev. B} {\bf 90}, 054514 (2014). 

\bibitem{blancocanosa2014} Blanco-Canosa, S., Frano, A., Schierle, E., Porras, J., Loew, T., Minola, M., Bluschke, M., Weschke, E., Keimer, B., Le Tacon, M., Resonant x-ray scattering study of charge-density wave correlations in YBa$_2$Cu$_3$O$_{6+x}$. {\it Phys. Rev. B} {\bf 90}, 054513 (2014).



%%%%%%%%%%%%%%%

\bibitem{chen2014} Chen, Q. and Wang, J., Pseudogap phenomena in ultracold atomic Fermi gases. {\it Front. Phys.} {\bf 9} 539-570 (2014).

%

\bibitem{timusk1999} Timusk, T., Statt, B. The pseudogap in high-temperature superconductors: an experimental survey. {\it Rep. Prog. Phys.} {\bf 62}, 61-122 (1999).

%

\bibitem{gaebler2010} Gaebler, J. P., Stewart, J. T., Drake, T. E., Jin, D. S., Perali, A., Pieri, P., Strinati, G. C., Observation of pseudogap behaviour in a strongly interacting Fermi gas. {\it Nat. Phys.}{\bf 6}, 569-573 (2010).

\bibitem{magierski2011} Magierski, P., Wlazłowski, G., Bulgac, A., Onset of a pseudogap regime in ultracold Fermi gases. {\it Phys. Rev. Lett.} {\bf 107}, 145304 (2011).

\bibitem{chin2004} Chin, C., Bartenstein, M., Altmeyer, A., Riedl, S., Jochim, S., Hecker Denschlag, J., Grimm, R., Observation of the pairing gap in a strongly interacting Fermi gas. {\it Science} {\bf 305}, 1128-1130 (2004).%pairing gap observation

\bibitem{perali2011} Perali, A., Palestini, F., Pieri, P., Strinati, G. C., Stewart, J. T., Gaebler, J. P., Drake, T. E., Jin, D. S., Evolution of the normal state of a strongly interacting Fermi gas from a pseudogap phase to a molecular Bose gas. {\it Phys. Rev. Lett.} {\bf 106}, 060402 (2011).

%

\bibitem{tsuchiya2009} Tsuchiya, S., Watanabe, R., Ohashi, Y., Single-particle properties and pseudogap effects in the BCS-BEC crossover regime of an ultracold Fermi gas above $T_{\rm c}$. {\it Phys. Rev. A} {\bf 80}, 033613 (2009).

\bibitem{jensen2020} Jensen, S., Gilbreth, C. N., Alhassid, Y. Pairing correlations across the superfluid phase transition in the unitary Fermi gas. {\it Phys. Rev. Lett.} {\bf 124}, 090604 (2020).

\bibitem{richie2020} Richie-Halford, A., Drut, J. E., Bugac, A., Emergence of a pseudogap in the BCS-BEC crossover. {\it Phys. Rev. Lett.} {\bf 125} 060403 (2020).

%

\bibitem{hufner2008} H\"{u}fner, S., Hossain, M.A., Damascelli, A., Sawatsky, G.A., Two gaps make a high-temperature superconductor? {\it Rep. Prog. Phys.} {\bf 71}, 062501 (2008).

%

\bibitem{chakravarty2001} Chakravarty, S., Laughlin, R.B., Morr, D.K., Nayak, C., Hidden order in the cuprates. {\it Phys. Rev. B} {\bf 63}, 094503 (2001).

\bibitem{lee2006} Lee, P. A., Nagaosa, N., Wen, X.-G. Doping a Mott insulator: Physics of high-temperature superconductivity. {\it Rev. Mod. Phys.} {\bf 78}, 17-85 (2006).

\bibitem{varma2006} Varma, C.M., Theory of the pseudogap state of the cuprates. {\it Phys. Rev. B} {\bf 73}, 155113 (2006).

\bibitem{rice2012} Rice, T.M., Yang, K.-Y., Zhang, F.C., A phenomenological theory of the anomalous pseudogap phase in underdoped cuprates. {\it Rep. Prog. Phys.} {\bf 75}, 016502 (2012).

\bibitem{nie2014} Nie, L., Tarjus, G., Kivelson, S.A., Quenched disorder and vestigial nematicity in the
pseudogap regime of the cuprates. {\it Proc. Nat. Acad. Sci. USA} {\bf 111}, 7980-7985 (2014).

\bibitem{schmalian1999} Schmalian, J., Pines, D., Stojkovi\'{c}, Microscopic theory of weak pseudogap behavior in the underdoped cuprate superconductors: General theory and quasiparticle properties. {\it Phys. Rev. B} {\bf 60}, 667-686 (1999).

%%%%%%%%%%%%%%%%%%%%

\bibitem{carbotte1990} Carbotte, J. P., Properties of boson-exchange superconductors. {\it Rev. Mod. Phys.} {\bf 62}, 1027-1157 (1990).

\bibitem{inosov2011} Inosov, D., S., Park, J. T., Charnukha, A., Yuan, L., Boris, A. V., Keimer, B., Hinkov, V., Crossover from weak to strong pairing in unconventional superconductors. {\it Phys. Rev. B} {\bf 83}, 214520 (20011).

\bibitem{schirotzek2008} Schirotzek, A., Shin, Y., Schunck, C. H., Ketterle, W., Determination of the superfluid gap in atomic Fermi gases by quasiparticle spectroscopy. {\it Phys. Rev. Lett.} {\bf 101}, 140403 (208). 

%

\bibitem{cuprategap} Since Fig.~\ref{gap}c includes spectroscopic data taken at low temperatures, we assume this to provide an estimate of $\Delta$~\cite{supplemental}. 

%



%%%%%%%%%%%%%%%%%%

\bibitem{leggett1980} Leggett, A. J. in {\it Modern Trends in the Theory of Condensed Matter} (eds Pekalski, A. \& Przystawa, J.) 13–27 (Proc. XVIth Karpacz Winter School of Theoretical Physics, Springer, Berlin, 1980).

%

\bibitem{zwerger2016} Zwerger, W. in {\it Proceedings of the International School of Physics ``Enrico Fermi'' --- Course 191 ``Quantum matter at ultralow temperatures''} (Inguscio, M., Ketterle, W., Roati, Q. eds.) 63-142 (IOS Press, Amsterdam; SIF Bologna, 2016).

%


%%%%%%%%%%%%%%%%%

\bibitem{loram1993} Loram, J. W., Mirza, K. A., Cooper, J. R., Liang, W. Y. Electronic specific heat of YBa$_2$Cu$_3$O$_{6+x}$ from 1.8 to 300~K. {\it Phys. Rev. Lett.} {\bf 71}, 1740-1743 (1993).

\bibitem{wade1994} Wade, J. M., Loram, J. W., Mirza, K. A., Cooper, J. R., Tallon, J. R., Electronic specific heat of Tl$_2$Ba$_2$CuO$_{6+\delta}$ from 2~K to 300~K for 0~$\leq\delta\leq$~0.1. {\it J. Superconductivity} {\bf 7}, 261-264 (1994).

\bibitem{loram1998} Loram, J. W., Mirza, K. A., Cooper, J. R., Tallon, J. L. Specific heat evidence of the normal state pseudogap. {\it J. Phys. Chem. Solids} {\bf 59}, 2091-2094 (1998).

\bibitem{loram2001} Loram, J. W., Luo, J., Cooper, J. R., Liang, W. Y., Tallon, J. L., Evidence on the pseudogap and condensate from the electronic specific heat, {\it J. Physics and Physical Chemistry of Solids} {\bf 62}, 59-64 (2001).

\bibitem{mirmelstein1995} Mirmelstein, A., Junod, A., Triscone, G., Wang, K.-Q., Muller, J., Specific heat of Tl$_2$Ba$_2$CuO$_6$ (``2201'') 90~K superconducting ceramics in magnetic fields up to 14~T. {\it Physica C} {\bf 248}, 225-342 (1995).

\bibitem{wen2009} Wen, H.-H., Mu, G., Luo, H., Yang, H., Shan, L., Ren, C., Cheng, P., Yan, J., Fan, L., Specific-heat measurement of a residual superconducting state in the normal state of underdoped Bi$_2$Sr$_{2-x}$La$_x$CuO$_{6+\delta}$ cuprate cuperconductors. {\it Phys. Rev. Lett.} {\bf 103}, 067002 (2009).


\bibitem{compositions} For the high $T_{\rm c}$ cuprates LSCO, YBCO, Ca-YBCO, BSCCO, BSLCO, TBCO, Nd-LSCO and HBCO refer to La$_{2-x}$Sr$_x$CuO$_4$~\cite{loram2001}, YBa$_2$Cu$_3$O$_{6+x}$~\cite{loram1993,loram2001} and YBa$_2$Cu$_4$O$_8$~\cite{curro1997}, Y$_{0.8}$Ca$_{0.2}$Ba$_2$Cu$_3$O$_{6+x}$~\cite{loram1998,michon2019}, 
%
Bi$_2$Sr$_2$CaCu$_2$O$_{8+\delta}$ doped with 20\% Pb or 15\% Y~\cite{loram2001}, Bi$_2$Sr$_{2-x}$La$_x$CuO$_{6+\delta}$~\cite{wen2009}, % and Pb$_{0.55}$Bi$_{1.5}$Sr$_{1.6}$La$_{0.4}$CuO$_{6+\delta}$\cite{he2011}, 
%
Tl$_2$Ba$_2$CuO$_{6+\delta}$~\cite{radcliffe1996,kubo1991} La$_{2-y-x}$Nd$_y$Sr$_x$CuO$_4$~\cite{michon2019}, and HgBa$_2$CuO$_{4+\delta}$~\cite{chan2020}, respectively.

%%%%%%%%%%%%%%%%%%%%%
%%%%%%%%%%%%%%%%%%%%%%
%%%%%%%%%%%%%%%%%%%%%%


\bibitem{curro1997} Curro, N.J., Imai, T., Slichter, C.P., Dabrowski, B., High-temperature $^{63}$Cu(2) nuclear quadrupole and magnetic resonance measurements of YBa$_2$Cu$_4$O$_8$. {\it Phys. Rev. B} {\bf 56}, 877-885 (1997). 

\bibitem{kubo1991} Kubo, Y., Shimakawa, Y.,  Manako, T., Igarashi, H. Transport and magnetic properties of Tl$_2$Ba$_2$Cu0$_{6+\delta}$ showing a $\delta$-dependent gradual transition from an 85~K superconductor to a nonsuperconducting metal. {\it Phys. Rev. B} {\bf 43}, 7875-7882 (1991).

\bibitem{chan2020} Chan, M. K., McDonald, R. D.,, Ramshaw, B. J., Betts, J. B., Shekhter, A., Bauer, E. D., Harrison, N. Extent of Fermi-surface reconstruction in the high-temperature superconductor HgBa2CuO4$_\delta$. {\it Proc. Nat. Acad. Sci. USA} {\bf 117},  9782-9786 (2020).

%%%%%%%%%%%%%%%%%%%%%%
%%%%%%%%%%%%%%%%%%%%%%
%%%%%%%%%%%%%%%%%%%%%%

\bibitem{curty2003} Curty, P., Beck, H., Thermodynamics and phase diagram of high temperature superconductors. {\it Phys. Rev. Lett.} {\bf 91}, 257002 (2003).

\bibitem{banerjee2011} Banerjee, S., Ramakrishnan, T. V., Dasgupta, C., Phenomenological Ginzburg-Landau-like theory for superconductivity in the cuprates. {\it Phys. Rev. B} {\bf 83}, 024510 (2011).

\bibitem{noat2021} Noat, Y., Mauger, A., Nohara, M., Eisaki, H., Sacks, W., How ‘pairons’ are revealed in the electronic specific heat of cuprates. {\it Solid State Commun.} {\bf 323}, 114109 (2021).

%%%%%%%%%%%%%%%%%%%%%

\bibitem{moshe2019} Moshe, A. G., Farber, E., Deutscher, G., Optical conductivity of granular aluminum films near the Mott metal-to-insulator transition. {\it Phys. Rev. B} {\bf 99}, 224503 (2019).

\bibitem{pisani2018} Pisani, L., Pieri, P., Strinati, G. C., Gap equation with pairing correlations beyond the mean-field approximation and its equivalence to a Hugenholtz-Pines condition for fermion pairs. {\it Phys. Rev. B} {\it Phys. Rev. B} {\bf 98}, 104507 (2018).

%%%%%%%%%%%%%%%%%


\bibitem{nakagawa2021} Nakagawa, Y., Kasahara, Y., Nomoto, T., Arita, R., Nojima, T., Iwasa, Gate-controlled BCS-BEC crossover in a two-dimensional superconductor. {\it Science} {\bf 372}, 190-195 (2021).

%%%%%%%%%%%%%%%%%


\bibitem{polynomial} In the cuprates, we have used the smaller scatter of the locations of the maxima in $\gamma$ extracted from experimental data~\cite{supplemental} (plotted in Fig.~\ref{gap}d) to constrain the functional form of $\Delta$ versus $p$ in Fig.~\ref{gap}c by fitting. 
Fitting yields $T_\gamma=T_0+T_1p+T_2p^2$, where $T_0=$~478~$\pm$~8~K, $T_1=$~-2970~$\pm$~110~K and $T_2=$~4590~$\pm$~380~K.

%

\bibitem{sutherland2003} Sutherland, M., Hawthorn, D.G., Hill, R.W., Ronning, F., Wakimoto, S., Zhang, H., Proust, C., Boaknin, E., Lupien, C., Taillefer, L., Liang, R., Bonn, D.A., Hardy, W.N., Gagnon, R., Hussey, N.E., Kimura, T., Nohara, M., Takagi, H., Thermal conductivity across the phase diagram of cuprates: Low-energy quasiparticles and doping dependence of the superconducting gap. {\it Phys. Rev. B} {bf 67}, 174520 (2003).

\bibitem{mukhopadhyay2019} Mukhopadhyay, S., Sharma, R., Kim, C.K., Edkins, S.D., Hamidian, M.H., Eisaki, H., Uchida, S.-I., Kim, E.-A., Lawler, M.J., Mackenzie, A.P., Davis, J.C.S., Fujita, K., Evidence for a vestigial nematic state in the cuprate pseudogap phase. {\it Proc. Nat. Acad. Sci. USA} {\bf 116}, 13249-13254 (2019).

%%%%%%%%%%%%%%%%

\bibitem{engelbrecht1998} Engelbrecht, J. R., Nazarenko, A., Randeria, M., Dagotto, E., Pseudogap above $T_{\rm c}$ in a model with $d_{x^2-y^2}$ pairing. {\it Phys. Rev. B} {\bf 57}, 13406-13409 (1998).

\bibitem{wyk2016} van Wyk, P., Tajima, H., Hanai, R., Ohashi, Y., Specific heat and effects of pairing fluctuations in the BCS-BEC-crossover regime of an ultracold Fermi gas. {\it Phys. Rev. A} {\bf 93}, 013621 (2016).


\bibitem{enss2012i} Enss, T., Haussmann, R., Quantum mechanical limitations to spin diffusion in the unitary Fermi gas. {\it Phys. Rev. Lett.} {\bf 109}, 195303 (2012).

\bibitem{tajima2014} Tajima, H., Kashimura, T., Hanai, R., Watanabe, R., Ohashi, Y., Uniform spin susceptibility and spin-gap phenomenon in the BCS-BEC-crossover regime of an ultracold Fermi gas. {\it Phys. Rev. A} {\bf 89}, 033617 (2014).

\bibitem{loram1994} Loram, J. W., Mirza, K. A., Wade, J. M., Cooper, J. R., Liang, W. Y., The electronic specific heat of cuprate superconductors. {\it Physica C} {\bf 235-240}, 134-137 (1994).

\bibitem{alloul1989} Alloul, H., Ohno, T., Mendels, P., $^{89}$ Y NMR evidence for a Fermi-liquid behavior in YBa$_2$Cu$_3$O$_{6+x}$. {\it Phys. Rev. Lett.} {\bf 63}, 1700-1703 (1989).

\bibitem{johnston1989} Johnston, D.C., Magnetic susceptibility scaling in La$_{2-x}$Sr$_x$CuO$_{4-\delta}$. {\it Phys. Rev. Lett.} {\bf 62}, 957-960 (1989).


\bibitem{nakano1994} Nakano, T., Oda, M., Manabe, C., Momono, N., Miura, Y., Ido, M., Magnetic properties and electronic conduction of superconducting La$_{2-x}$Sr$_x$CuO$_4$. {\it Phys. Rev. B} {\bf 49}, 16000-16008 (1994).


\bibitem{vyaselev1994} Vyaselev, O. M., Kolesnikov, N. N., Schegolev, I. F., Transition from strong to weak coupling regime with lowering $T_{\rm c}$ in Tl$_2$Ba$_2$CuO$_{6+x}$. {\it Physica C} {\bf 235-240}, 1613-1614 (1994).


\bibitem{crocker2011} Crocker, J., Dioguardi, A. P., apRoberts-Warren, N., Shockley, A. C., Grafe, H.-J., Xu, Z., Wen, J., Gu, G., Curro, N. J., NMR studies of pseudogap and electronic inhomogeneity in Bi$_2$Sr$_2$CaCu$_2$O$_{8+\delta}$. {\it Phys. Rev. B} {\bf 84}, 224502 (20110.

\bibitem{alloul2016} Alloul, H. in {\it Quantum Materials: Experiments and Theory} (Pavarini, E., Koch, E., van den Brink, J., Sawatsky, G. eds.) 13.1-13.30 (Forschungszentrum J\"{u}lich GmbH Institute for Advanced Simulation, 2016). at $<$https://juser.fz-juelich.de/record/819465/files/correl16.pdf$>$

\bibitem{susceptibilitynote} In the cuprates, the $T$-dependences of $\chi_{\rm m}$ and $K$ are dominated by the spin susceptibility $\chi$ at a sufficiently high $T$ compared to $T_{\rm c}$, and in a sufficiently strong magnetic field (in the case of $K$), enabling $\chi_{\rm m}$ and $K$ to be considered as representative of the $T$-dependence of $\chi$. 


\bibitem{gapnote} In the unitary and BEC regimes of a Fermi gas, the pseudogap at $T>T_{\rm c}$ consists primarily of a gap of comparable energy to the low $T$ pairing gap that is extensively smeared by $T$-dependent line broadening effects~\cite{strinati2018,chen2014,zwerger2016}. The line broadening is primarily associated with the loss of phase coherence above $T_{\rm c}$. 
%
Extensive line broadening leads to a minimum in the density of states instead of a well defined gap, causing the maxima in $\gamma$ and $\chi$ to become less pronounced. 

%%%%%%%%%%%%%%%%

\bibitem{chen2000} Chen, Q. J. {\it Generalization of BCS theory to short coherence length superconductors: A BCS-Bose-Einstein crossover scenario}. Ph.D. thesis, University of Chicago (2000). (freely accessible in the ProQuest Dissertations \& Theses Database online).

\bibitem{chen2001} Chen, Q, Levin, K., Kosztin, I., Superconducting phase coherence in the presence of a pseudogap: Relation to specific heat, tunneling, and vortex core spectroscopies. {\it Phys. Rev. B} {\bf 63}, 184519 (2001). 

%

%\bibitem{transition}
%The peak in $\delta\gamma(T_{\rm c})$ at $T_{\rm c}=T_\gamma$ cannot easily be understood if the normal state gap has a non pairing origin~\cite{norman2014}. In such a case, we would expect the peak in $\delta\gamma(T_{\rm c})$ to occur instead when the normal state gap vanishes, as has been found experimentally at the antiferromagnetic quantum phase transition in heavy fermion superconductors~\cite{park2008} and in thermodynamic simulations of a non-pairing pseudogap state critical point~\cite{leblanc2009,rice2012}.

%%%



%%%

\bibitem{cooper2000} Cooper, J. R., Loram, J. W., The normal state gap and other strange properties of cuprate superconductors. {\it  J. Phys. IV France} {\bf 10}, Pr3-213-224 (2000).

%%%%%%%%%%%%%%%%

\bibitem{bozovic2016} Bo\u{z}ovi\'{c}, I., He, X., Wu, J., Bollinger, A. T., Dependence of the critical temperature in overdoped copper oxides on superfluid density. {\it Nature} {\bf 536}, 309-311 (2016).

%

\bibitem{ma2021} Although it may also be affected by its co-location with stripe-ordering in this system; 
Ma, Q., Rule, K. C., Cronkwright, Z. W., Dragomir, M., Mitchell, G., Smith, E. M., Chi, S., Kolesnikov, A. I., Stone, M. B., Gaulin, B. D., Parallel spin stripes and their coexistence with superconducting ground states at optimal and high doping in La$_{1.6-x}$Nd$_{0.4}$Sr$_x$CuO$_4$. {\it Phys. Rev. Research} {\bf 3}, 023151 (2021).

%%%%%%%%%%%%%%%%

\bibitem{cooper2009} Cooper, R. A., Wang, Y., Vignolle, B., Lipscombe, O. J., Hayden, S. M., Tanabe, Y., Adachi, T., Koike, Y., Nohara, M., Takagi, H., Proust, C., Hussey, N. E. Anomalous criticality in the electrical resistivity of La$_{\rm 2-x}$Sr$_{\rm x}$CuO$_4$. {\it Science} {\bf 323}, 603 (2009).

\bibitem{giraldogallo2018} Giraldo-Gallo, P., Galvis, J. A., Stegen, Z., Modic, K. A., Balakirev, F. F., Betts, J. B., Lian, X., Moir, C., Riggs, S. C., Wu, J., Bollinger, A. T., He, X., Bo\u{z}ovi\'{c}, I., Ramshaw, B. J., McDonald, R. D., Boebinger, G. S., Shekhter, A., Scale-invariant magnetoresistancein a cuprate superconductor. {\it Science} {\bf 361}, 479-481 (2018).

\bibitem{legros2019} Legros, A., Benhabib, S., Tabis, W., Lalibert\'{e}, F., Dion, M., Lizaire, M., Vignolle, B., Vignolles, D., Raffy, H., Li, Z. Z., Auban-Senzier, P., Doiron-Leyraud, N., Fournier, P., Colso, D., Taillefer, L., Proust, C., Universal $T$-linear resistivity and Planckian dissipation in overdoped cuprates. {\it Nature Phys.} {\bf 15}, 142-147 (2019).

%

\bibitem{zaanen2019} Zaanen, J., Planckian dissipation, minimal viscosity and the transport in cuprate strange metals. {\it SciPost Phys.} {\bf 6}, 061 (2019).

%

\bibitem{cao2010} Cao, C., Elliott, E., Joseph, J., Wu, H., Petricka,, J., Schafer, T., Thomas, J. E., Universal quantum viscosity in a unitary Fermi gas, {\it Science} {\bf 331}, 58-61 (2010).

\bibitem{enss2012ii} Enss, T., Quantum critical transport in the unitary Fermi gas. {\it Phys, Rev. A} {\bf 86}, 013616 (2012).

%

\bibitem{pistolesi1996} Pistolesi, F., Strinati, G. C., Evolution from BCS superconductivity to Bose condensation:
Calculation of the zero-temperature phase coherence length. {\it Phys. Rev. B} {\bf 53}, 15168-15192 (1996).

\bibitem{engelbrecht1997} Engelbrecht, J. R., Randeria, M., S\'{a} de Melo, C. A. R., BCS to Bose crossover: Broken-symmetry state. {\it Phys. Rev. B} {\bf 55}, 15153-15156 (1997). 

%

\bibitem{grissonnanche2014} Grissonnanche, G., Cyr-Choini\`{e}re, O., Lalibert\'{e}, F., Ren\'{e} de Cotret, S., Juneau-Fecteau, A., Dufour-Beaus\'{e}jour, S., Delage, M.-\`{E}, LeBoeuf, D., Chang, J., Ramshaw, B. J., Bonn, D. A., Hardy, W. N., Liang, R., Adachi, S., Hussey, N. E., Vignolle, B., Proust, C., Sutherland, M., Kr\"{a}mer, S., Park, J.-H., Graf, D., Doiron-Leyraud, N., Taillefer, L. Direct measurement of the upper critical field in cuprate superconductors. {\it Nature Commun.} {\bf 5}, 3280 (2014).

%




%


%\bibitem{cyrchoiniere2018} Cyr-Choini\`{e}re, O., Daou, R., Lalibert\'{e}, F., Collignon, O., Badoux, S., LeBoeuf, D., Chang, J., Ramshaw, B. J., Bonn, D. A., Hardy, W. N., Liang, R., Yan, J.-Q., Chang, J.-G., Zhou, J.-S., Goodenough, J. B., Pyon, S., Takayama, T.,, Takagi, H., Doiron-Leyraud, N., Taillefer, L., Pseudogap temperature $T^\ast$ of cuprate superconductors from the Nernst effect. {\it Physs. Rev. B} {\bf 97}, 0674502 (2018).

%

%\bibitem{pincus1973} Pincus, P., Chaikin, P., Coll, C. F., Correlated pairs in attractive Hubbard model. {\it Solid State Commun.} {\bf 12}, 1265-1268 (1973). 
%
%\bibitem{domanski2002} T.~Doma\'{n}ski, Effect of on-site Coulomb repulsion on superconductivity in the boson-fermion model. {\it Phys. Rev. B} {\bf 66}, 134512 (2002).
%
%\bibitem{altman2002} Altman, E., Auerbach, A., Plaquette boson-fermion model of cuprates. {\it Phys. Rev. B} {\bf 65}, 104508 (2002). 
%
%\bibitem{domanski2003} T.~Doma\'{n}ski, Feshbach resonance described by boson-fermion coupling. {\it Phys. Rev. A} {\bf 68} 013603 (2003).
%
%\bibitem{garg2005} Garg, A., Krishnamurthy, H. R., Randeria, M., BCS-BEC crossover at $T=0$: A dynamical mean-field theory approach. {\it Phys. Rev. B} {\bf 72} 024517 (2005).


%%%%%%%%%%%%%%%%


%


%


%


%\bibitem{kaminski2014} Kaminski, A., Kondo, T., Takeuchi, T., Gu., G. Pairing, pseudogap and Fermi arcs in cuprates. {\it Phil. Mag.} {\bf 95}, 453-466 (2014).


%
%\bibitem{startseva1999} Startseva, T., Timusk, T., Puchkov, A.V., Basov, D.N., Mook, H.A., Okuya, M., Kimura, T., Kishio. K., Temperature evolution of the pseudogap state in the infrared response of underdoped La${2-x}$Sr$_x$CuO$_4$. {\it Phys. Rev. B} {\bf 59}, 7184-7190 (1999). 
%
%\bibitem{sato1999} Sato, T., Yokoya, T., Naitoh, Y., Takahashi, T., Yamada, K., Endo, Y., Pseudogap of optimally doped La$_{1.85}$Sr$_{0.15}$CuO$_4$ observed by ultrahigh-resolution photoemission spectroscopy. {\it Phys. Rev. Lett.} {\bf 83}, 2254-2257 (1999).




%%%%%%%%%%




%\bibitem{chen2000} Chen, Q. J. {\it Generalization of BCS theory to short coherence length superconductors: A BCS-Bose-Einstein crossover scenario}. Ph.D. thesis, University of Chicago (2000). (freely accessible in the ProQuest Dissertations \& Theses Database online).


%\bibitem{domanski2003} Domanski, T., Ranninger, J., Bogoliubov shadow bands in the normal state of superconducting systems with strong pair fluctuation. {\it Phys. Rev. Lett.} {\bf 91}, 255301 (2003). 


%%%%%%%%%%%%%%%%


%


%



%


%%%%%%%%%%%%%%%%


%





%%%%%%%%%%%%%%%%



%%%%%%%%%%%%%%%%



%%%%%%%%%%%%%%%%%%%


%

%\bibitem{suzuki1991} Suzuki M., Hikita, M., Resistive transition, magnetoresistance, and anisotropy in La$_{2-x}$Sr$_x$CuO$_4$ single-crystal thin films. {\it Phys. Rev. B} {\bf 44}, 249-261 (1991).

%

%%%%%%%%%%%%%%%%%%%




%


%



%%%%%%%%%%%%%%%%%%%



%






%%%%%%%%%%%%%%%%%%%






%%%%%%%%%%%%%%%%%%%



%%%%%%%%%%%%%%%%%%%

%%%%%%%%%%%%%%%%%%%




%%%%%%%%%%%%%%%%%%%


%



%


%%%%%%%%%%%%%%%%%%%



%
%
%
%
%
%\bibitem{zhao2017} Zhao, L., Belvin, C.A., Liang, R., Bonn, D.A., Hardy, W.N., Armitage, N.P., Hsieh, D., A global inversion-symmetry-broken phase inside the pseudogap region of YBa$_2$Cu$_3$O$_y$. {\it Nature Phys.} {\bf 13}, 250-254 (2017).
%
%\bibitem{sato2017} Sato, Y., Kasahara, S., Murayama, H., Kasahara, Y., Moon, E.G., Nishizaki, T., Loew, T., Porras, J., Keimer, B., Shibauchi, T., Matsuda, Y.,  Thermodynamic evidence for a nematic phase transition at the onset of the pseudogap in YBa$_2$Cu$_3$O$_y$. {\it Nature Phys.} {\bf 13}, 1974-1979 (2017).
%
%
%
%\bibitem{he2011} He, R.-H., Hashimoto, M., Karapetyan, H., Koralek, D., Hinton, J.P., Testaud, J.P., Nathan, V., Yoshida, Y., Yao, H., Tanaka, K., Meevasana, W., Moore, R.G., Lu, D.H., Mo, S.-K., Ishikado, M., Eisaki, H., Hussain, Z., Devereaux, T.P., Kivelson, S.A., Orenstein, J., Kapitulnik, A., Shen, Z.-X., From a single-band metal to a
%high-temperature superconductor via two thermal phase transitions. {\it Science} {331}, 1579-1581 (2011).
%



%%%%%%%%%%%%%  methods
%
%\bibitem{takagi1989} Takagi, H., Ido, T., Ishibashi, S., Uota, M., Uchida, S., Tokura, Y., Superconductor-to-nonsuperconductor transition in (La$_{1-x}$Sr$_x$)$_2$CuO$_4$ as investigated by transport and magnetic measurements. {\it Phys. Rev. B} {\bf 40}, 2254-2261 (1989).
%
%\bibitem{radcliffe1996} Radcliffe, J.W., Loram, J.W., Wade, J.M., Wltschek, G., Tallon, J.W., Electronic specific heat of overdoped TI$_2$Ba$_2$CuO$_{6+\delta}$ in a magnetic field. {\it J. Low. Temp. Phys.} {\bf 105}, 903-908 (1996).
%
%
%\bibitem{liang2006} Liang, R., Bonn, D. A., Hardy, W. N. Evaluation of CuO$_2$ plane hole doping in YBa$_2$Cu$_3$O$_{6+x}$ single crystals. {\it Phys. Rev. B} {\bf 73}, 180505 (2006).
%
%
%\bibitem{zhou1996} Zhou, J.-S., Goodenough, J.B., Dabrowski, B., Rogacki, K., 
%Transport properties of a YBa$_2$Cu$_4$O$_8$ crystal under high pressure. {\it Phys. Rev. Lett.} {\bf 77}, 4253-4256 (1996)
%
%\bibitem{tallon1995} Tallon, J.L., Bernhard, C., Shaked, H., Hitterman, R.L., Jorgensen, J.D., Generic superconducting phase-behavior in high-$T_{\rm c}$ variation with hole concentration in YBa$_2$Cu$_3$O$_{7-\delta}$, {\it Phys. Rev. B} {\bf 51}, 12911-12914 (1995). 
%
%
%%\bibitem{niedermayer1993} Niedermayer,Ch., Bernhard, C., Binninger, U., Gl\"{u}ckler, H., Tallon, J.L., Ansaldo, E.J., Budnick, J.I., Muon spin rotation study of the correlation between $T_{\rm c}$, and $n_{\rm s}/rn^\ast$ in overdoped Tl$_2$Ba$_2$CuO$_{6+\delta}$. {\it Phys. Rev. Lett.} {\bf 71}, 1764-1767 (1993).
%
%\bibitem{carrington1994} Carrington, A., Mackenzie, A. P., Sinclair, D. C., Cooper, J. R., Field dependence of the resistive transition in Tl$_2$Ba$_2$CuO$_{6+\delta}$. {\it Phys. Rev. B} {\bf 49}, 13243-13246 (1994).
%
%\bibitem{kittel2004} Kittle, C.,  Introduction to Solid State Physics, eighth ed. (Wiley, New York, 2004).
%
%
%\bibitem{hazra2019} Hazra, T., Verma, N., Randeria, M., Bounds on the superconducting transition temperature:
%Applications to twisted bilayer graphene and cold atoms. {\it Phys. Rev. X} {\bf 9}, 031049 (2019). 
%
%\bibitem{rourke2010} Rourke, P. M. C., Bangura, A. F., Benseman, T. M., Matusiak, M., Cooper, J. R., Carrington, A., Hussey, N. E., A detailed de Haas–van Alphen effect study of the overdoped cuprate Tl$_2$Ba$_2$CuO$_{6+\delta}$, {\it New J. Phys.} {\bf 12}, 105009 (2010).
%
%
%\bibitem{leboeuf2007} LeBoeuf, D., Doiron-Leyraud, N., Levallois, J., Daou, R., Bonnemaison, J. B., Hussey, N. E., Balicas, L., Ramshaw, B. J., Liang, R. X., Bonn, D. A., Hardy, W. N., Adachi, S., Proust, C., Taillefer, L. Electron pockets in the Fermi surface of hole-doped high-$T_{\rm c}$ superconductors. {\it Nature} {\bf 450}, 533-536 (2007).
%
%
%
%\bibitem{nolthing2009} Nolthing, W., Ramakanth, A., Quantum Theory of Magnetism (Springer, New York, 2009).
%
%\bibitem{markiewicz1997} Markiewicz, R. S., A survey of the van Hove scenario for high-$T_{\rm c}$ superconductivity with special emphasis on pseudogaps and striped phases. {\it J. Phys. Chem. Solids} {\bf 58}, 1179-1310 (1997).
%
%%
%
%\bibitem{norman1998} Norman, M.R., Ding, H., Randeria, M., Campuzano, J.C., Yokoya, T., Takeuchik, T, Takahashi, T., Mochiku, T., Kadowaki, K., Guptasarma, P., Hinks, D.G., Destruction of the Fermi surface in underdoped high-$T_{\rm c}$ superconductors. {\it Nature} {\bf 392}, 157-160 (1998).
%
%%
%
%
%%
%
%\bibitem{chubukov2007} Chubukov, A.V., Norman, M.R., Millis, A.J., Abrahams, E., Gapless pairing and the Fermi arc in the cuprates. {\it Phys. Rev. B} {\bf 76}, 180501 (2007).
%
%\bibitem{demello2012} de~Mello, E.V.L., Disordered-based theory of pseudogap, superconducting gap, and Fermi arc of cuprates. {\it Euro. Phys. Lett.} {\bf 99}, 37003 (2012).
%
%%
%
%\bibitem{hucker2014} H\"{u}cker, Christensen, N.B., Holmes, A.T., Blackburn, E., Forgan, E.M., Liang, R., Bonn, D.A., Hardy, W.N., Gutowski, O., Zimmermann, M.v., Hayden, S.M., Chang, J., Competing charge, spin, and superconducting orders in underdoped YBa$_2$Cu$_3$O$_y$.  {\it Phys. Rev. B} {\bf 90}, 054514 (2014). 
%
%\bibitem{blancocanosa2014} Blanco-Canosa, S., Frano, A., Schierle, E., Porras, J., Loew, T., Minola, M., Bluschke, M., Weschke, E., Keimer, B., Le Tacon, M., Resonant x-ray scattering study of charge-density wave correlations in YBa$_2$Cu$_3$O$_{6+x}$. {\it Phys. Rev. B} {\bf 90}, 054513 (2014).
%
%%
%
%\bibitem{fradkin2015} Fradkin, E., Kivelson, S.A., Tranquada, J.M., Colloquium: Theory of intertwined orders in high temperature superconductors. {\it Rev. Mod. Phys.} {\bf 87}, 457-482 (2015).
%
%\bibitem{agterberg2020} Agterberg, D.F., Davis, J.C.S., Edkins, S.D., Fradkin, E.,Van~Harlingen, D.J., Kivelson, S.A., Lee, P.A., Radzihovsky, L., The physics of pair-density waves: cuprate superconductors and beyond. {\it Annu. Rev. Condens. Matter Phys.} {\bf 11}, 231–270 (2020).
%
%%%%%%%%%%%%%%%%%%%%
%
%
%\bibitem{damascelli2003} Damascelli, A., Hussain, Z., Shen, Z.-X. Angle-resolved photoemission studies of the cuprate superconductors. {\it Rev. Mod. Phys.} {\bf 75}, 473-541 (2003).
%
%
%%
%
%\bibitem{rullieralbenque2008} Rullier-Albenque, F., Alloul, H., Balakirev, F., Proust, C., Disorder, metal-insulator crossover and phase diagram in high-$T_{\rm c}$ cuprates. {\it EPL} {\bf 81}, 37008 (2008).
%
%\bibitem{fukuzumi1996} Fukuzumi, Y., Mizuhashi, K., Takenaka, K., Uchida, S., Universal superconductor-insulator transition and $T_{\rm c}$ depression in Zn-substituted high-$T_{\rm c}$ cuprates in the underdoped regime. {\it Phys. Rev. Lett.} {\bf 76}, 684-687 (1996).
%
%\bibitem{nachumi1996} Nachumi, B., Keren, A., Kojima, K., Larkin, M., Luke, G. M., Merrin, J., Tchernysh\"{o}v, O., Uemura, Y. J., Ichikawa, N., Goto, M., Uchida, S., Muon spin relaxation studies of Zn-substitution effects in high-$T_{\rm c}$ cuprate superconductors. {\it Phys. Rev. Lett.} {\bf 77}, 5421-5424 (1996).
%
%%%%%%%%%%%%%%%%%%%%
%
%


%
%
%
%
%
%
%
%
%%
%
%
%%
%
%
%
%%%%%%%%%%%%%%%%%%%%
%
%
%%
%
%
%\bibitem{anderson1995} Anderson, M. H., Ensher, J. R., Matthews, M. R., Wieman, C. E., Cornell, E.. Observation of Bose-Einstein condensation in a dilute atomic vapor. {\it Science} {\bf 269}, 198-201 (1995). 

%


%



%




%

%
%\bibitem{vignolle2008} Vignolle. B., Carrington, A., Cooper, R. A., French, M. M., Mackenzie, A. P., Jaudet, C., Vignolles, D., Proust, C., Hussey, N. E., Quantum oscillations in an overdoped high-Z$T_{\rm c}$ superconductor, {\it Nature} {\bf 455}, 952-955 (2008). 
%%
%
%
%%
%
%
%
%%%%%%%%%%%%%%%%%%%%%%%
%
%
%
%
%
%\bibitem{shekhter2013} Shekhter, A., Ramshaw, B.J., Liang, R.X., Hardy, W.N., Bonn, D.A., Balakirev, F.F., McDonald, R.D., Betts, J.B., Riggs, S.C., Migliori, A., Bounding the pseudogap with a line of phase transitions in YBa$_2$Cu$_3$O$_{6+\delta}$. {\it Nature} {\bf 498},75-77 (2013).
%

%
%\bibitem{daou2010} Daou, R., Chang, J., LeBoeuf, D., Cyr-Choinie\`{e}re, O., Lalibert\'{e}, F., Doiron-Leyraud, N., Ramshaw, B.J., Liang, R., Bonn, D.A., Hardy, W.N., Taillefer, L., Broken rotational symmetry in the pseudogap phase of a high-$T_{\rm c}$ superconductor. {\it Nature} {\bf 463}, 519-522 (2010).
%
%\bibitem{xia2008} Xia, J., Schemm, E., Deutscher, G., Kivelson, S.A., Bonn, D.A., Hardy, W.N., Liang, R., Siemons, W., Koster, G., Fejer, M.M., Kapitulnik, A., Polar Kerr-effect measurements of the high-temperature YBa$_2$Cu$_3$O$_{6+x}$ Superconductor: Evidence for broken symmetry near the pseudogap temperature. {\it Phys. Rev. Lett.} {\bf 100}, 127002 (2008).
%
%\bibitem{fauque2006} Fauqu\'{e}, B., Sidis, Y., Hinkov, V., Path\`{e}s, S., Lin, C.T., Chaud, X., Bourges, P., Magnetic order in the pseudogap phase of high-$T_{\rm c}$ superconductors. {\it Phys. Rev. Lett.} {\bf 96}, 197001 (2006).
%
%%
%
%\bibitem{sordi2012} Sordi, G., S\'{e}mon, P., Haule, K., Tremblay, A.-M. S., Pseudogap temperature as a Widom line in doped Mott insulators. {\it Scientific Reports} {\bf 2}, 547 (2012).
%
%%
%
%\bibitem{norman2005} Norman, M. R., Pines, D., Kallin, C. The pseudogap: friend or foe of high $T_{\rm c}$? {\it Adv. Phys.} {\bf 54}, 715-733 (2005).
%
%%%%%%%%%%%%%%%%%%%%
%
%
%
%
%
%
%%%%%%%%%%%%%%%%%%%%
%
%\bibitem{varma1987} Varma, C. M., Schmitt-Rink, S., Abrahams, E., Charge transfer excitations and superconductivity in ``ionic'' metals. {\it Solid. State Commun.} {\bf 62}, 681-685 (1987). 
%

%\bibitem{dejongh1974} de~Jongh, L.J., Miedema, A.R., Experiments on simple magnetic model systems. {\it Adv. Phys.} {\bf 23}, 1-260 (1974).
%
%
%
%\bibitem{leggett1980} Leggett, A. J., in {\it Modern Trends in the Theory of Condensed Matter.} (Pekalski, J. A., Przystawa, A. eds.) 13-27 (Springer, Berlin, Heidelberg 1980).
%
%\bibitem{nozieres1985} Nozi\`{e}res, P., Schmitt-Rink, S., Bose condensation in an attractive fermion gas: From weak to strong coupling superconductivity. {\it J. Low Temp. Phys.} {\bf 59}, 195-211 (1985). 
%
%%
%
%\bibitem{dagotto1996} Dagotto, E., Rice, T.M., Surprises on the way from one- to two-dimensional quantum magnets: The ladder materials. {\it Science} {\bf 271}, 618-623 (1996).
%
%\bibitem{madrus1995} Mandrus, D., Sarrao, J.L., Migliori,A., Thompson, J.D., Fisk, Z., Thermodynamics of FeSi. {\it Phys. Rev. B} {\bf 51}, 4763-4767 (1995).
%
%\bibitem{coleman2007} Coleman, P., in {\it Handbook of Magnetism and Advanced Magnetic Materials} (Helmut Kronm\"{u}ller, H., and Parkin, S. eds.) {\bf 1}, 95-148 (Wiley, New York, 2007). %, Vol. 1, pp. 95–148
%
%%
%
%\bibitem{read1989} Read, N., Sachdev, S., Valence-bond and spin-Peierls ground states of low-dimensional quantum antiferromagnets. {\it Phys. Rev. Lett.} {\bf 62}, 1694-1697 (1989).
%
%\bibitem{anderson2004} Anderson, P.W., Lee, P.A., Randeria, M., Rice, T.M., Trivedi, N., Zhang, F.C., The physics behind high-temperature superconducting cuprates: the `plain vanilla' version of RVB. {\it J. Phys.: Condens. Matter} {\bf 16}, R755-R769 (2004). 


%%%%%%%%%%%%%%%%%%

%\bibitem{singh2007} Singh, Y., Johnston, D.C., Singlet ground state in the spin-$\frac{1}{2}$ dimer compound Sr$_3$Cr$_2$O$_8$. {\it Phys. Rev. B} {\bf 76}, 012407 (2007).
%
%\bibitem{mito2002} Mito, M., Akama, H., Kawae, T., Takeda, K., Deguchi, H., Takagi, S., Pressure effects on an $S=\frac{1}{2}$ Heisenberg two-leg ladder antiferromagnet Cu$_2$(C$_5$H$_{12}$N$_2$)$_2$Cl$_4$. {\it Phys. Rev. B} {\bf 65}, 104405 (2002).
%
%\bibitem{kageyama1999} Kageyama, H., Yoshimura, K., Stern, R., Mushnikov, N.V., Onizuka, K., Kato, M., Kosuge, K., Slichter, C.P., Goto, T., Ueda, Y., Exact dimer ground state and quantized magnetization plateaus
%in the two-dimensional spin system SrCu$_2$(BO$_3$)$_2$. {\it Phys. Rev. Lett.} {\bf 82}, 3168-3171 (1999).
%
%\bibitem{manson2009} Manson, J.L., Schlueter, J.A., Funk, K.A., Southerland, H.I., Twamley, B., Lancaster, T., Blundell, S.J., Baker, P.J., Pratt, F.L., Singleton, J., McDonald, R.D., Goddard, P.A., Sengupta, P., Batista, C.D., Ding L., Lee, C., Whangbo, M.-H., Cox, S., Baines, C., Trial, D., Strong H~$\cdot\cdot\cdot$~F hydrogen bonds as synthons in polymeric quantum magnets: structural, magnetic, and theoretical characterization of [Cu(HF$_2$)(pyrazine)$_2$]SbF$_6$, [Cu$_2$F(HF)(HF$_2$)(pyrazine)$_4$](SbF$_6$)$_2$, and [CuAg(H$_3$F$_4$)(pyrazine)$_5$](SbF$_6$)$_2$. {\it J.~AM.~Chem.~Soc.} {\bf 131}, 6733-6747 (2009).
%
%%
%
%\bibitem{riseborough2000} Riseborough, P.S., Heavy fermion semiconductors. {\it Adv. Phys.} {\bf 49}, 257-320 (2000).

%






%
%
%
%
%%
%

%
%%%%%%%%%%%%%%%%%
%
%\bibitem{kanigel2006} Kanigel, A., Norman, M.R., Randeria, M., Chatterjee, U., Souma, S., Kaminski, A., Fretwell, H.M., Rosenkranz, S., Shi, M., Sato, T., Takahashi, T., Li, Z.Z., Raffy, H., Kadowaki, K., Hinks, D., Ozyuzer, L., Campuzano, J.C., Evolution of the pseudogap from Fermi arcs to the nodal liquid. {\it Nature Phys.} {\bf 2}, 447-451 (2006).
%%


%%%%%%%%%%%%%%%%%%%

%%%%%%%%%%%%%%%%%%%


%
%
%
%%%%%%%%%%%%%%%%%%%%
%\bibitem{bulgac2006} Bulgac, A., Drut, J. E., Magierski, P., Spin $1/2$ fermions in the unitary regime: a superfluid of a new type. {\it Phys. Rev. Lett.} {\bf 96}, 090404 (2006).
%
%%%%%%%%%%%%%%%%%%%%
%
%



%


%%%%%%%%%%%%%%%%%%%

%%%%%%%%%%%%%%%%


%%%%%%%%%%%%%%%%%%%



%%%%%%%%%%%%%%%%%%



%%%%%%%%%%%%%%%%%%%

%
%
%\bibitem{kanigel2006} Kanigel, A., Norman, M.R., Randeria, M., Chatterjee, U., Souma, S., Kaminski, A., Fretwell, H.M., Rosenkranz, S., Shi, M., Sato, T., Takahashi, T., Li, Z.Z., Raffy, H., Kadowaki, K., Hinks, D., Ozyuzer, L., Campuzano, J.C., Evolution of the pseudogap from Fermi arcs to the nodal liquid. {\it Nature Phys.} {\bf 2}, 447-451 (2006).
%
%\bibitem{chubukov2007} Chubukov, A.V., Norman, M.R., Millis, A.J., Abrahams, E., Gapless pairing and the Fermi arc in the cuprates. {\it Phys. Rev. B} {\bf 76}, 180501 (2007).
%
%\bibitem{demello2012} de~Mello, E.V.L., Disordered-based theory of pseudogap, superconducting gap, and Fermi arc of cuprates. {\it Euro. Phys. Lett.} {\bf 99}, 37003 (2012).

%%%%%%%%%%%%%%%%%%%

%
%
%\bibitem{takahashi1994} Takahashi, H., Shaked, H., Hunter, B.A., Radaelli, P.G., Hitterman, R.L., Hinks, D.G., Jorgensen, J.D., Structural effects of hydrostatic pressure in orthorhombic La$_{2-x}$Sr$_x$CuO$_4$. {\it Phys. Rev. B} {\bf 50}, 3221-3229 (1994).
%
%%%%%%%%%%%%%%%%%%%
%
%\bibitem{lee2014} Lee, P.A., Amperean pairing and the pseudogap phase of cuprate superconductors. {\it Phys. Rev. X} {\it Phys. Rev. X} {\bf 4}, 031017 (2014).
%
%
%
%







%



%

%


%





%%%%%%%%%%%%%%%%


%

%

%%%%%%%%%%%%%%%%%%%




%%%%%%%%%%%%%%%%%%%

%%%%%%%%%%%%%%%%%%%



%%%%%%%%%%%%%%%%%%%



%%%%%%%%%%%%%%%%%%%


%%%%%%%%%%%%%%%%%%%


%%%%%%%%%%%%%%%%%%%

%%%%%%%%%%%%%%%%%%%


%%%%%%%%%%%%%%%%%%%









%
%\bibitem{emery1997} Emery, V.J., Kivelson, S.A., Zachar, O., Spin-gap proximity effect mechanism of high-temperature superconductivity. {\it Phys. Rev. B} {\bf 56}, 6120-6147 (1997).
%




%
%
%
%

%%%%%%%%%%%%%%%%%
%
%
%%
%
%%
%
%\bibitem{bednorz1986} Bednorz, J. G., M\"{u}ller, K. A. Possible high $T_{\rm c}$ superconductivity in the Ba-La-Cu-O system. {\it Zeitschrift f\"{u}r Physik B} {\bf 64}, 189-193 (1986).
%
%
%%
%
%
%
%
%
%%
%
%
%
%
%%
%
%
%%%%%%%%%%%%%%%%%%%%%
%
%
%
%%
%
%
%
%%%%%%%%%%%%%%%%%%%%%
%
%
%%%%%%%%%%%%%%%%%%%%
%
%
%%%%%%%%%%%%%%%%%%%%%
%
%
%\bibitem{hamidian2016} Hamidian, M. H., Edkins, S. D., Joo, S. H., Kostin, A., Eisaki, H., Uchida, S., Lawler, M. J., Kim, E. A., Mackenzie, A. P., Fujita, K., Lee, J., Davis, J. C. S. Detection of a Cooper-pair density wave in Bi$_2$Sr$_2$CaCu$_2$O$_{8+x}$. {\it Nature} {\bf 532}, 343-347 (2016). 
%
%\bibitem{edkins2019} Edkins, S. D., Kostin, A., Fujita, K., Mackenzie, A. P., Eisaki, H., Uchida, S., Sachdev, S., Lawler, M. J., Kim, E. A., Davis, J. C. S., Hamidian, M H. Magnetic field-induced pair density wave state in the cuprate vortex halo. {\it Science} {\bf 364}, 976-980 (2019).
%
%%
%
%
%%
%
%
%
%%




%


%

%%%%%%%%%%%%%%%%%%


%%%%%%%%%%%%%%%%%%%%%



%


%\bibitem{randeria1998} Randeria, M., Trivedi, N., Pairing correlations above $T_{\rm c}$ and pseudogaps in underdoped cuprates. {\it J. Phys. Chem. Solids} {\bf 59}, 1754-1758 (1998).




%%%%%%%%%%%%%%%%%%%%%




%




%








%%%%%%%%%%%%%%%%%%


%
%\bibitem{coleman2015} Coleman, P., in {\it Many-body physics: from Kondo to Hubbard} (Pavarini, E., Koch, E., Coleman, P. eds.) {\bf 5}, 2-30 (Forschungszentrum J\"{u}lich, 2015). at $<$https://www.cond-mat.de/events/correl15/manuscripts/correl15.pdf$>$.





%


%

%



%





%





%

%\bibitem{tallon2001} Tallon, J. L. and Loram, J. W. The doping dependence of $T^\ast$ – what is the real high-$T_{\rm c}$ phase diagram? {\it Physica C} {\bf 349}, 53-68 (2001).

%




%

%


%




%

%


%


%\bibitem{wang2002} Wang, Y., Ong, N. P., Xu, Z. A., Kakeshita, T., Uchida, S., Bonn, D. A., Liang, R., Hardy, W. N., High field phase diagram of cuprates derived from the Nernst effect. {\it Phys. Rev. Lett.} {\bf 88}, 257003 (2002).



%%
%\bibitem{leridon2009} Leridon, B., Monod, P., Colson, D., Forget, A., Thermodynamic signature of a phase transition in the pseudogap phase of YBa$_2$Cu$_3$O$_x$ high-$T_{\rm c}$ superconductor. {\it Europhys. Lett.} {\bf 87}, 17011 (2009).
%
%\bibitem{kawasaki2010} Kawasaki, S., Lin, C., Kuhns, P.L., Reyes, A.P., Zheng, G.-Q., Carrier-concentration dependence of the pseudogap ground state of superconducting Bi$_2$Sr$_{2-x}$La$_x$CuO$_{6+\delta}$ revealed by $^{63,65}$Cu-nuclear magnetic resonance in very high magnetic fields. {\it Phys. Rev. Lett.} {\bf 105}, 137002 (2010).
%
%
%\bibitem{storey2015} Storey, J.G., Closing the pseudogap quietly. {\it EPL} {\bf 111}, 57004 (2015).
%
%\bibitem{tallon2020} Tallon, J.L., Storey, J.G., Cooper, J.R., Loram, J.W., Locating the pseudogap closing point in cuprate superconductors: Absence of entrant or reentrant behavior. {\it Phys. Rev. B} {\bf 101}, 174512 (2020).
%





%%


%

%


%

%
%\bibitem{nakatsuji2004} Nakatsuji, S., Pines, D., Fisk, Z., Two fluid description of the Kondo lattice. {\it Phys. Rev. Lett.} {\bf 92}, 016401 (2004).
%
%\bibitem{jang2020} Yang, S., Denlinger, J. D., Allen, J. W., Zapf, V. S., Maple, M. B., Kim, J. N., Jang, B. G., Shim, J. H., Evolution of the Kondo lattice electronic structure above the transport coherence temperature.  {\it Proc. Nat. Acad. Sci. USA} {\bf 117}, 23467-23476 (2020).
%
%%
%
%
%
%

%


%



%

%




%

%\bibitem{chen2001} Chen, Q., Levin, K., Kosztin, I., Superconducting phase coherence in the presence of a pseudogap: Relation to specific heat, tunneling, and vortex core spectroscopies. {\it Phys. Rev. B} {\bf 63}, 184519 (2001).





%



%%%%%%%%


%

%


%



%
%
%
%
%
%%
%
%
%
%%\bibitem{erten2017} Erten, O., Chang, P.-Y., Coleman, P., Tsvelik, A. M., Skyrme Insulators: Insulators at the Brink of Superconductivity. {\it Phys. Rev. Lett.} {\bf 119}, 057603 (2017). 
%
%%\bibitem{vanillar2004} Anderson, P. W., Lee, P. A., Randeria, M., Rice, T. M., Trivedi, N., Zhang, F. C., The physics behind high-temperature superconducting cuprates: the `plain vanilla' version of RVB {\it Journal of Phys: Cond. Matt.} {\bf 16}, R755-R769 (2004).
%
%
%%
%
%%
%
%
%%%%%
%
%\bibitem{ono2007} Ono, S., Komiya, S., Ando, Y., Strong charge fluctuations manifested in the high-temperature Hall coefficient of high-$T_{\rm c}$ cuprates, {\it Phys. Rev. B} {\bf 75}, 024515 (2007).
%
%
%
%
%\bibitem{fedorov1999} Fedorov, A.V., Valla, T., Johnson, P.D., Li, Q., Gu, G.D., Koshizuka, N., Temperature dependent photoemission studies of optimally doped Bi$-$Sr$_2$CaCu$_2$O$_8$. {\it Phys. Rev. Lett.} {\bf 82}, 2179-2182 (1999). 
%
%\bibitem{sato2002} Sato, T., Matsui, H., Nishina, S., Takahashi, T., Fujii, T, Watanabe, T., Matsuda, A., Low energy excitation and scaling in Bi$_2$Sr$_2$Ca${n-1}$Cu$_n$O$_{2n+4}$ ($n=$~1-3):angle-resolved
%photoemission spectroscopy. {\it Phys. Rev. Lett.} {\bf 89}, 067005 (2002).
%
%


%%%%%%%%%%%%%%%%%%


%



%






%



%%%%%%%%%%%%%%%%%%%

%
%
%\bibitem{comin2014} Comin, R., Frano, A., Yee, M. M., Yoshida, Y., Eisaki, H., Schierle, E., Weschke, E., Sutarto, R., He, F., Soumyanarayanan, A., He, Y., Le Tacon, M., Elfimov, I. S., Hoffman, J. E., Sawatzky, G. A., Keimer, B., Damascelli, A. Charge order driven by Fermi-arc instability in Bi$_2$Sr$_{2-x}$La$_x$CuO$_{6+\delta}$. {\it Science} {\bf 343}, 390-392 (2014).
%
%
%%
%
%
%
%
%%%%%%%%%%%%%%%%%%%
%
%
%
%\bibitem{coleman2015} Coleman, P., in {\it Many-body physics: from Kondo to Hubbard} (Pavarini, E., Koch, E., Coleman, P. eds.) {\bf 5}, 2-30 (Forschungszentrum J\"{u}lich, 2015). at $<$https://www.cond-mat.de/events/correl15/manuscripts/correl15.pdf$>$.
%
%%
%

%%%%%%%%%%%%%%%
%
%
%
%
%
%
%
%
%
%
%
%
%
%
%\bibitem{gruner1994} Gr\"{u}ner, G., Density waves in solids (CRC Press, Boca Raton 2000).
%
%\bibitem{padamsee1973} Padamsee, H., Neighbor, J.E., Shiffman, C.A., Quasiparticle phenomenology for thermodynamics of strong-coupling superconductors. {\it J. Low Temp. Phys.} {\bf 12}, 387-410 (1973).
%
%\bibitem{yosida1958} Yosida, K., Paramagnetic susceptibility in superconductors. {\it Phys. Rev.} {\bf 110}, 769-770 (1958).
%
%%%%%%%%%%%%%%%%%%%
%
%
%
%
%
%%%%%%%%%%%%%%%%%%%%%
%
%\bibitem{shastry1988} B.~S.~Shastry, Exact solution of an $S=\frac{1}{2}$ Heisenberg antiferromagnetic chain with long-ranged interactions. {\it Phys. Rev. Lett.} {\bf 60}, 639-642 (1988).
%
%\bibitem{haldane1988} Haldane F.D.M.,Exact Jastrow-Gutzwiller resonating-valence-bond ground state of the spin-$\frac{1}{2}$ antiferromagnetic Heisenberg chain with $1/r^2$ exchange. {\it Phys. Rev. Lett.} {\bf 60}, 635-638 (1988).
%
%\bibitem{liang1988} Liang, S., Doucot, B., Anderson, P.W., Some new variational resonating-valence-bond-type wave functions for the spin-$\frac{1}{2}$ antiferromagnetic Heisenberg model on a square lattice. {\it Phys. Rev. Lett.} {\bf 61}, 365-368 (1988).
%
%\bibitem{jiang2012} Jiang, H.-C., Yao, H., Balents, L., Spin liquid ground state of the spin-$\frac{1}{2}$ square $J_1$ - $J_2$ Heisenberg model. {\it Phys. Rev. B} {\bf 86}, 024424 (2012).
%
%%%%%%%%%%%%%%%%%%%%
%
%
%%%%%%%%%%%%%%%%%%%%%%
%
%%%%%%%%%%%%%%%%%%%%%%
%
%
%%%%%%%%%%%%%%%%%%%%%
%
%
%
%\bibitem{zhang1988i} Zhang, F.C., Gros, C., Rice, T.M., Shiba, H., A renormalised Hamiltonian approach to a resonant valence bond wavefunction, {\it Supercond.Sci.Technol.} {\bf 1}, 36-46 (1988).
%
%%\bibitem{kotliar1988} Kotliar, G., Liu, J., Superexchange mechanism and $d$-wave superconductivity. {\it Phys. Rev. B} {\bf 38}, 5142-5145 (1988).
%
%%\bibitem{lee1992} Lee, P.A., Nagaosa, N., Gauge theory of the normal state of high-$T_{\rm c}$ superconductors. {\it Phys. Rev. B} {\bf 46}, 5621-5639 (1992).
%
%
%%
%
%
%\bibitem{leigh2007} Leigh, R.G., Phillips, P., Choy, T.-P., Hidden charge $2e$ boson in doped Mott insulators. {\it Phys. Rev. Lett.} {\bf 99}, 046404 (2007).
%
%%%%%%%%%%%%%%%%%%%
%
%\bibitem{ruckenstein1991} Ruckenstein, A.E., Varma, C.M., A theory of marginal Fermi-liquids. {\it Physica C} {\bf 185-189}, 134-140 (1991).
%
%
%
%%%%%%%%%%%%%%%%%%%%%%%%%
%
%
%
%
%
%
%
%%%%%%%%%%%%%%%%%%%%%%%%%
%
%
%

%
%
%
%
%%
%%
%%
%%%\bibitem{jerome1967} Jerome, D., Rice, T.M., Kohn, W., Excitonic insulator, {\it Phys. Rev.} {\bf 158}, 462-475 (1967). 
%%%
%%%\bibitem{halperin1968} Halperin, B.I., Rice, T.M., Possible anomalies at a semimetal-semiconductor transition. {\bf 40}, 755-766 (1968).
%%
%%\bibitem{mott1961} Mott, N.F.,  The transition to the metallic state. {\it Philosophical Magazine} {\bf 6}, 287–309 (1961).
%%
%%
%%\bibitem{mott1974} Mott, N.F., Rare-earth compounds with mixed valencies. {\it Philosophical Magazine} {\bf 30}, 403-416 (1974). 
%%
%%
%%
%%
%%
%%%Volume 1: Fundamentals and Theory.  2007 John Wiley & Sons, Ltd. ISBN: 978-0-470- 02217-7.
%%
%%%\bibitem{alloul2016} Alloul, H. in {\it Quantum Materials: Experiments and Theory} (Pavarini, E., Koch, E., van den Brink, J., Sawatsky, G. ed.) 13.1-13.30 (Forschungszentrum J\"{u}lich GmbH Institute for Advanced Simulation, 2016). at $<$https://juser.fz-juelich.de/record/819465/files/correl16.pdf$>$
%%
%%%\bibitem{aeppli1992} Aeppli, G., Fisk, Z., Kondo Insulators. {\it Comments Cond. Mat. Phys.} {\bf 16}, 155-170 (1992)). 
%%
%%
%%
%%
%%%
%%
%%
%
%%
%%
%%\bibitem{demler2004} Demler, E., Hanke, W., Zhang, S.-C., {\it {\bf SO(5)}} theory of antiferromagnetism and superconductivity. {\it Rev. Mod. Phys.} {\bf 76}, 909-974 (2004).
%%
%%\bibitem{chen2005} Chen, Q., Stajic, J., Tan, S., Leven, K., BCS–BEC crossover: From high temperature superconductors to ultracold superfluids. {\it Phys. Rep.} {\bf 312}, 1-88 (2005).
%%
%%\bibitem{randeria2014} Randeria, M., Taylor, E., Crossover from Bardeen-Cooper-Schrieffer to Bose-Einstein condensation and the unitary Fermi gas. {\it Annu. Rev. Condens. Matter Phys.} {\bf 5}, 209–232 (2014).
%%
%%\bibitem{kivelson2003} Kivelson, S.A., Fradkin, E., Oganesyan, V., Bindloss, I.P., Tranquada, J.M., Kapitulnik, A., Howald, C., How to detect fluctuating stripes in the high-temperature superconductors. {\it Rev. Mod. Phys.} {\bf 75}, 1201-1241 (2003).
%%
%
%
%
%
%
%%\bibitem{jaccarino1967} Jaccarino, V., Wertheim, G.K., Wernick, J.H., Walker, L.R., Paramagnetic excited state of FeSi. {\it Phys. Rev.} {\bf 160}, 476-482 (1967).
%
%
%
%
%
%
%
%
%
%
%
%
%
%
%
%
%
%
%
%
%
%
%%%%%%%%%%%%%%%%%%%new paragraph
%
%
%
%
%
%%%%%%%%%%%%%%%%%%%new paragraph
%
%
%
%
%
%%%%%%%%%%%%%%%%%%%new paragraph
%
%%%%%%%%%%%%%%%%%%%%
%
%
%
%%\bibitem{tari2003} A.~Tari, The heat capacity of matter at low temperatures (Imperial College Press, 1$^{\rm st}$ ed., 2003).
%
%%\bibitem{phelan2014} Phelan, W.A., Koohpayeh S.M., Cottingham, P., Freeland, J.W., Leiner, J.C., Broholm, C.L., McQueen, T.M., Correlation between bulk thermodynamic measurements
%%and the low-temperature-resistance plateau in SmB$_6$. {\it Phys. Rev. X} {\bf 4}, 031012 (2014).
%%
%%\bibitem{akintola2018} Akintola, K., Dunsiger, S.R., Fang, A.C.Y. Fang, Potma, M., Saha, S.R., Wang, X., Paglione, J., Sonier, J.E., Freezing out of a low-energy bulk spin exciton in SmB$_6$. {\it npg Quantum Materials} {\bf 3}, 36 (2018); doi:10.1038/s41535-018-0110-7
%%
%%\bibitem{kageyama2000} Kageyama, H., Onizuka, K., Ueda, Y., Nohara, M., Suzuki, H., Takagi, H., Low-temperature specific heat study of SrCuZ$_2$(BO$_3$)$_2$ with an exactly solvable ground state. {\it Journal of Experimental and Theoretical Physics} {\bf 90}, 129-132 (2000).
%
%%%%%%%%%%%% 3rd paragraph
%
%
%%%%%%%%%%%%%%%%%%
%
%
%
%
%%%%%%%%%%%%%% 4th paragraph
%
%
%%%%%%%%%%%%%% 5th paragraph
%
%%%%%%%%%%%%%% 6th paragraph
%
%
%
%%\bibitem{shen2018} Shen, H., Fu, L., Quantum oscillation from in-gap states and a non-Hermitian Landau level problem. {\it Phys. Rev. Lett.} {\bf 121}, 026403 (2018).
%%
%%\bibitem{ding2001} Ding, H., Engelbrecht, J.R., Wang, Z., Campuzano, J.C., Wang, S.-C., Yang, H.-B., Rogan, R., Takahashi, T., Kadowaki, K., Hinks, D.G., Coherent quasiparticle weight and its connection to high-$T_{\rm c}$ superconductivity from angle-resolved photoemission. {\it Phys. Rev. Lett.} {\bf 87}, 227001 (2001).
%
%
%
%
%
%
%
%%%%%%%%%%%%%% 7th paragraph
%
%
%%
%%
%
%%\bibitem{zhang1988} Zhang, F.C., Rice, T.M., Effective Hamiltonian for the superconducting Cu oxides. {\it Phys. Rev. B} {\bf 37}, 3759-3761 (1988).
%
%%%%%%%%%%%%%% 8th paragraph
%
%
%
%%%%%%%%%%%%%% 9th paragraph
%
%
%
%
%
%





%\bibitem{imada2019} Imada, M., Suzuki, J., Excitons and dark fermions as origins of Mott gap, pseudogap and superconductivity in cuprate superconductors --- general concept and basic formalism based on gap physics. {\it J. Phys. Soc. Japan} {\bf 88}, 024710 (2019).

%%%%%%%%%%%%% 10th paragraph





%%%%%%%




%
%\bibitem{zaanen1989} Zaanen, J., Gunnarsson, O., Charged magnetic domain lines and the magnetism of high-$T_{\rm c}$ oxides. {\it Phys. Rev. B} {\bf 40}, 7391-7394 (1989).
%
%
%
%
%
%
%
%
%
%
%
%%%%%%%%%%%%%%% 2nd paragraph
%
%\bibitem{yarmohammadi2017} Yarmohammadi, M., Electronic miniband structure, heat capacity and magnetic susceptibility of monolayer and bilayer silicene in TI, VSPM and BI regimes. {\it Physics Letters A} {\bf 381}, 1261-1267 (2017).
%
%
% \bibitem{jarrell1992} Jarrel, M., Hubbard-model in infinite dimensions --- a quantum Monte-Carlo study. {\it Phys. Rev. Lett.} {\bf 69}, 168-171 (1992).
%
%
%
%
%%%%%%%%%%%%%%% 7th paragraph
%
%
%
%
%\bibitem{kivelson2005} Kivelson, S.A., Fradkin, E. in {\it Handbook of high-temperature superconductivity. Theory and experiment} (Schrieffer J.R., Brookss, J.S. ed.)  570-596 (Springer, 2007). %How optimal inhomogeneity produces high temperature superconductivity
%
%

%
%
%
%
%
%
%
%
%
%
%
%
%
%
%\bibitem{norman1998} Norman, M. R., Ding, H., Randeria, M., Campuzano, J. C., Yokoya, T.,  Takeuchi, T., Takahashi, T., Mochiku, T., Kadowaki, K., Guptasarma, P., Hinks, D. G., Destruction of the Fermi surface underdoped high-$T_{\rm c}$ superconductors. {\it Nature} {\bf 392}, 157-160 (1998).
%
%
%
%\bibitem{lee2007} Lee, W.~S., Vishik, I.~M., Tanaka, K., Lu, D.H., Sasagawa, T., Nagaosa, N., Devereaux, T.P., Hussain, Z., Shen, Z.-X., Abrupt onset of a second energy gap at the superconducting transition of underdoped Bi2212. {\bf Nature} {\bf 450}, 81-84 (2007).
%
%\bibitem{yoshida2012} Yoshida, T., Hashimoto, M., Vishik, I. M., Shen, Z.-X., Fujimori, A., Pseudogap, superconducting gap, and Fermi Arc in high-$T_{\rm c}$ cuprates revealed by angle-resolved photoemission spectroscopy. {\it J. Phys. Soc. Japan} {\bf 81}, 011006 (2012).
%
%
%\bibitem{hayden1991} Hayden, S.M., Aeppli, G., Osborn, R., Taylor, A.D., Perring, T.G., Cheong, S.-W., Fis, Z., High-energy spin waves in La$_2$CuO$_4$, {\it Phys. Rev. Lett.} {\bf 67}, 3622-3625 (1991).
%
%\bibitem{hayden1996} Hayden, S.M., Aeppli, G., Perring, T.G., Mook, H.A., Do\u{g}an, High-frequency spin waves in YBa$_2$Cu$_3$O$_{6.15}$. {\it Phys. Rev. B} {\bf 54}, R6905-R6908 (1996).
%
%\bibitem{bourges1997}  Bourges, P., Casalta, H., Ivanov, A.S., Petitgrand, D., Superexchange coupling and spin susceptibility spectral weight in undoped monolayer cuprates. {\it Phys. Rev. Lett.} {\bf 79}, 4906-4909 (1997).
%
%
%\bibitem{paramekanti2004} Paramekanti, A., Randeria, M., Trivedi, M., High-$T_{\rm c}$ superconductors: A variational theory of the superconducting state. {\it Phys. Rev. B} {\bf 70}, 054504 (2004).
%
%\bibitem{mattheiss1987} Mattheiss, L.F., Electronic band properties and superconductivity in La$_{2-y}X_y$CuO$_4$., {\it Phys. Rev. Lett.} {\bf 58}, 1028-1030 (1987).
%
%\bibitem{andersen1995} Andersen, O. K., Liechtenstein, A. I., Jepsen, O. \& Paulsen, F. LDA energy bands, low energy Hamiltonians, $t^\prime$, $t^{\prime\prime}$, $t_\perp (k)$ and $J_\perp$, {\it J. Phys. Chem. Solids} {\bf 56}, 1573-1591 (1995).
%
%\bibitem{yoshida2006}  Yoshida, T., Zhou, X.J., Tanaka, K., Yang, W.L., Hussain, Z., Shen, Z.-X., Fujimori, A., Sahrakorpi, S., Lindroos, M., Markiewicz, R.S., Bansil, A., Komiya, S., Ando, Y., Eisaki, H., Kakeshita, T., Uchida, S., Systematic doping evolution of the underlying Fermi surface of La$_{2-x}$Sr$_x$CuO$_4$. {\it Phys. Rev. B} {\bf 74}, 224510 (2006).
%
%\bibitem{horio2018} Horio, M., Hauser, K., Sassa, Y., Mingazheva, Z., Sutter, D., Kramer, K., Cook, A., Nocerino, E., Forslund, O. K., Tjernberg, O., Kobayashi, M., Chikina, A., Schr\"{o}ter, N. B. M., Krieger, J. A., Schmitt, T., Strocov, V. N., Pyon, S., Takayama, T., Takagi, H., Lipscombe, O.J., Hayden, S.M., Ishikado, M., Eisaki, H. Neupert, T., M\o{a}nsson, M., Matt, C. E., Chang, J., Three-dimensional Fermi surface of overdoped La-based cuprates. {\it Phys. Rev. Lett.} {\bf 121} 077004 (2018).
%
%\bibitem{varma1997} Varma, C.M., Non-Fermi-liquid states and pairing instability of a general model of copper oxide metals. {\it Phys. Rev. B} {\bf 55}, 14554-14580 (1997).
%
%\bibitem{hussey2008} Hussey, N.E., Phenomenology of the normal state in-plane transport properties of high-$T_{\rm c}$ cuprates. {\it J. Phys.: Cond.Matter} {\bf 20}, 123201 (2008).
%
%
%

%
%

%
%\bibitem{kaminski2015} Kaminski, A., Kondo, T., Takeuchi, T., Gug, G., Pairing, pseudogap and Fermi arcs in cuprates. {\it Phil. Mag.} {\bf 95}, 453-466 (2015).
%
%
%\bibitem{basov2005} Basov, D.N., Timusk, T. Electrodynamics of high-$T_{\rm c}$ superconductors. {\it Rev. Mod. Phys.} {\bf 77}, 721-779 (2005).
%
%
%\bibitem{sebastian2010} Sebastian, S. E., Harrison, N., Altarawneh, M. M., Liang, R.-X., Bonn, D. A., Hardy, W. N., Lonzarich, G. G. Fermi-liquid behavior in an underdoped high-$T_{\rm c}$ superconductor. {\it Phys. Rev. B} {\bf 81}, 140505 (2010).
%
%
%\bibitem{radaelli1994} Radaelli, P.G., Hinks, D.G., Mitchell, A.W., Hunter, B.A., Wagner, J.L., Dabrowski, B., Vandervoort, K.G., Viswanathan, H.K., Jorgensen, J.D., Structure and superconducting properties of La$_{2-x}$Sr$_x$CuO$_4$ as a function of Sr content, {\it Phys. Rev. B} {\bf 49}, 4163-4175 (1994).
%
%
%\bibitem{zhou2003} Zhou, X.J., Yoshida, T.,  Lanzara, A., Bogdanov, P. V., Kellar, S. A., Shen, K. M., Yang, W. L., Ronning, F., Sasagawa, T.,  Kakeshita, T., Noda, T., Eisaki, H., Uchida, S., Lin, C. T., Zhou, F., Xiong, J. W.,  Ti, W. X., Zhao, Z. X., Fujimori, A., Hussain, Z., Shen, Z.-X., Universal nodal Fermi velocity. {\it Nature} {\bf 491}, 398 (2003).
%
%\bibitem{vishik2010} Vishik, I.M., Lee, W.S., Schmitt, F., Moritz, B., Sasagawa, T., Uchida, S., Fujita, K., Ishida, S., Zhang, C., Devereaux, T.P., Shen, Z.X., Doping-dependent nodal Fermi velocity of the high-temperature superconductor Bi$_2$Sr$_2$CaCu$_2$O$_{8+\delta}$ revealed using high-resolution angle-resolved photoemission spectroscopy. {\it Phys. Rev. Lett.} {\bf 104}, 207002 (2010).
%
%\bibitem{doiron2007} Doiron-Leyraud, N., Proust, C. LeBoeuf, D., Levallois, J., Bonnemaison, B., Liang, R.-X. Hardy, W. N., Bonn, D. A., Taillefer, L. Quantum oscillations and the Fermi surface in an underdoped high-$T_{\rm c}$ superconductor. {\it Nature} {\bf 447}, 565-568 (2007).
%
%
%\bibitem{barisic2013} Bari\u{s}i\'{c}, N., Badoux, S., Chan, M.K., Dorow, C., Tabis, W., Vignolle, B., Yu, G., B\'{e}ard, J., Zhao, X., Proust, C., Greven, M., Universal quantum oscillations in the underdoped cuprate superconductors. {\bf Nature Phys.} {\bf 9}, 761-764 (2013).
%
%\bibitem{chan2016} Chan, M. K., Harrison, N., McDonald, R. D., Ramshaw, B. J., Modic, K. A., Barisic, N., Greven, M. Single reconstructed Fermi surface pocket in an underdoped single-layer cuprate superconductor. {\it Nature Commun.} {\bf 7}, 12244 (2016).
%
%
%\bibitem{yang2008} Yang, H.-B., Rameau, J.D., Johnson, P.D., Valla, T., Tsvelik, A., Gu, G.D., Emergence of preformed Cooper pairs from the doped Mott insulating state in Bi$_2$Sr$_2$CaCu$_2$O$_8$, {\it Nature} {\bf 456}, 77-80 (2008).
%
%\bibitem{loram1998} Loram, J.W., Mirza, K.A., Cooper, J.R., Tallon, J.L., Specific heat evidence of the normal state pseudogap, {\it J. Phys. Chem Solids} {\bf 59}, 2091-2095 (1998).
%
%
%
%
%\bibitem{kacmarcik2018}  Ka\u{c}mar\u{c}\'{i}k, J., Vinograd, I., Michon, B., Rydh, A., Demeur, A., Zhou, R., Mayaffre, H., Liang, R., Hardy, W.N., Bonn, D.A., Doiron-Leyraud, N., Taillefer, L., Julien, M.-H., Marcenat, C., Klein, T., Unusual interplay between superconductivity and field-induced charge order in YBa$_2$Cu$_3$O$_y$. {\it Phys. Rev. Lett.}{\bf 121}, 167002 (2018).
%
%\bibitem{tallon2020} Tallon, J.L., Storey, J.G., Cooper, J.R., Loram, J.W., Locating the pseudogap closing point in cuprate superconductors: Absence of entrant or reentrant behavior. {\it Phys. Rev. B} {\bf 101}, 174512 (2020).
%
%\bibitem{storey2008} Storey, J.L., Tallon, J.L., Williams, G.V.M., Pseudogap ground state in high-temperature superconductors. {\it Phys. Rev. B} {\bf 78}, 140506 (2008).
%
%
%\bibitem{ghiringhelli2012} Ghiringhelli, G., Le Tacon, M., Minola, M., Blanco-Canosa, S., Mazzoli, C., Brookes, N.B., De Luca, G.M., Frano, A., Hawthorn, D.G., He, F., Loew, T., Sala, M.M., Peets, D.C., Salluzzo, M., Schierle, E., Sutarto, R., Sawatzky, G.A., Weschke, E., Keimer, B., Braicovich, L., Long-range incommensurate charge fluctuations in (Y,Nd)Ba$_2$Cu$_3$O$_{6+x}$. {\it Science} {\bf 337}, 821-825 (2212).
%
%\bibitem{chubukov2007} Chubukov, A.V., Norman, M.R., Millis, A.J., Abrahams, E., Gapless pairing and the Fermi arc in the cuprates. {\it Phys. Rev. B} {\bf 76}, 180501 (2007).
%
%
%
%
%\bibitem{miyakawa1999} Miyakawa, N., Zasadzinski, J.F., Ozyuzer, L., Guptasarma, P., Hinks, D.G,  Kendziora, C., Gray, K.E., Predominantly superconducting origin of large energy gaps in underdoped Bi$_2$Sr$_2$CaCu$_2$O$8+\delta$ from tunneling spectroscopy. {\it Phys. Rev. Lett.} {\bf 83}, 1018-1021 (1999).
%
%\bibitem{chiao2000} Chiao, M., Hill, R., Lupien, C., Taillefer, L., Lambert, P., Gagnon, R., Fournier, P., Low-energy quasiparticles in cuprate superconductors: A quantitative analysis. {\it Phys. Rev. B} {\bf 62}, 3554-3558 (2000).
%
%
%\bibitem{dai2020} Dai, Z., Senthil, T., Lee, P.A., Modeling the pseudogap metallic state in cuprates: Quantum disordered pair density wave. {\it Phys. Rev. B} {\bf 101}, 064502 (2020).
%
%\bibitem{kawasaki2010} kawasaki, K., Lin, C., Kuhns, P.L., Reyes, A.P., Zheng, G.-Q., Carrier-concentration dependence of the pseudogap ground state of superconducting Bi$_2$Sr$_{2-x}$La$_x$CuO$_{6+\delta}$ revealed by $^{63,65}$Cu-nuclear magnetic resonance in very high magnetic fields. {\it Phys. Rev. Lett.} {\bf 105}, 137002 (2010).
%
%
%
%
%
%
%
%
%






%\bibitem{peregbarnea2009} Pereg-Barnea, T., Weber, H., Rafael, G., Franz, M., Quantum oscillations from Fermi arcs. {\it Nature Phys.} {\bf 6}, 44-49 (2009).
%



%
%
%
%
%
%
%\bibitem{zhang1988} Zhang, F.C., Gros, C., Rice, T.M., Shibat, H., A renormalised Hamiltonian approach to a resonant valence bond wavefunction. {\it Supercond. Sci. Technol.} {\bf 1}, 36-46  (1988).
%
%\bibitem{shoenberg1984} Shoenberg, D. {\it Magnetic oscillations in metals}. (Cambridge Univ. Press, Cambridge, 1984).
%
%
%\bibitem{hartstein2020} Hartstein, M., Hsu, Y.-T., Modic, K.A., Porras, J., Loew, T., Le Tacon, M., Zuo, H., Wang, J., Zhu, Z., Chan, M.K., McDonald, R.D., Lonzarich, G.G., Keimer, B., Sebastian, S.E., Harrison, N., Hard antinodal gap revealed by quantum oscillations in the pseudogap regime of underdoped high-$T_{\rm c}$ superconductors. {\it Nature Phys.} (in press 2020).
%
%
%
%\bibitem{sebastian2014} Sebastian, S. E., Harrison, N., Balakirev, F. F., Altarawneh, M. M., Goddard, P. A., Liang, R.-X. Bonn, D. A., Hardy, W. N., Lonzarich, G. G. Normal-state nodal electronic structure in underdoped high-$T_{\rm c}$ copper oxides. {\it Nature} {\bf 511}, 61-64 (2014).
%
%
%
%
%
%
%
%%53
%\bibitem{mackenzie1} Carrington, A., Mackenzie, A. P., Sinclair D. C., Cooper, J. R. Field Dependence of the resistive transition in Tl$_2$Ba$_2$CuO$_{6+\delta}$. {\it Phys. Rev. B} {\bf 49}, 13243 (1994).
%
%%1
%
%%2
%
%%14
%%18
%\bibitem{wu1} Wu, T., Mayaffre, H., Kramer, S., Horvatic, M., Berthier, C., Hardy, W. N., Liang, R.-X., Bonn, D. A., Julien, M.-H. Magnetic-field-induced charge-stripe order in the high-temperature superconductor YBa$_2$Cu$_3$O$_y$. {\it Nature} {\bf 477}, 1910194 (2011).
%
%%20
%\bibitem{chang1} Chang, J., Blackburn, E., Holmes, A. T., Christensen, N. B., Larsen, J., Mesot, J., Liang, R. X., Bonn, D. A., Hardy, W. N., Watenphul, A., von Zimmermann, M., Forgan, E. M., Hayden, S. M. Direct observation of competition between superconductivity and charge density wave order in YBa$_2$Cu$_3$O$_{6.67}$. {\it Nature Phys.} {\bf 8}, 871-876 (2012).
%
%\bibitem{ghiringhelli1} Ghiringhelli, G., Le Tacon, M., Minola, M., Blanco-Canosa, S., Mazzoli, C., Brookes, N. B., De Luca, G. M., Frano, A., Hawthorn, D. G., He, F., Loew, T., Sala, M. M. Peets, D. C., Salluzzo, M., Schierle, E., Sutarto, R., Sawatzky, G. A., Weschke, E., Keimer, B., Braicovich, L. Long-range incommensurate charge fluctuations in (Y,Nd)Ba$_2$Cu$_3$O$_{6+x}$. {\it Science} {\bf 337}, 821-825 (2012).
%
%%19
%\bibitem{gerber1} Gerber, S., Jang, H., Nojiri, H., Matsuzawa, S., Yasumura, H., Bonn, D. A., Liang, R., Hardy, W. N., Islam, Z., Mehta, A., Song, S., Sikorski, M., Stefanescu, D., Feng, Y., Kivelson, S. A., Devereaux, T. P., Shen, Z.-X., Kao, C.-C., Lee, W.-S., Zhu, D., Lee, J.-S. Three-dimensional charge density wave order in YBa$_2$Cu$_3$O$_{6.67}$ at high magnetic fields. {\it Science} {\bf 350}, 949-952 (2015).
%
%\bibitem{gammel1} Gammel, P. L., Schneemeyer, L. F., Waszczak, J. V., Bishop, D. J., Evidemce from mechanical measurements for flux-lattice melting in single-crystal YBa$_2$Cu$_3$O$_7$ and Bi$_{2.2}$Sr$_2$Ca$_{0.8}$Cu$_2$O$_8$. {\it Phys. Rev. Lett.} {\bf 61}, 1666-1669 (1988).
%%26
%\bibitem{kwok1} Kwok, W. K., Fleshler, S., Welp, U., Vionkur, V. M., Downey, J., Crabtree, G. W., Miller, M. M. Vortex lattice melting in untwinned and twinned single-crystals of YBa$_2$Cu$_3$O$_{7-\delta}$. {\it Phys. Rev. Lett.} {\bf 69}, 3370-3373 (1992).
%
%\bibitem{blatter1} Blatter, G., Feigel'man, M. Y., Geshkenbein, Y. B., Larkin, A. I., Vinokur, V. M. Vortices in high-temperature superconductors
%. {\it Rev. Mod. Phys.} {\bf 66}, 1125-1388 (1994).
%
%\bibitem{zeldov1} Zeldov, E., Majer, D., Konczykowski, M., Geshkenbein, V. B., Vinokur, V. M., Shtrikman, H., Thermodynamic observation of first-order vortex lattice melting transition in Bi$_2$Sr$_2$CaCu$_2$O$_8$, {\it Nature} {\bf 375}, 373-376 (1995).
%
%%55
%\bibitem{ando1} Ando, Y., Komiya, S., Segawa, K., Ono, S., Kurita, Y. Electronic phase diagram of high-$T_{\rm c}$ cuprate superconductors from a mapping of the in-plane resistivity curvature. {\it Phys. Rev. Lett.} {\bf 93}, 267001 (2004).
%
%%56
%\bibitem{takagi1} Takagi, H., Batlogg, B., Kao, H. L., Kwo, J., Cava, R. J., Krajewski, J. J., Peck, W. F. Jr. Systematic evolution of temperature-dependent resistivity in La$_{\rm 2-x}$Sr$_{\rm x}$CuO$_4$. {\it Phys. Rev. Lett.} {\bf 69}, 2975 (1992).
%
%%57
%
%%27
%\bibitem{rullier1} Rullier-Albenque, F., Alloul, H., Proust, C., Lejay, P., Forget, A., Colson, D. Total suppression of superconductivity by high magnetic fields in YBa$_2$Cu$_3$O$_{6.6}$. {\it Phys. Rev. Lett.} {\bf 99}, 027003 (2007).
%
%\bibitem{greven1} Chan, M.?K., Veit, M.?J., Dorow, C.?J., Ge, Y., Li, Y., Tabis, W., Tang, Y., Zhao, X., Bari\u{s}i\'{c}, N., Greven, M., In-plane magnetoresistance obeys Kohler?s rule in the pseudogap phase of cuprate superconductors, {\it Phys. Rev. Lett.} {\bf 113}, 177005 (2014).
%
%%4
%
%%15
%\bibitem{proust1} Proust, C., Vignolle, B., Levallois, J., Adachi, S., Hussey, N. E. Fermi liquid behavior of the in-plane resistivity in the pseudogap state of YBa$_2$Cu$_4$O$_8$. {\it Proc. Nat. Acad. Sci. USA} {\bf 113} 13654-13659 (2016).
%
%\bibitem{tinkham2} Tinkham, M. Resistive transition of high-temperature superconductors. {\it Phys. Rev. Lett.} {\bf 61}, 1658-1661 (1988).
%
%%7
%\bibitem{grissonnanche2} Grissonnanche, G., Laliberte, F., Dufour-Beausejour, S., Matusiak, M., Badoux, S., Tafti, F. F., Michon, B., Riopel, A., Cyr-Choiniere, O.,  Baglo, J. C., Ramshaw, B. J., Liang, R., Bonn, D. A., Hardy, W. N., Kramer, S., LeBoeuf, D., Graf, D.,  Doiron-Leyraud, N., Taillefer, L. Wiedemann-Franz law in the underdoped cuprate superconductor YBa$_2$Cu$_3$O$_y$. {\it Phys. Rev. B} {\bf 93}, 064513 (2016).
%
%%10
%\bibitem{xu1} Xu, Z. A., Ong, N. P., Wang, Y., Kakeshita, T., Uchida, S. Vortex-like excitations and the onset of superconducting phase fluctuation in underdoped La$_{2-x}$Sr$_x$CuO$_4$. {\it Nature} {\bf 406}, 486-488 (2000).
%
%%9
%\bibitem{fu1} Yu, F., Hirschberger, M., Loew, T., Li, G., Lawson, B. J., Asaba, T., Kemper, J. B., Liang, T., Porras, J., Boebinger, G. S., Singleton, J., Keimer, B., Li, L., Ong, N. P. Magnetic phase diagram of underdoped YBa$_2$Cu$_3$O$_y$ inferred from torque magnetization and thermal conductivity. {\it Proc. Nat. Acad. Sci. USA} {\bf 113}, 12667-12672 (2016).
%
%%11
%\bibitem{wang1} Wang, Y. Y., Ong, N. P., Xu, Z. A., Kakeshita, T., Uchida, S., Bonn, D. A., Liang, R., Hardy, W. N. High field phase diagram of cuprates derived from the Nernst effect. {\it Phys. Rev. Lett.} {\bf 88}, 257003 (2002).
%
%%12
%\bibitem{ong_ref1} Wang, Y., Li, L., Ong, N. P. Nernst effect in high-$T_{\rm c}$ superconductors. {\it Phys. Rev. B} {\bf 73}, 024510 (2006).
%
%%13
%\bibitem{ong_ref2} Li, L., Wang, Y., Komiya, S., Ono, S., Ando, Y., Gu, G. D., Ong, N. P. Diamagnetism and Cooper pairing above $T_{\rm c}$ in cuprates. {\it Phys. Rev. B} {\bf 81}, 054510 (2010).
%
%%5
%\bibitem{ramshaw7} Ramshaw, B. J., Day, J., Vignolle, B., LeBoeuf, D., Dosanjh, P., Proust, C.,  Taillefer, L.,  Liang, R. X., Hardy, W. N.,  Bonn D. A.Vortex lattice melting and $H_{\rm c2}$ in underdoped YBa$_2$Cu$_3$O$_y$. {\it Phys. Rev. B} {\bf 86}, 174501 (2012).
%
%%6

%
%%8
%\bibitem{badoux1} Badoux, S., Tabis, W., Lalibert\'{e}, F., Grissonnanche, G., Vignolle, B., 	Vignolles,	D., B\'{e}ard, J., Bonn, D. A., Hardy, W. N., Liang, R., Doiron-Leyraud, N., Taillefer, L.,	Proust, C. Change of carrier density at the pseudogap critical point of a cuprate superconductor. {\it Nature} {\bf 531}, 210-214 (2016).
%
%%25
%\bibitem{senoussi1} Senoussi, S. Review of the critical current densities and magnetic irreversibilities in high $T_{\rm c}$ superconductors. {\it J. Phys. III France} {\bf 2}, 1041-1257 (1992).
%
%
%
%
%
%
%
%
%
%
%
%
%
%
%%16
%
%%17
%
%
%
%
%
%
%
%%21
%\bibitem{zhou1} Zhou, R., Hirata, M., Wu, T.,  Vinograd, I. , Mayaffre, H., Kr\"{a}mer, S. Horvati\'{c}, M.,  Berthier, C., Reyes, A. P., Kuhns, P. L., Liang, R., Hardy, W. N., Bonn, D. A.,  Julien, M.-H. Quasiparticle scattering off defects and possible bound states in charge-ordered 
%YBa$_2$Cu$_3$O$_y$. {\it Phys. Rev. Lett.} {\bf 118}, 017001 (2017).
%
%%22
%\bibitem{blanco1} Blanco-Canosa, S.,  Frano, A. , Schierle, E., Porras, J., Loew, T., Minola, M., Bluschke, M., Weschke, E., Keimer, B., Le~Tacon, M. Resonant X-ray scattering study of charge-density wave correlations in YBa$_2$Cu$_3$O$_{6+x}$. {\it Phys. Rev. B} {\bf 90}, 054513 (2014).
%
%%23
%\bibitem{letacon1} Le~Tacon, M., Sacuto, A., Georges, A., Kotliar, G., Gallais, Y., Colson, D. Two energy scales and two distinct quasiparticle dynamics in the superconducting state of underdoped cuprates. {\it Nature Phys.} {\bf 2}, 537-543 (2006).
%
%%24
%\bibitem{hufner1} H\"{u}fner, S., Hossain, M. A., Damascelli, A., Sawatsky, G. A. Two gaps make a high-temperature superconductor? {\it Rep. Prog. Phys.} {\bf 71}, 062501 (2008).
%
%
%
%
%%28
%\bibitem{zhou2} Zhou, R., Hirata, M., Wu, T., Vinograd, I., Mayaffre, H., Kr{\"a}mer, S., Reyes, A. P., Kuhns, P. L., Liang, R., Hardy, W. N., Bonn, D. A., and Julien, M. -H. Spin susceptibility of charge-ordered YBa$_2$Cu$_3$O$_{\rm y}$ across the upper critical field. {\it Proc. Natl. Acad. Sci.} {\bf 114}, 13148 - 13153 (2017).
%
%%29
%\bibitem{marcenat1} Marcenat, C., Demuer, A. Beauvois, K., Michon, B., Grockowiak, A., Liang, R., Hardy, W., Bonn, D.~A., Klein, T. Calorimetric determination of the magnetic phase diagram of underdoped ortho-II YBa$_2$Cu$_3$O$_{6.54}$ single crystals. {\it Nature Commun.} {\bf 6}, 7927 (2015).
%
%%30
%\bibitem{cyr1}  Cyr-Choini\`{e}re, O., Daou, R.,  Lalibert\'{e}, F., LeBoeuf, D., Doiron-Leyraud, N., Chang, J., Yan, J.-Q., Cheng, J.-G., Zhou, J.-S.,Goodenough, J. B., Pyon, S. Takayama, T., Takagi, H., Tanaka, Y.,Taillefer, L. Enhancement of the Nernst effect by stripe order in a high-$T_{\rm c}$ superconductor. {\it Nature} {\bf 458}, 743-745 (2009).
%
%%30
%\bibitem{riggs1} Riggs, S.C., Vafek, O., Kemper, J. B., Betts, J. B., Migliori, A., Balakirev, F. F., Hardy, W. N., Liang, R.-X., Bonn, D. A., Boebinger, G. S. Heat capacity through the magnetic-field-induced resistive transition in an underdoped high-temperature superconductor. {\it Nature Phys.} {\bf 7}, 332-335 (2011).
%
%%31
%\bibitem{sebastian3} Sebastian, S. E., Harrison, N., Palm, E., Murphy, T. P., Mielke, C. H.,  Liang, R.-X., Bonn, D. A., Hardy, W. N., Lonzarich, G. G. A multi-component Fermi surface in the vortex state of an underdoped high-$T_{\rm c}$ superconductor. {\it Nature} {\bf 454}, 200-203 (2008).
%
%%32
%\bibitem{druyvesteyn1} Druyvesteyn, W. F., Van Ooijen, D. J., Berben, T. J. Variation of the Critical Fields of Superconducting Lead with the Residual Resistivity. {\it Rev. Mod. Phys.} {\bf 36}, 58 (1964).
%
%%33
%\bibitem{lyard1} Lyard, L., Samuely, P., Szabo, P., Klein, T., Marcenat, C., Paulius, L., Kim, K. H. P.,  Jung, C. U.,   Lee, H.-S..
%Kang, B.,  Choi, S.,   Lee, S.-I.,  Marcus, J.  Blanchard, S.,   Jansen, A. G. M., Welp, U., Karapetrov, G., Kwok, W. K. Anisotropy of the upper critical field and critical current in single crystal MgB$_2$. {\it Phys. Rev. B} {\bf 66}, 180502(R) (2002).
%
%%34
%
%
%%35
%\bibitem{sebastian100} Sebastian, S. E., Harrison, N.,  Liang, R., Bonn, D. A., Hardy, W. N., Mielke, C. H., Lonzarich, G. G. Quantum oscillations from nodal bilayer magnetic breakdown in the underdoped high temperature superconductor YBa$_2$Cu$_3$O$_{6+x}$. {\it Phys. Rev. Lett.} {\bf 108}, 196403 (2012).
%
%%36
%\bibitem{corcoran1} Corcoran, R., Harrison, N., Hayden, S. M., Meeson, P., Springford, M., Vanderwel, P. J. Quasiparticles in the vortex state of V$_3$Si. {\it Phys. Rev. Lett.} {\bf 72}, 701-704 (1994).
%
%%37
%\bibitem{maniv1} Maniv, T., Zhuravlev, V., Vagner, I., Wyder, P. Vortex states and quantum magnetic oscillations in conventional type-II
%superconductors. {\it Rev. Mod. Phys.} {\bf 73} 867-911 (2001).
%
%%38
%\bibitem{yasui1} Yasui, K., Kita, T. Theory of the de Haas-van Alphen effect in type-II superconductors. {\it Phys. Rev. B} {\bf 66}, 184516 (2002).
%
%%39
%\bibitem{senthil1} Senthil, T., Lee, P. A. Synthesis of the phenomenology of the underdoped cuprates. {\it Phys. Rev. B} {\bf 79}, 245116 (2009).
%
%%40
%\bibitem{micklitz1} Micklitz, T., Norman, M. R. Nature of spectral gaps due to pair formation in superconductors. {\it Phys. Rev. B} {\bf 80}, 220513(R) (2009).
%
%%41
%\bibitem{banerjee1} Banerjee, S., Zhang, S., Randeria, M. Theory of quantum oscillations in the vortex-liquid state of high-$T_{\rm c}$ superconductors. {\it Nature Commun.} {\bf 4}, 1700 (2013).
%
%%42
%\bibitem{spivak1} Spivak,~B., Oreto,~P., Kivelson, S.~A. Theory of quantum metal to superconductor transitions in highly conducting systems. {\it Phys. Rev. B} {\bf 77}, 214523 (2008).
%
%%43
%\bibitem{kallin1} Zelli, M., Kallin, C., and Berlinsky, A. J. Quantum oscillations in a $\pi$-striped superconductor. {\it Phys. Rev. B} {\bf 86}, 104507 (2012).
%
%%44
%\bibitem{varma1} Varma, C. M. Magneto-oscillations in underdoped cuprates. {\it Phys. Rev. B} {\bf 79}, 085110 (2009).
%
%%45
%\bibitem{vafek1} Melikyan, A., Vafek, O. Quantum oscillations in the mixed state of d-wave superconductors. {\it Phys. Rev. B} {\bf 78}, 020502(R) (2008).
%
%%46
%\bibitem{miyake1} Miyake, K., de Haas-van Alphen oscillations in superconducting states as a probe of gap anisotropy. {\it Physica B} {\bf 186}-{\bf 188}, 115-117 (1993).
%
%%47
%\bibitem{terashima1} Terashima, T., Haworth, C., Takeya, H., Uji, S., Aoki, H. Kadowaki, K. Small superconducting gap on part of the Fermi surface of YNi$_2$B$_2$C from the de Haas-van Alphen effect. {\it Phys. Rev. B} {\bf 56}, 5120-5123 (1997).
%
%%48
%\bibitem{franz1} Franz, M., Tesanovic, Z. Self-consistent electronic structure of a $d_{x^2-y^2}$ and a $d_{x^2-y^2}+id_{xy}$ vortex. {\it Phys. Rev. Lett.} {\bf 80}, 4763-4766 (1998). 
%
%%49
%\bibitem{houghton1} Houghton, A., Pelcovits, R. A., Sudbo, A. Flux lattice melting in high-$T_{\rm c}$ superconductors. {\it Phys. Rev. B} {\bf 40}, 6763-6770 (1989).
%
%%50
%\bibitem{safar1} Safar, H., Gammel, P. L., Huse, D. A., Bishop, D. J. Experimental evidence for a multicritical point in the magnetic phase diagram for the mixed state of clean, untvvinned YBa$_2$Cu$_3$O$_7$. {\it Phys. Rev. Lett.} {\bf 70}, 3800-3803 (1993).
%
%%51
%\bibitem{tinkham1} Tinkham, M., {\it Introduction to superconductivity} (McGraw-Hill, Inc., New York, 1996).
%
%%52
%\bibitem{oh1} Oh, B., Char, K., Kent, A. D., Naito, M., Beasley, M. R., Geballe, T. H., Hammond, R. H., Kapitulnik, A., Graybeal, J. M. Upper critical field, fluctuation conductivity, and dimensionality of YBa$_2$Cu$_3$O$_{7-x}$. {\it Phys. Rev. B} {\bf 37}, 7861-7864 (1988).
%
%
%%54
%\bibitem{ito1} Ito, T., Takenaka, K., Uchida, S. Systematic deviation from $T$-linear behavior in the in-plane resistivity of YBa$_2$Cu$_3$O$_{7-y}$: Evidence for dominant spin scattering. {\it Phys. Rev. Lett.} {\bf 70}, 3995-3998 (1993).
%
%
%
%%58
%\bibitem{naqib1} Naqib, S. H., Cooper, J. R., Tallon, J. L., Islam, R. S., Chakalov, R. A. Doping phase diagram of Y$_{1-x}$Ca$_x$Ba$_2$(Cu$_{1-y}$Zn$_y$)$_3$O$_{7-\delta}$ from transport measurements: Tracking the pseudogap below $T_{\rm c}$. {\it Phys. Rev. B} {\bf 71}, 054502 (2005).
%
%%59
%\bibitem{alloul1} Rullier-Albenque, F., Alloul, H., Rikken, G. High-field studies of superconducting fluctuations in high-$T_{\rm c}$ cuprates:
%Evidence for a small gap distinct from the large pseudogap. {\it Phys. Rev. B} {\bf 84}, 014522 (2011).
%
%%60
%%\bibitem{welp2} Welp, U., Fleshier, S., Kwok, W. K., Klemm, R. A., Vinkur, V. M., Downey, J., Veal, B., Crabtree, G. W. High-field scaling behavior of thermodynamic and transport quantities of YBa$_2$Cu$_3$O$_{7-\delta}$ near the superconducting transition. {\it Phys. Rev. Lett.} {\bf 67}, 3180-3183 (1991).
%
%%60
%\bibitem{randeria1} Randeria, M., Trivedi, N., Moreo, A., Scalettar, R. T. Pairing and spin gap in the normal state of short coherence length superconductors. {\it Phys. Rev. Lett.} {\bf 69}, 2001-2004 (1992).
%
%%61
%
%%62
%
%%63
%\bibitem{emery1} Emery, V. J., Kivelson, S. A. Importance of phase fluctuations in superconductors with small superfluid density. {\it Nature} {\bf 374}, 434-437 (1995).
%
%%64
%\bibitem{ong_ref3}Li, L., Wang, Y., Naughton, M. J., Ono, S., Ando, Y., Ong, N. P. Strongly nonlinear magnetization above $T_{\rm c}$ in Bi$_2$Sr$_2$CaCu$_2$O$_{8+\delta}$. {\it EPL} {\bf 72}, 451 (2005).
%
%%65
%\bibitem{ong_ref4} Wang, Y., Li, L., Naughton, M. J., Gu, G. D., Uchida, S., Ong, N. P. Field-Enhanced Diamagnetism in the Pseudogap State of the Cuprate Bi$_2$Sr$_2$CaCu$_2$O$_{8+\delta}$ Superconductor in an Intense Magnetic Field. {\it Phys. Rev. Lett.} {\bf 95}, 247002 (2005).
%
%%66
%\bibitem{ong_ref5} Wang, Y., Li, L., Ong, N. P. Nernst effect in high-$T_{\rm c}$ superconductors. {\it Phys. Rev. B}, {\bf 73}, 024510 (2006).
%
%%67
%\bibitem{bernhard1} Dubroka, A., R\"{o}ssle, M., Kim, K.W., Malik, V. K., Munzar, D., Basov, D. N., Schafgans, A. A., Moon, S. J., Lin, C. T.,
%Haug, D., Hinkov, V., Keimer, B., Wolf, Th., Storey, J. G., Tallon, J. L., Bernhard, C. Evidence of a precursor superconducting phase at Temperatures as high as 180~K in $R$Ba$_2$Cu$_3$O$_{7-\delta}$ ($R=$~Y, Gd, Eu) superconducting crystals from infrared spectroscopy. {\it Phys. Rev. Lett.} {\bf 106}, 047006 (2011).
%
%%68
%\bibitem{bilbro1} Beilbro, L.S., Valde\'{e}s Aguilar, R., Logvenov, G., Pelleg, O., Bo\u{z}ovi\'{c}, I., Armitage, N.~P. Temporal correlations of superconductivity above the transition temperature in La$_{2-x}$Sr$_x$CuO$_4$ probed by terahertz spectroscopy. {\it Nature Phys.} {\bf 7}, 
%298-302 (2011).
%
%%69
%\bibitem{vishik1} Vishik, I.M., Lee, W.S., Schmitt, F., Moritz, B., Sasagawa, T., Uchida, S., Fujita, K., Ishida, S., Zhang, C., Devereaux, T.P., Shen, Z.X. Doping-dependent nodal Fermi velocity of the high-temperature superconductor Bi$_2$Sr$_2$CaCu$_2$O$_{8+\delta}$ revealed using high-resolution angle-resolved photoemission spectroscopy. {\it Phys. Rev. Lett.} {\bf 104}, 207002 (2010).
% 
%%70
%\bibitem{pushp1} Pushp, A., Parker, C.V., Pasupathy, A.N., Gomes, K.K., Ono, S., Wen, J., Xu, Z., Gu, G., Yazdani, A. Extending universal nodal excitations optimizes superconductivity in Bi$_2$Sr$_2$CaCu$_2$O$_{8+\delta}$. {\it Science} {\bf 324}, 1689-1693 (2009).
%
%%
%%\bibitem{franz} Pereg-Barnea, T., Weber, H., Refael, G., Franz, M. Quantum oscillations from Fermi arcs. {\it Nature Physics} {\bf 6}, 44â49 (2010).
%
%%71
%\bibitem{xia1} Xia, J., Schemm, E.,  Deutscher, G., Kivelson, S. A., Bonn, D. A., Hardy, W. N., Liang, R., Siemons, W., Koster, G., Fejer, M. M.,  Kapitulnik, A. Polar Kerr-effect measurements of the high-temperature YBa$_2$Cu$_3$O$_{6+x}$ superconductor: Evidence for broken symmetry near the pseudogap temperature. {\it Phys. Rev. Lett.} {\bf 100}, 127002 (2008).
%
%%72
%\bibitem{matsuda1} Sato, Y., Kasahara, S., Murayama, H., Kasahara, Y., Moon, E.G., Nishizaki, T., Loew, T., Porras, J., Keimer, B., Shibauchi, T., Matsuda, Y. Thermodynamic evidence for nematic phase transition at the onset of pseudogap in YBa$_2$Cu$_3$O$_y$. {\it Nature Phys.} online publication nphys4205 (2017).
%
%%73
%\bibitem{shekhter1} Shekhter, A., Ramshaw, B. J., Liang, R.-X., Hardy, W. N., Bonn, D. A., Balakirev, F.~F., McDonald, R.~D., Betts, J.~B., 	Riggs, S.~C., Migliori, A. Bounding the pseudogap with a line of phase transitions in YBa$_2$Cu$_3$O$_{6+\delta}$. {\it Nature} {\bf 498}, 75-77 (2013).
%
%%74
%\bibitem{hsieh1} Zhao, L., Belvin, C. A., Liang, R.  Bonn, D. A., Hardy, W. N., Armitage, N. P., Hsieh, D. A global inversion-symmetry-broken phase inside the pseudogap region of YBa$_2$Cu$_3$O$_y$. {\it Nature Phys.} {\bf 13}, 250-255 (2017).
%
%%75
%\bibitem{bourges1} Fauqu$\acute{e}$, B, Sidis, Y., Hinkov, V., PailhÃšs, S., Lin, C. T., Chaud, X., and Bourges, P. Magnetic Order in the Pseudogap Phase of High-$T_{\rm c}$ Superconductors. {\it Phys. Rev. Lett.} {\bf 96}, 197001 (2006).
%
%%76
%\bibitem{baskaran1} Baskaran, G., Zou, Z., Anderson, P. W. The resonating valence bond state and high-$T_{\rm c}$ superconductivity a mean field theory. {\it Solid State Comms.} {\bf 63}, 973-976 (1987).
%
%%77
%\bibitem{kotliar1} Kotliar, G., Liu, J. Superexchange mechanism and d-wave superconductivity. {\it Phys. Rev. B} {\bf 38}, 5142 (1988).
%
%%78
%\bibitem{fukuyama1} Fukuyama, H. On magnetic properties of high $T_{\rm c}$ oxides. {\it Progress of Theoretical Physics Supplement} {\bf 108}, 287-295 (1992).
%
%%79
%\bibitem{doniach1} Doniach, S., Inui, M. Long-range Coulomb interactions and the onset of superconductivity in the high $T_{\rm c}$ materials. {\it Phys. Rev. B} {\bf 41}, 6668 (1990).
%
%%81
%\bibitem{larkin1} Geshkenbein, V. B., Ioffe, L. B., Larkin, A. I. Superconductivity in a system with preformed pairs. {\it Phys. Rev. B} {\bf 55}, 3173 (1997).
%
%%82
%\bibitem{millis1} Ioffe L. B., Millis A. J. Big fast vortices in the d-wave resonating valence bond theory of high-temperature superconductivity. {\it Phys. Rev. B} {\bf 66}, 094513 (2002).
%
%%83
%\bibitem{auerbach1} Altman, E., Auerbach, A. Plaquette boson-fermion model of cuprates. {\it Phys. Rev. B} {\bf 65}, 104508 (2002).
%
%%84
%\bibitem{chatterjee3} Chatterjee, S., Sachdev, S. Insulators and metals with topological order and discrete symmetry breaking. {\it Phys. Rev. B} {\bf 95}, 205133 (2017). 
%
%%85
%\bibitem{pepin6} Kloss, T., Montiel, X., 1 P\'{e}pin, C. SU(2) symmetry in a realistic spin-fermion model for cuprate superconductors. {\it Phys. Rev. B} {\bf 91}, 205124 (2015).
%
%%86
%\bibitem{lee3} Lee, J., Fujita, K., Schmidt, A. R., Kim, C. K., Eisaki, H., Uchida, S., Davis J. C. Spectroscopic fingerprint of phase-incoherent superconductivity in the underdoped Bi$_2$Sr$_2$CaCu$_2$O$_{8+\delta}$. {\it Science} {\bf 325}, 1099-1103 (2009).
%
%%87
%\bibitem{fradkin1} Fradkin, E., Kivelson, S. A., Tranquada, J. M. Colloquium: Theory of intertwined orders in high temperature superconductors. {\it Rev. Mod. Phys.} {\bf 87} 457-482 (2015).
%
%%88
%\bibitem{wang2} Wang, Y., Agterberg, D. F., Chubukov, A. Coexistence of charge-density-wave and pair-density-wave orders in underdoped cuprates. {\it Phys. Rev. Lett.} {\bf 114}, 197001 (2015).
%
%%89
%\bibitem{lee1} Lee, P.~A. Amperean pairing and the pseudogap phase of cuprate superconductors. {\it Phys. Rev. X} {\bf 4}, 031017 (2014).
%
%%these references below are only in the methods....
%%
%%
%
%%90
%\bibitem{lin2} Lin, C.T., Liang, B., Chen, H.C. Top-seeded solution growth of Ca-doped YBCO single crystals. {\it J. Crys. Grow.} {\bf 237-239}, 778-782 (2002).
%
%%91
%\bibitem{borisenko2} Borisenko, S.V., Kordyuk, A. A., Zabolotnyy, V. Geck, J.,  Inosov, D.,  Koitzsch, A., Fink, J., Knupfer, M., B\"{u}chner, B., Hinkov, V., Lin, C. T.,  Keimer, B., Wolf, T., Chiuzbaian, S. G.,  Patthey, L., and Follath, R. Kinks, nodal bilayer splitting, and interband scattering in YBa$_2$Cu$_3$O$_{6+x}$. {\it Phys. Rev. Lett.} {\bf 96} 117004 (2006).
%
%%references below are included in Figure caption
%%%%%%%%%%%%%%%%%%%%%%%%%%%%%
%
%
%
%%91
%\bibitem{sebastianqcp} Sebastian, S. E., Harrison, N., Altarawneh, M. M., Mielke, C. H., Liang, R., Bonn, D. A., Hardy, W. N., Lonzarich, G. G. Metal-insulator quantum critical point beneath the high $T_{\rm c}$ superconducting dome. {\it Proc. Natl. Acad. Sci.} {\bf 107}, 6175 (2010).
%
%%92
%\bibitem{beanmodel} Bean, C. P. Magnetization of high-field superconductors. {\it Reviews of Modern Physics} {\bf 36}(1), 31-39 (1964).

%\bibitem{anderson1} Anderson, P. W., The resonating valence bond in La$_2$CuO$_4$ and superconductivity. {\it Science} {\bf 235}, 1196-1198 (1987).

%\bibitem{varma1} Varma, C. M., Pseudogap phase and the quantum-critical point in copper-oxide metals. {\it Phys. Rev. Lett.} {\bf 83}, 3538-3541 (1999).

%\bibitem{chakravarty1} Chakravarty, S., Laughlin, R. B., D. K. Morr, Nayak, C. Hidden order in the cuprates. {\it Phys. Rev. B} {\bf 63}, 094503 (2001).

%\bibitem{kivelson2} Kivelson, S. A., Bindloss, I. P., Fradkin, E., Oganesyan, V., Tranquada, J. M., Kapitulnik, A., Howald, C. How to detect fluctuating stripes in the high-temperature superconductors. {\it Rev. Mod. Phys.} {\bf 75}, 1201-1241 (2003).

%\bibitem{cyr3} Cyr-Choiniere, O., LeBoeuf, D., Badoux, S., Dufour-Beausejour, S., Bonn, D. A., Hardy, W. N.,  Liang, R., Doiron-Leyraud, N.,   Taillefer, L. Suppression of charge order by pressure in the cuprate superconductor YBa$_2$Cu$_3$O$_y$: Restoring the full superconducting dome. arXiv:1503.02033 (2015).
%%
%\bibitem{letacon2} Le Tacon, M., Ghiringhelli, G., Chaloupka, J., Moretti Sala, M., Hinkov, V., Haverkort, M. W., Minola, M., Bakr, M., Zhou, K. J., Blanco-Canosa, S., Monney, C., Song, Y. T.,	Sun, G. L., Lin, C. T., De Luca, G. M., Salluzzo, M., Khaliullin, G., Schmitt, T., Braicovich, L., Keimer, B. Intense paramagnon excitations in a large family of high-temperature superconductors. {\it Nature Phys.} {\bf 7}, 725?730 (2011).
%
%\bibitem{ghiringhelli1} Ghiringhelli, G., Le Tacon, M., Minola, M., Blanco-Canosa, S., Mazzoli, C., Brookes, N. B., De Luca, G. M., Frano, A., Hawthorn, D. G., He, F., Loew, T., Sala, M. M. Peets, D. C., Salluzzo, M., Schierle, E., Sutarto, R., Sawatzky, G. A., Weschke, E., Keimer, B., Braicovich, L. Long-range incommensurate charge fluctuations in (Y,Nd)Ba$_2$Cu$_3$O$_{6+x}$. {\it Science} {\bf 337}, 821-825 (2012).
%
%\bibitem{sordi1} Sordi, G., S\'{e}mon, P., Haule, K., Tremblay, A.-M. S. Pseudogap temperature as a Widom line
%in doped Mott insulators. {\it Sci. Rep.} {\bf 2}, 547 (2012).
%

%\bibitem{taylor1} Taylor, B. J., Scanderbeg, D. J.,  Maple, M. B., Kwon, C., Jia, Q. X. Role of quantum fluctuations in the vortex solid to vortex liquid transition of type-II superconductors. {\it Phys. Rev. B} {\bf 76}, 014518 (2007).

%\bibitem{sademelo1} S\'{a} de Melo, C. A. R., Randeria, M., Engelbrecht, J. R., Crossover from BCS to Bose superconductivity: transition- temparature and time-dependent Ginzburg-Landau theory {\it Phys. Rev. Lett.} {\bf 71}, 3202-3205 (1993).

%

%
%
%\bibitem{gomes1} Gomes, K. K., Pasupathy, A. N., Pushp, A., Ono, S., Ando, Y., Yazdani, A. Visualizing pair formation on the atomic scale in the high-$T_{\rm c}$ superconductor Bi$_2$Sr$_2$CaCu$_2$O$_{8+\delta}$. {\it Nature} {\bf 447}, 569-572 (2007). 
%

%The pseudogap in high-temperature superconductors: an experimental survey
%By: Timusk, T; Statt, B
%REPORTS ON PROGRESS IN PHYSICS  Volume: 62   Issue: 1   Pages: 61-122   Published: JAN 1999

%arXiv:1503.07572 (2015). PHYSICAL REVIEW B


%\bibitem{cubitt1} Cubitt, R., Forgan, E. M., Yang, G., Lee, S. L., Paul, D. M., Mook, H. A., Yethiraj, M., Kes, P. H., Li, T. W., Menovsky A. A., Tarnawski, Z., Mortensen, K. Direct observation of magnetic-flux-line lattice melting and decomposition in the high-$T_{\rm c}$ superconductor Bi$_{2.15}$Sr$_{1.95}$CaCu$_2$O$_{8+x}$. {\it Nature} {\bf 365}, 407-411 (1993).
%
%\bibitem{safar2} Safar, H., Gammel, P. L., Huse, D. A., Alers, G. B., Bishop, D. J., Lee, W. C., Giapintzakis, J., Ginsberg, D. M. Vortex dynamics below the flux-lattice melting transition in YBa$_2$Cu$_3$0$_{7-\delta}$. {\it Phys. Rev. B} {\bf 52}, 6211-6214 (1995).
%
%\bibitem{charalambous1} Charalambous, M., Koch, R. H., Masselink, T., Doany, T., Feild, C.,  Holtzberg, F. Subpicovolt resolution measurements of the current-voltage characteristics of twinned crystalline YBa$_2$Cu$_3$O$_{7-x}$: new evidence for a vortex-glass phase. {\it Phys. Rev. Lett.} {\bf 75}, 2578-2581 (1995).
%
%
%\bibitem{senoussi1} Senoussi, S. Review of the critical current densities and magnetic irreversibilities in high-$T_{\rm c}$ superconductors. {\it J. Phys. III France} {\bf 2},  1041-1257 (1992).
%
%
%\bibitem{anderson2} Anderson, P. W. Two new vortex liquids. {\it Nature Physics} {\bf 3}, 160-162 (2007).
%
%\bibitem{doiron4} Doiron-Leyraud, N., Badoux, S., Ren\'{e} de Cotret, S., Lepault, S., LeBoeuf, D., Lalibert\'{e}, F., Hassinger, E., Ramshaw, B. J., Bonn, D. A., Hardy, W. N., Liang, R., Park, J.-H., Vignolles, D., Vignolle, B., Taillefer, L., Proust, C. Evidence for a small hole pocket in the Fermi surface of underdoped YBa$_2$Cu$_3$O$_y$. {\it Nature Commun.} {\bf 6}, 6034 (2015).
%
%\bibitem{harrison7} Harrison, N., Hayden, S. M.,  Meeson, P.,  Springford, M.,  van~der~Wel, P. J. de Haas-van Alphen effect in the vortex state of Nb$_3$Sn. {\it Phys. Rev. B} {\bf 50}, 4208-4211 (1994).
%
%\bibitem{norman3} Norman, M. R., MacDonald, A. H., Akera, H. Magnetic oscillations and quasiparticle band structure in the mixed state of type-II superconductors. {\it Phys. Rev. B} {\bf 51}, 5927-5942 (1995).
%
%s\bibitem{prange1} Prange, R. E.,  Girvin, S. E. (eds.). Elementary theory : the incompressible quantum fluid, in The quantum Hall effect.(Springer, Heidelberg, 1987).

%\bibitem{sasaki1} Sasaki, T.,  Fukuda, T., Nishizaki, T., Fujita,T., Yoneyama, N., Kobayashi,N. Low-temperature vortex liquid states induced by quantum fluctuations in the quasi-two-dimensional organic superconductor $\kappa$-ï¿œ(BEDT-TTF)$_2$Cu(NCS)$_2$. {\it Phys. Rev. B} {\bf 66}, 224513 (2002).

%\bibitem{higgins1} Higgins, M. J., Bhattacharya. Varieties of dynamics in a disordered flux-line lattice. Physica C {\bf 257}, 232-254 (1996).

%\bibitem{sasaki1} Sasaki, T.,  Fukuda, T., Nishizaki, T., Fujita,T., Yoneyama, N., Kobayashi,N. Low-temperature vortex liquid states induced by quantum fluctuations in the quasi-two-dimensional organic superconductor $\kappa$-ï¿œ(BEDT-TTF)$_2$Cu(NCS)$_2$. {\it Phys. Rev. B} {\bf 66}, 224513 (2002).
%
%\bibitem{blatter1} Blatter, G., Feigel'man, M. Y., Geshkenbein, Y. B., Larkin, A. I., Vinokur, V. M. Vortices in high-temperature superconductors
%. {\it Rev. Mod. Phys.} {\bf 66}, 1125-1388 (1994).
%
%\bibitem{narlikar1} Narlikar, A. V. (ed.). Field penetration and magnetization of high temperature superconductors. (Nova Science Publishers, New York, 1995).
%
%\bibitem{fournier1} Fournier, D., Levy, G., Pennec, Y., McChesney, J. L., Bostwick, A., Rotenberg, E., Liang, R., Hardy, W. N., Bonn, D. A., Elfimov, I. S., Damascelli, A. Loss of nodal quasiparticle integrity in underdoped YBa$_2$Cu$_3$O$_{6+x}$. {\it Nature Phys.} {\bf 6}, 906-911 (2010).

%\bibitem{volovik1} Volovik, G. E., Superconductivity with lines of gap nodes -- density-of-states in the vortex cores. {\it J. Exp. Theor. Phys. Lett.} {\bf 58}, 469-473 (1993). %   Pages: 457   Published: 1993
%
%\bibitem{nakai1} Nakai, N., Miranovi\'{c}, P., Ichioka, M., Machida, K. Field dependence of the zero-field density of states around vortices in an anisotropic-gap superconductor. Phys. Rev. B {\bf 70}, 100503(R) (2004).
%
%\bibitem{yang1} Yang, K., Sondhi, S. l. Response of a $d_{x^2-y^2}$ superconductor to a Zeeman magnetic field. Phys. Rev. B {\bf 57}, 8566-8570 (1998). % Published 1 April 1998
%
%\bibitem{chen1} Chen, H. D., Vafek, O., Yazdani, A., Zhang, S. C. Pair density wave in the pseudogap state of high temperature superconductors. Phys. Rev. Lett. {\bf 93}, 187002 (2004).
%
%\bibitem{agterberg1} Agterberg, D. F., Sigrist, M., Tsunetsugu, H. Order parameter and vortices in the superconducting $Q$ phase of CeCoIn$_5$. {\it Phys. Rev. Lett.} {\bf 102}, 207004 (2009).
%

%\bibitem{pereg1}  Pereg-Barnea, T., Weber, H., Refael, G., Franz, M. Quantum oscillations from Fermi arcs. Nature Phys. {\bf 6} 44-49 (2010).
%
%\bibitem{rice1} Rice, T. M., Yang, K.-Y.,  Zhang, F. C. A phenomenological theory of the anomalous pseudogap phase in underdoped cuprates. Rep. Prog. Phys. {\bf 75}, 016502 (2012).
%
%\bibitem{norman2} Norman, M. R., Kanigel, A., Randeria, M. Chatterjee, U.,  Campuzano, J. C. Modeling the Fermi arc in underdoped cuprates. Phys. Rev. B {\bf 76}, 174501 (2007).
%
%\bibitem{reber1} Reber, T. J., Plumb, N. C., Sun, Z., Cao, Y., Wang, Q., McElroy, K., Iwasawa, H., Arita, M., Wen, J. S., Xu, Z. J., Gu, G., Yoshida, Y., Eisaki, H., Aiura, Y., Dessau, D. S. The origin and non-quasiparticle nature of Fermi arcs in Bi$_2$Sr$_2$CaCu$_2$O$_{8+\delta}$. Nature Phys. {\bf 8}, 606-610 (2006).
%

%%Article Number: 7927
%%DOI: 10.1038/ncomms8927
%%Published: AUG 2015

%


%Volume: 5     Article Number: 3280   Published: FEB 2014

%\bibitem{millis1} Millis, A. J., Norman, M. R. Antiphase stripe order as the origin of electron pockets observed in 1/8-hole-doped cuprates. {\it Phys. Rev. B} {\bf 76}, 220503 (2007).
%
%\bibitem{chakravarty2} Chakravarty, S., Kee, H.-Y. Fermi pockets and quantum oscillations of the Hall coefficient in high-temperature superconductors. {\it Proc. Nat. Acad. Sci. USA} {\bf 105}, 8835-8839 (2008).
%
%\bibitem{allais1} Allais, A., Chowdhury, D., Sachdev, S. Connecting high-field quantum oscillations to zero-field electron spectral functions in the underdoped cuprates. {\it Nature Commun.} {\bf 5}, 5771 (2014).
%
%\bibitem{chang2} Chang, J., Doiron-Leyraud, N., Cyr-Choini\`{e}re, O., Grissonnanche, G., Laliberte, F., Hassinger, E., Reid, J. P., Daou, R., Pyon, S., Takayama, T., Takagi, H., Taillefer, L. Decrease of upper critical field with underdoping in cuprate superconductors. {\it Nature Phys.} {\bf 8}, 751-756 (2012).
%
%\bibitem{syshchenko1} Syshchenko, O., Day, J., Beamish, J. Frequency dependence and dissipation in the dynamics of solid helium. {\it Phys. Rev. Lett.} {\bf 104}, 195301 (2010).
%
%\bibitem{kenzelmann1} Kenzelmann, M., Gerber, S., Egetenmeyer, N., Gavilano, J. L., Str\"{a}ssle, Th., Bianchi, A. D.,  Ressouche, E.,
%Movshovich, R., Bauer, E. D., Sarrao, J. L., Thompson, J. D. Evidence for a magnetically driven superconducting $Q$ Phase of CeCoIn$_5$. {\it Phys. Rev. Lett.} {\bf 104}, 127001 (2010).

%%

%\bibitem{ding1} Ding, H., Yokoya, T., Campuzano, J. C., Takahashi, T., Randeria, M., Norman, M. R., Mochiku, T., Kadowaki, K., Giapintzakis, J. Spectroscopic evidence for a pseudogap in the normal state of underdoped high-$T_{\rm c}$ superconductors. {\it Nature} {\bf 382}, 51-54 (1996).

%\bibitem{valla1} Valla, T., Fedorov, A. V., Lee, J., Davis, J. C. The ground state of the pseudogap in cuprate superconductors. {\it Science} {\bf 314}, 1914-1916 (2006).
%
%\bibitem{gomes1} Gomes, K. K., Pasupathy, A.N., Pushp, A., Ono, S., Ando, Y., Yazdani, A. Visualizing pair formation on the atomic scale in the high-$T_{\rm c}$ superconductor Bi$_2$Sr$_2$CaCu$_2$O$_{8+\delta}$. {\it Nature} {\bf 447}, 569-572 (2007).
%
%\bibitem{shen1} Shen, K. M., Ronning, F., Lu, D. H., Baumbergerm Ingle, N. J. C., Lee, W. S., Meevasana, W., Kohsaka, Y., Azuma, M., Takano, M., Takagi, H., Shen, Z. X. Nodal quasiparticles and antinodal charge ordering in Ca$_{2-x}$Na$_x$CuO$_2$Cl$_2$. {\it Science} {\bf 307}, 901-904 (2005).
%
%\bibitem{wise1} Wise, W. D., Boyer, M. C., Chatterjee, K., Kondo, T., Takeuchi, T., Ikuta, H., Wang, Y. Y. Hudson, E. W. Charge-density-wave origin of cuprate checkerboard visualized by scanning tunnelling microscopy. {\it Nature Phys.} {\bf 4}, 696-699 (2008).
%
%\bibitem{senthil1} Senthil, T., Sachdev, S., Vojta, M. Fractionalized Fermi liquids. {\it Phys. Rev. Lett.} {\bf 90}, 216403 (2003).
%
%\bibitem{lee1} Lee, P. A., Nagaosa, N., Wen, X. G. Doping a Mott insulator: Physics of high-temperature superconductivity. {\it Rev. Mod. Phys.} {\bf 78}, 17-85 (2006).
%
%\bibitem{yang1} Yang, K.-Y., Rice, T. M., Zhang, F.-C. Phenomenological theory of the pseudogap state. {\it Phys. Rev. B} {\bf 73}, 174501 (2006).
%
%\bibitem{kaul1} Kaul, R. K., Kolezhuk, A., Levin, M., Sachdev, S., Senthil, T. Hole dynamics in an antiferromagnet across a deconfined quantum critical point. {\it Phys. Rev. B} {\bf 75}, 235122 (2007).
%
%\bibitem{kohsaka1} Kohsaka, Y., , C., Wahl, P., Schmidt, A., Lee, J., Fujita, K., Alldredge, J. W., McElroy, K., Lee, J., Eisaki, H., Uchida, S., Lee, D.-H. Davis, J. C. How cooper pairs vanish approaching the Mott insulator in Bi$_2$Sr$_2$CaCu$_2$O$_{8+\delta}$. {\it Nature} {\bf 454}, 1072-1078 (2008).
%
%\bibitem{shoenberg1} Shoenberg, D. Magnetization of a two-dimensional electron-gas. {\it J. Low Temp. Phys.} {\bf 56}, 417-440 (1984).
%
%\bibitem{eisenstein1} Eisenstein, J. P., Stormer, H. L., Narayanamurti, V., Cho, A. Y., Gossard, A. C., Tu, C. W. Density of states and de~Haas-van~Alphen effect in two-dimensional electron systems. {\it Phys. Rev. Lett.} {\bf 55} 875-878 (1985).
%
%\bibitem{jauregui1} Jauregui, K., Marchenko, V. I., Vagner, I. D. Magnetization of a two-dimensional electron gas. {\it Phys. Rev. B} {\bf 41}, 12922-12925 (1990).
%
%\bibitem{harrison1} Harrison, N., Bogaerts, R., Reinders, P. H. P., Singleton, J., Blundell, S. J., Herlach, F. Numerical model of quantum oscillations in quasi-two-dimensional organic metals in high magnetic fields.{\it Phys. Rev. B} {\bf 54}, 9977-9987 (1996).
%
%\bibitem{potts1} Potts, A., Shepherd, R., Herrenden-Harker, W. G., Elliott, M, Jones, C. L.,
%Usher A., Jones, G. A. C., Ritchie, D. A., Linfield, E. H., Grimshaw, M. Magnetization studies of Landau level broadening in two-dimensional electron systems. {\it J. Phys.: Condens. Matter} {\bf 8}, 5189-5207 (1996).
%
%\bibitem{itskovsky1} Itskovsky, M. A., Maniv, T., Vagner, I. D. Wave form of de~Haas-van~Alphen oscillations in a two-dimensional metal. {\it Phys. Rev. B} {\bf 61}, 14616-14627 (2000).
%
%\bibitem{champel1} Champel, T. Chemical potential oscillations and de Haas-van Alphen effect. {\it Phys. Rev. B} {\bf 64}, 054407 (2001).
%
%\bibitem{wilder1} Wilde, M. A., Schwartz, M. P., Heyn, Ch., Heitmann, D., Grundler, D. Experimental evidence of the ideal de~Haas-van~Alphen effect in a two-dimensional system. {\it Phys. Rev. B} {\bf 73}, 125325 (2006).
%

%
%\bibitem{wosnitza1} Wosnitza, J., Wanka, S., Hagel,  J., Balthes, E., Harrison, N., Schlueter, J. A., Kini, A. M., Geiser, U., Mohtasham, J., Winter, R. W., Gard, G. L. Two-dimensional Fermi liquid with fixed chemical potential. {\it Phys. Rev. B} {\bf 61}, 7383-7387 (2000).
%
%\bibitem{mineev1} Champel, T., Mineev, V. P. 
%de Haas-van Alphen effect in two- and quasi-two-dimensional metals and superconductors. {\it Phil. Mag. B} {\bf 81}, 55-74 (2001).
%%By: Champel, T; Mineev, VP
%%PHILOSOPHICAL MAGAZINE B-PHYSICS OF CONDENSED MATTER STATISTICAL MECHANICS ELECTRONIC OPTICAL AND MAGNETIC PROPERTIES   Volume: 81   Issue: 1   Pages: 55-74   Published: JAN 2001
%

%
%\bibitem{vignolle1} Vignolle, B., Carrington, A ., Cooper, R. A., French, M. M. J., Mackenzie, A. P., Jaudet, C., Vignolles, D., Proust, C., Hussey, N. E. Quantum oscillations in an overdoped high-$T_{\rm c}$ superconductor. {\it Nature} {\bf 455}, 952-955 (2008).
%
%\bibitem{yelland1} Yelland, E. A., Singleton, J., Mielke, C. H., Harrison, N., Balakirev, F. F., Dabrowski, B., Cooper, J. R. Quantum oscillations in the underdoped cuprate YBa$_2$Cu$_4$O$_8$. {\it Phys. Rev. Lett.} {\bf 100}, 047003 (2008).
%
%\bibitem{bangura1} Bangura, A. F., Fletcher, J. D., Carrington, A., Levallois, J., Nardone, M., Vignolle, B., Heard, P. J., Doiron-Leyraud, N., LeBoeuf, D., Taillefer, L. Small fermi surface pockets in underdoped high temperature superconductors: Observation of Shubnikov-de Haas oscillations in YBa$_2$Cu$_4$O$_8$. {\it Phys. Rev. Lett.} {\bf 100}, 047004 (2008).
%
%\bibitem{barisic1} Bari\u{s}i\'{c}, N., Badoux, S., Chan, M. K., Dorow, C., Tabis, W., Vignolle, B., Guichan, Y., Beard, J., Zhao, X., Proust, C. Universal quantum oscillations in the underdoped cuprate superconductors, {\it Nature Phys.} {\bf 9}, 761-764 (2013).
%
%\bibitem{doiron2} Doiron-Leyraud, N., Badoux, de~Cotret, S. R., Lepault, S., LeBoeuf, D., Laliberte, F., Hassinger, E., Ramshaw, B. J., Bonn, D. A., Hardy, W. N., Liang, R., Park, J. -H., Vignolles, D., Vignolle, B., Taillefer, L., Proust, C. Evidence for a small hole pocket in the Fermi surface of underdoped YBa$_2$Cu$_3$O$_y$. {\it Nature Commun.} {\bf 6}, 6034 (2015).
%
%\bibitem{tan1} Tan, B. S., Harrison, N., Zhu, Z., Balakirev, F., Ramshaw, B. J., Srivastava, A., Sabok, S. A., Dabrowski, B., Lonzarich, G. G., Sebastian, S. E. Fragile charge order in the nonsuperconducting
%ground state of the underdoped high-temperature superconductors. {\it Proc. Nat. Acad. Sci. USA} {\bf 112}, 9568-9572 (2015).
%
%\bibitem{maharaj1} Maharaj, A.~V.,Hosur, P., Raghu, S. Crisscrossed stripe order from interlayer tunneling in hole-doped cuprates. {\it Phys. Rev. B} {\bf 90}, 125108 (2014).
%
%\bibitem{tabis1} Tabis, W., Le Tacon, M., Braicovich, L., Kreyssig, A., Minola, M., Dellea, G., Weschke, E., Veit, M. J., Ramazanoglu, M., Goldman, A. I., Schmitt, T., Ghiringhelli, G., Barisic, N., Chan, M. K., Dorow, C. J., Yu, G., Zhao, X., Keimer, B., Greven, M. Charge order and its connection with Fermi-liquid charge transport in a pristine high-$T_{\rm c}$ cuprate. {\it Nature Commun.} {\bf 5}, 5875 (2014).
%
%\bibitem{alexandrov1} Alexandrov, A.~S., Bratkovsky, A.~M. de Haas-van Alphen effect in canonical and grand canonical multiband Fermi liquid. {\it Phys. Rev. Lett.} {\bf 76}, 1308-1311 (1996).
%
%\bibitem{grissonnanche2} Grissonnanche, G., Laliberte, F., Dufour-Beausejour, S., Matusiak, M., Badoux, S., Tafti, F. F., Michon, B., Riopel, A., Cyr-Choiniere, O.,  Baglo, J. C., Ramshaw, B. J., Liang, R., Bonn, D. A., Hardy, W. N., Kramer, S., LeBoeuf, D., Graf, D.,  Doiron-Leyraud, N., Taillefer, L. 
%Wiedemann-Franz law in the underdoped cuprate superconductor YBa$_2$Cu$_3$O$_y$.
%arXiv:1503.07572 (2015).
%
%\bibitem{harrison2} Harrison, N., Sebastian, S.~E. Magnetotransport signatures of a single nodal electron pocket constructed from Fermi arcs. {\it Phys. Rev. B} {\bf 92}, 224505 (2015).
%
%%\bibitem{helm1} Helm, T., Kartsovnik, M. V., Bartkowiak, M., Bittner, N., Lambacher, M., Erb, A., Wosnitza, J., Gross, R. Nd$_{2-x}$Ce$_x$CuO$_4$ revealed by Shubnikov-de Haas oscillations. {\it Phys. Rev. Lett.} {\bf 103}, 157002 (2009).
%%
%%\bibitem{vonklitzing1} Von~Klitzing, K., Dorda, G., Pepper, M. New method for high-accuracy determination of the fine-structure constant based on quantized Hall resistance. {\it Phys. Rev. Lett.} {\bf 45}, 494-497 (1980).
%%
%%\bibitem{tsui1} Tsui, D. C., Stormer, H. L., Gossard, A. C. Two-dimensional magnetotransport in the extreme quantum limit. {\it Phys. Rev. Lett.} {\bf 48} 1559-1562 (1982).
%%
%%\bibitem{doiron3} Doiron-Leyraud, N., Lepault, S., Cyr-Choiniere, O., Vignolle, B., Grissonnanche, G., Laliberte, F., Chang, J., Barisic, N., Chan, M. K., Ji, L., Zhao, X., Li, Y., Greven, M., Proust, C., Taillefer, L. Hall, Seebeck, and Nernst coefficients of underdoped HgBa$_2$CuO$_{4+\delta}$: Fermi-surface reconstruction in an archetypal cuprate superconductor. {\it Phys. Rev. X} {\bf 3}, 021019 (2013).
%
%%G. Grissonnanche, F. Laliberte, S. Dufour-Beausejour, M. Matusiak, S. Badoux, F. F. Tafti, B. Michon, A. Riopel, O. Cyr-Choiniere, J. C. Baglo, B. J. Ramshaw, R. Liang, D. A. Bonn, W. N. Hardy, S. Kramer, D. LeBoeuf, D. Graf, N. Doiron-Leyraud, L. Taillefer
%
%%
%%\bibitem{leboeuf2} LeBoeuf, D., Kramer, S., Hardy, W. N., Liang, R. X., Bonn, D. A., Proust, C. Thermodynamic phase diagram of static charge order in underdoped YBa$_2$Cu$_3$O$_y$. {\it Nature Phys.} {\bf 9}, 79-83 (2013).
%


%\bibitem{anderson1987} Anderson, P.W., The resonating valence bond state in La$_2$CuO$_4$ and superconductivity. {\it Science} {\bf 235}, 1196-1198 (1987).
%
%\bibitem{schrieffer1988} Schrieffer, J. R., Wen, X.~G., Zhang, S. S., Spin-gap mechanism of high-temperature superconductivity. {\it Phys. Rev. Lett.} {\bf 60}, 944-947 (1988). 
%
%\bibitem{monthoux1991} Monthoux, P., Balatsky, A. V., Pines, D., Toward a theory of high-temperature superconductivity in the antiferromagnetically correlated cuprate oxides. {\it Phys. Rev. Lett.} {\bf 67}, 3448-3451 (1991). 

%\bibitem{alford2008} Alford, M. G., Schmitt, A., Rajagopal, K., Sch\"{a}fer, T., Color superconductivity in dense quark matter. {\it Rev. Mod. Phys.} {\bf 80}, 1455-1515 (2008).

%\bibitem{gapnote} While the magnitude of the excitation gap within the pseudogap region is strongly suppressed at $T>T_{\rm c}$, the maxima in $\gamma$ and $\chi$ nevertheless continue to be quantitatively consistent in energy gap with $T_\gamma=2\Delta/6.5k_{\rm B}$ and $T_\chi=2\Delta/3k_{\rm B}$. 


%\bibitem{miller2007} Miller, D. E., Chin, J. K., Stan, C. A., Liu, Y., Setiawan, W., Sanner, C., Ketterie, W., Critical velocity for superfluid flow across the BEC-BCS crossover. {\it Phys. Rev. Lett.} {\bf 99}, 070402 (2007).

%


%


%\bibitem{eagles1969} Eagles, D.M., Possible pairing without superconductivity at low carrier concentrations in bulk and thin-film superconducting semiconductors. {\it Phys. Rev.} {\bf 186}, 456-463 (1969).


%%%%%%%%%%%%%%%%

%
%
%\bibitem{tsai2006} Tsai, W.-F., Kivelson, S. A, Superconductivity in inhomogeneous Hubbard models, {\it Phys. Rev. B} {\bf 73}, 214510 (2006).
%
%\bibitem{micnas2007} Micnas, R., Superfluid transition temperature of the boson-fermion model on a lattice. {\it Phys. Rev. B} {\bf 76}, 184507 (2007).
%
%\bibitem{stauber2007} Stauber, T., Ranninger, J., First-order transition from superfluid to Bose-metal state in systems with resonant pairing. {\it Phys. Rev. Lett.} {\bf 99}, 045301 (2007).
%
%\bibitem{kagan2019} Kagan, M. Y., Bianconi, A., Fermi-Bose mixtures and BCS-BEC crossover in high-$T_{\rm c}$ superconductors. {\it Condens. Matter} {\bf 4}, 4020051 (2019).


\end{thebibliography}

\section{Supplementary Information}

\subsection{Introduction}
This Supplementary Information document contains the following: data pertaining to $\delta\gamma(T_{\rm c})$ in conventional BCS superconductors and Fe- and Ni-based superconductors; information about cuprate hole dopings $p$ used in constructing graphs in the main paper; raw $\gamma$ and $\chi$ data including the locations of $T_\gamma$ and $T_\chi$ used in the main paper; Knight shift data on BLSCO; a discussion pertaining to prior modeling of $\delta\gamma(T_{\rm c})$ versus $p$ by Chen {\it et al.}; details concerning cold atomic gas data used in the main paper; estimates of the gap ratio at the unitary point of a cold atomic gas; a discussion of the thermodynamics of maxima in $\gamma$ and $\chi$; coherence length estimates in various cuprates based on $H_{\rm c2}$ or the upper extent in magnetic field for the vortex solid phase; estimates of the Fermi energy in the cuprates;  a discussion of the higher values of $p^\ast$ in cuprates with lower $T_{\rm c}$'s; the origin of the `Fermi arcs;' the $T$-dependence of the antinodal gap; lifetime effects on $N_0$ in a cold atomic Fermi gas; the choice of a smaller gap option for $\Delta$ in the cuprates; evidence for a hidden maximum in the normal state heat capacity data of a cold atomic Fermi gas; and reports of peaks in $\delta\gamma(T_{\rm c}$, $\gamma$ and $m^\ast$ associated with quantum criticality.

\subsection{$\delta\gamma(T_{\rm c})/\bar{\gamma}$ for conventional BCS and Fe- and Ni-based superconductors.}

\begin{figure}[h]
\begin{center}
\includegraphics[width=0.9\linewidth]{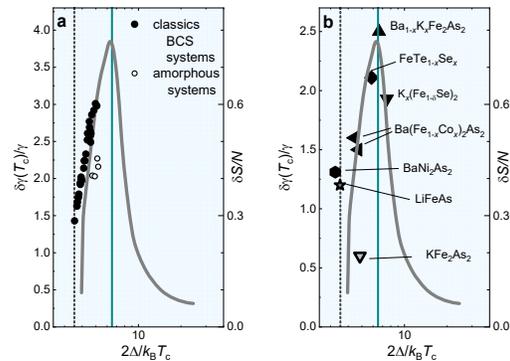}
\textsf{\caption{{\bf $\delta\gamma(T_{\rm c})$ data in other superconductors.} {\bf a}, $\delta\gamma(T_{\rm c})/\bar{\gamma}$ values for conventional BCS superconductors from Ref.~\cite{carbotte1990}. Open symbols represent amorphous superconductors for which $\delta\gamma(T_{\rm c})/\bar{\gamma}$ is reduced. {\bf b}, $\delta\gamma(T_{\rm c})/\bar{\gamma}$ values for various Fe- and Ni-based superconductors from Ref.~\cite{inosov2011}. The grey lines correspond to $\delta S$ for a cold atomic Fermi gas, as plotted in the main paper.
}
\label{fe}}
\end{center}
\end{figure}

Figure~\ref{fe}a shows $\delta\gamma(T_{\rm c})/\bar{\gamma}$ (where $\bar{\gamma}$ is an assumed constant normal state Sommerfeld coefficient) of various conventional BCS superconductors as a function of $2\Delta/k_{\rm B}T_{\rm c}$ from Ref.~\cite{carbotte1990}. The outliers with lower values of $\delta\gamma(T_{\rm c})/\bar{\gamma}$ correspond to amorphous materials. Apart from the outliers, $\delta\gamma(T_{\rm c})/\bar{\gamma}$ can be seen to steadily increase with $2\Delta/k_{\rm B}T_{\rm c}$, but never reaching the magic gap rationvalue of $\approx$~6.5.

Figure~\ref{fe}b shows $\delta\gamma(T_{\rm c})/\bar{\gamma}$ versus $2\Delta/k_{\rm B}T_{\rm c}$ for various Fe- and Ni-based superconductors, using data from Ref.~\cite{inosov2011}. Here, the system with the largest $\delta\gamma(T_{\rm c})/\bar{\gamma}$ value has a gap ratio $2\Delta/k_{\rm B}T_{\rm c}=$~6.6, which is therefore consistent with an optimally robust superconducting state.

A caveat with renormalization by $\bar{\gamma}$ is that this quantity is unlikely to be accurately determined by experiment once $2\Delta/k_{\rm B}T_{\rm c}\gtrsim$ 6.5, due to the formation of a pseudogap, which causes $\gamma$ to be a non monotonic function of $T$.

In the cuprates, owing to a significant deviation of $\gamma$ from a constant value at $T>T_{\rm c}$, due primarily to the pseudogap, we plot the absolute value of $\delta\gamma(T_{\rm c})$ rather than the ratio $\delta\gamma(T_{\rm c})/\bar{\gamma}$. This avoids incurring errors associated with estimating a constant $\bar{\gamma}$, as is typically done in BCS superconductors. We plot only the absolute value of $\delta\gamma(T_{\rm c})$ also for the case of a cold atomic gas, where this may also be the case. 

\subsection{Cuprate hole dopings}

For LSCO, the hole dopings $p$ are determined using $p=x$\cite{loram2001}. For YBCO, in the case of YBa$_2$Cu$_3$O$_{6+x}$ they are determined from the value of $T_{\rm c}$ using Ref.\cite{liang2006}, in the case of Ca-YBCO (YBCO in which 20\% of the Y is substituted with Ca), we have used $p$ values from Ref.\cite{michon2019}, while in the case of YBa$_2$Cu$_4$O$_8$ (data for $p=$~0.125) we have used $p$ from Ref.\cite{zhou1996}. For BSCCO, $p$ values are provided in Ref.\cite{loram2001}, for BSLCO $p$ is provided in Ref.\cite{wen2009}, while for TBCO, $p$ is obtained by applying the Tallon formula\cite{tallon1995} to the data in Ref.\cite{wade1994}, while it is assumed that $p=$~0.16 in Ref.\cite{mirmelstein1995}.
For YBCO in Figs.~2 and 3 of the main paper, the highest two compositions ($p=$~0.152 and 0.158) are for samples in which 2\% of the planar Cu has been substituted with Zn. 

For LSCO and Nd-LSCO, additional $T_{\rm c}$ values in Fig.~1c of the main paper are obtained from Refs.~\cite{takagi1989,ma2021}.

For TBCO, $\gamma$ data pertains to a sample in which $T_{\rm c}$ is suppressed by a magnetic field\cite{radcliffe1996}. Although it remains unclear to what extent the field suppresses the phase stiffness of the superconducting state relative to the pairing gap, the maximum in $\gamma$ presented at $B=$~13~T pertains to the resistive regime\cite{carrington1994} (the zero resistance state being almost completely suppressed). There is also relatively little change in the maximum between 10 and 13~T. Measurements on a non superconducting composition nearby in doping have enabled the phonon contribution to the heat capacity to be subtracted with reasonable accuracy\cite{radcliffe1996}.

\subsection{Determination of $T_\gamma$ and $T_\chi$ from the original cuprate data.} 

\begin{figure}[h]
\begin{center}
\includegraphics[width=0.9\linewidth]{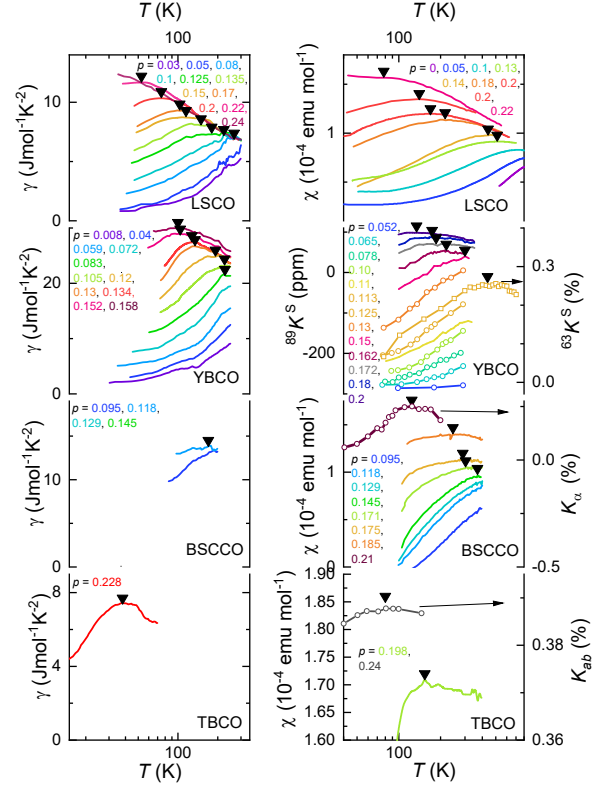}
\textsf{\caption{{\bf Raw data.} Data used  in Fig.~3 of the main paper plotted as a function of the actual temperature $T$. The inferred maximum are indicated by inverted triangles.  
}
\label{rawdata}}
\end{center}
\end{figure}

Figure~\ref{rawdata} shows the data from Fig.~1 of the main paper\cite{nakano1994,loram2001,alloul1989,loram1993,curro1997,alloul2016,radcliffe1996,kubo1991,johnston1989}, here plotted as a function of the actual temperature $T$. $T_\gamma$ and $T_\chi$ are obtained from the location of the maxima in $T$. In the case of noisier datasets, or datasets with fewer points, the location of the maxima is assisted by the fitting of a third order polynomial in $\ln T$.   We make an exception for the $p=$~0.22 composition of LSCO, where we read off the temperature at which $T_\chi$ exhibits a shoulder feature. The temperatures ($T_\gamma$ and $T_\chi$), at which maxima are observed, are plotted in Fig.~2d of the main paper. Owing to the broad widths of the maxima, the error bars are 10\% and 20\% for $T_\gamma$ and $T_\chi$, respectively. 

By using published temperature-dependent $\gamma$ and $\chi$ data, we rely on background subtractions made by the authors of the respective works. For $\gamma$, systematic errors are likely to be introduced owing to the complicated phonon spectra and the reliance on non-superconducting reference samples. Owing to assumptions that need to be made to account for changes in the concentration of O doped into the CuO chains that are necessary to produce changes in hole doping\cite{loram1993}, these systematic errors become largest at higher dopings of YBCO. Meingast {\it et al.} have shown that it is challenging to accurately model the phonon modes in this system using {\it AB-initio} methods\cite{meingast2009}. However, at the same O dopings where systematic errors in accounting for the phonon contribution are expected to become most significant, a peak in $\gamma$ is no longer visible owing to its concealment by superconductivity. The jump $\delta\gamma(T_{\rm c})$ in $\gamma$ at $T_{\rm c}$, by contrast, is unaffected by the phonon subtraction.

In the case of $\chi$, measurements of $K$ are made in the presence of a uniform magnetic field for which the effect of incipient diamagnetism above $T_{\rm c}$ is greatly reduced relative to measurements of the uniform susceptibility.  The temperature dependence of $K$ is therefore primarily attributed to changes in the spin susceptibility.  Incipient diamagnetism, mostly in the overdoped cuprates, can in some cases produce a maximum in $\chi$ that is unrelated to pair amplitude fluctuations or the pseudogap. Owing to its lower $T_{\rm c}$ values, measurements of the uniform susceptibility are less prone to the effects of incipient diamagnetism in LSCO. There is evidence for a diamagnetic contribution to $\chi_{\rm m}$ of BSCCO and TBCO at the lowest temperatures\cite{kubo1991,loram2001}, which we have cutoff in cases where this contribution becomes obtrusively large.

\subsection{BSLCO Knight shift data}

\begin{figure}[h]
\begin{center}
\includegraphics[width=0.9\linewidth]{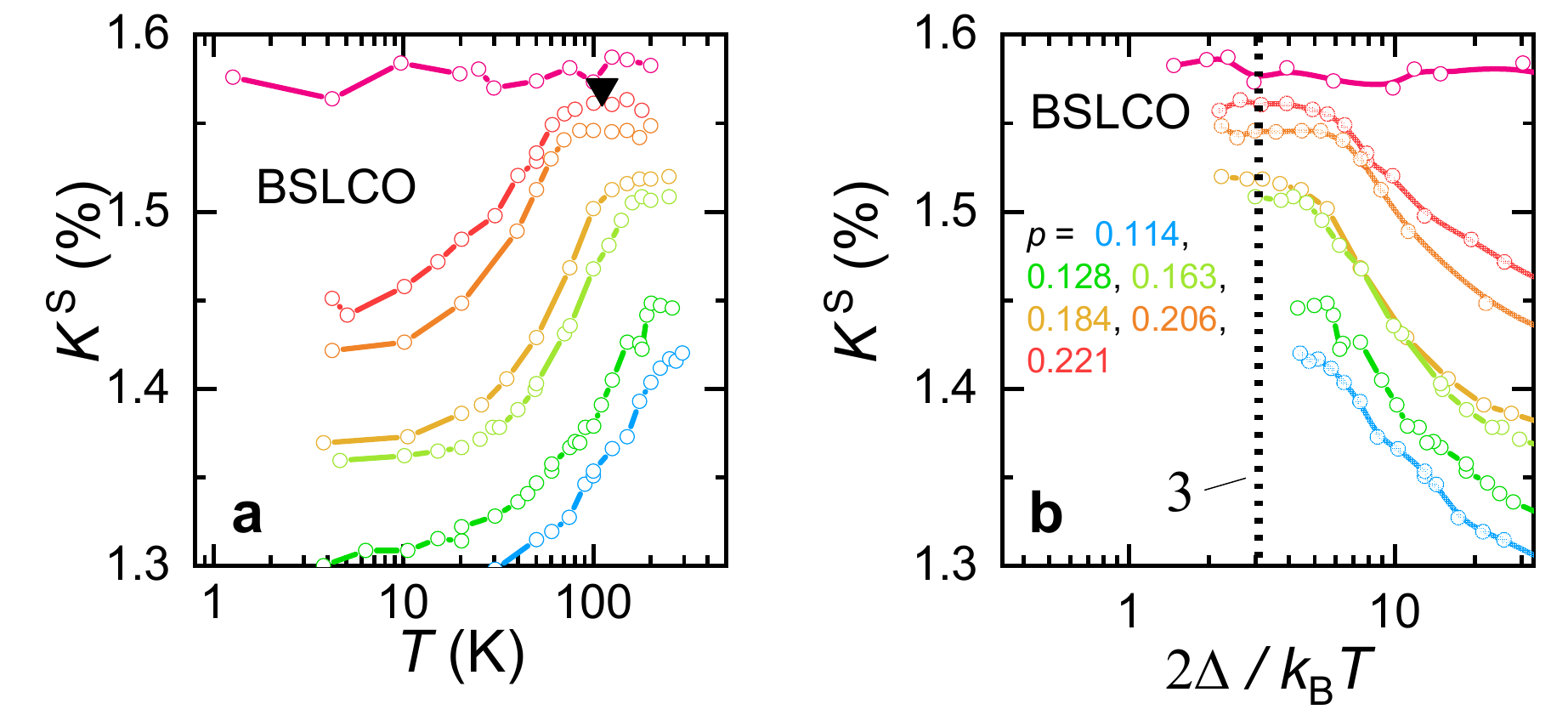}
\textsf{\caption{{\bf Raw data in BSLCO.} {\bf a}, Knight shift in BSLCO. {\bf b}, Knight shift in BSLCO plotted versus $2\Delta/k_{\rm B}T$, using doping values from Ref.~\cite{kawasaki2010}.
}
\label{BSLCO}}
\end{center}
\end{figure}

Knight shift data also exists for BSLCO (or Bi$_2$Sr$_{2-x}$La$_x$CuO$_{6+\delta}$), in which a downturn has previously been reported to occur at temperatures below $T^\ast$\cite{kawasaki2010} (see Fig.~\ref{BSLCO}). This  downturn is most clearly apparent on plotting the data versus $2\Delta/k_{\rm B}T$. %With a keen eye, a similar down turn appears to be occurring in YBCO at high $p$ in Fig.~2b of the main text.
While $K$ in BSLCO supports the existence of a maximum at $2\Delta/k_{\rm B}T\approx$~3, it was not measured to sufficiently high temperature in this experiment to determine this unambiguously. One possible exception is the sample with a hole doping of $p\approx$~0.206.

\subsection{Prior modeling of $\delta\gamma(T_{\rm c})$ versus $p$}

One strong coupling theory of superconductivity that has been shown to qualitatively account for the strong variation of $\delta\gamma(T_{\rm c})$ with $p$\cite{chen2000,chen2001}. In this model, the total jump in $\delta\gamma(T_{\rm c})$ is the sum of a vertical jump resulting from the onset of a superconducting order parameter $\Delta_{\rm s}$ and a lambda anomaly-like increase immediately above $T_{\rm c}$ resulting from line broadening effects relating to the pseudogap $\Delta_{\rm pg}$ on approaching $T_{\rm c}$. The two are assumed to add in quadrature: i.e. $\Delta=\sqrt{\Delta_{\rm s}^2+\Delta_{\rm pg}^2}$. In order to produce a variation in $\delta\gamma(T_{\rm c})$ with $p$ similar to that in experiments, it was necessary to assume an arbitrary functional form of the line broadening $\gamma_{\rm l}$ near $T_{\rm c}$ of the pseudogap on $T$ and on the overall magnitude of $\Delta_{\rm pg}$ as a function of $p$  in this theory. This is in contrast to the modeling of Haussmann {\it et al.}\cite{haussmann2007}, for which we find consistency with $\delta\gamma(T_{\rm c})$ in the cuprates  despite no attempt having been made to adjust the parameters of the theory to fit cuprate data.

\subsection{Cold atomic Fermi gas data}

In Fig.~1 of the main paper, $T_{\rm c}$, $\delta\gamma(T_{\rm c})$, $\delta S$ and $\Delta$ are extracted from Figs. 4, 5, 6 and 8 of Haussmann {\it et al.}\cite{haussmann2007}. The multivaluedness of $S(T)$ in the vicinity of $T_{\rm c}$ is known to be an artifact of the numerical method. We make two estimates of $\delta\gamma(T_{\rm c})$ in order to obtain a lower and upper bound.

\begin{figure}[h]
\begin{center}
\includegraphics[width=0.9\linewidth]{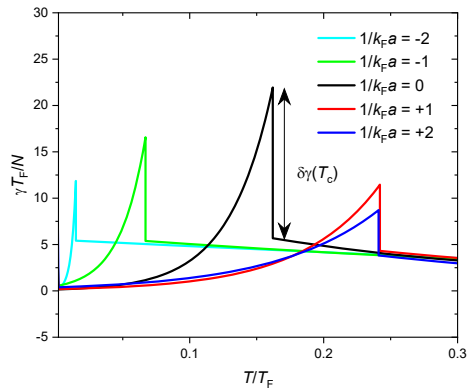}
\textsf{\caption{{\bf Phase transition anomaly in the unitary regime of a Fermi gas.} $\gamma$ is extracted from $S(T)$ in Ref.~\cite{haussmann2007} as described in the text.
}
\label{haussmann}}
\end{center}
\end{figure}

In order to obtain a lower bound estimate for $\delta\gamma(T_{\rm c})$ (in Fig.~1b of the main paper) from Ref.~\cite{haussmann2007}, $S(T)$ is differentiated with respect to $T$ in the superfluid state, where the functional form is exponential in temperature, and in the normal state, where it is monotonically varying with $T$. $\delta\gamma(T_{\rm c})$ is then taken as the difference in $\partial S/\partial T$  between the superfluid and normal states at $T_{\rm c}$ (see Fig.~\ref{haussmann}). The main limitation of this lower bound approximation is that that  does not fully account for entropy of the normal state for $1/k_{\rm F}a\geq0$. The problem is that $S(T)$ within the superfluid phase rises very steeply on the approach to $T_{\rm c}$, deviating from exponential behavior, very close to the transition. 
 
For an upper bound estimate of $\delta\gamma(T_{\rm c})$ (in Fig.~1b of the main paper), we use the approximation $\delta\gamma(T_{\rm c})\approx\frac{\delta S}{\delta T}-\frac{\partial S_{\rm n}}{\partial T}$, where $\delta T$ is the interval in $T$ over which $S$ is multivalued, $\delta S$ is the full extent of the variation in $S$ within the interval and $\frac{\partial S_{\rm n}}{\partial T}$ is the slope of $S$ in the normal state. 

In Fig.~3 of the main paper, $\gamma$ is obtained by differentiating $S$ in Fig.~19 of Chen and Wang\cite{chen2014}. Owing to the intersection of $T_\gamma$ and $T_{\rm c}$ at $1/k_{\rm F}a=0$, tuning of $1/k_{\rm F}a$ into the BEC regime is required in order to observe the maxima in $\gamma$\cite{chen2014}.

\subsection{Estimates of $2\Delta/k_{\rm B}T_{\rm c}$ at the unitary point of a cold atomic gas}

From the various theoretical methods tabulated in Table~1 of Ref.~\cite{randeria2014} and obtained from Ref.~\cite{moshe2019}, we obtain $2\Delta/k_{\rm B}T_{\rm c}=$~6.6~$\pm$~0.1. 
The experimental estimate of the gap ratio at the unitary point is $2\Delta/k_{\rm B}T_{\rm c}=$~5.3~$\pm$~0.8\cite{randeria2014}, on combining quoted experimental errors in $\Delta$ and $T_{\rm c}$. Combining experimental and numerical estimates, we obtain $2\Delta/k_{\rm B}T_{\rm c}=$~6.5~$\pm$~0.2.

\subsection{Thermodynamics of the maxima in $\gamma$ and $\chi$.}

\subsubsection{Gap in an electronic band.} 
For a gap opening in an electronic band obeying Fermi-Dirac statistics, we consider the electronic contribution to the Helmholtz free energy
\begin{widetext}
 \[F_{\rm el}=-N_{\rm A}\sum_{\sigma=\pm s}\int_{-\infty}^\infty k_{\rm B}TD(\varepsilon+sg\mu_{\rm B}B)\ln(1+{\rm e}^{-\frac{\varepsilon-\mu}{k_{\rm B}T}}){\rm d}\varepsilon+N_{\rm  A}N\mu,\] 
\end{widetext}
as an integral over states, where \[D(\varepsilon+sg\mu_{\rm B}B)\] is the density of states subject to Zeeman splitting $sg\mu_{\rm B}B$ (where $s$ is the spin, $g$ is the {\it g}-factor and $\mu_{\bf B}$ is the Bohr magneton). 
If we consider the energy of a single electronic band subject to Zeeman splitting, then spin singlet gap formation couples electron and hole states of like spin (or electrons or holes of opposite spin) whereas spin triplet formation couples electrons and holes of opposite spin.

Under the assumption of a symmetric gap centered on the chemical potential (defined as $\mu=0$) and a constant number of particles $N$, the electronic coefficient of the heat capacity and the spin susceptibility are given by\cite{kittel2004,nolthing2009}
\begin{widetext}
\begin{equation}\label{heatcapacity}
\gamma=\frac{C_{\rm el}}{T}=\frac{\partial^2 F_{\rm el}}{\partial T^2}\bigg|_{V,N}
=N_{\rm A}
\int_{-\infty}^{\infty} D(\varepsilon)\frac{\varepsilon^2}{k_{\rm B}T^3 }
\frac{{\rm e}^{\varepsilon/k_{\rm B}T}}{({\rm e}^{\varepsilon/k_{\rm B}T}+1)^2}{\rm d}\varepsilon
\end{equation}
and
\begin{equation}\label{susceptibility}
\chi=\mu_0\frac{\partial^2 F_{\rm el}}{\partial B^2}\bigg|_{V,N}
=\mu_0(sg\mu_{\rm B})^2N_{\rm A}
\int_{-\infty}^{\infty} D(\varepsilon)\frac{1}{k_{\rm B}T }
\frac{{\rm e}^{\varepsilon/k_{\rm B}T}}{({\rm e}^{\varepsilon/k_{\rm B}T}+1)^2}{\rm d}\varepsilon,
\end{equation}
\end{widetext}
respectively, in the limit $B\rightarrow0$ where $s=\frac{1}{2}$.

\subsubsection{Trivial gap in a narrow electronic band ($t\rightarrow0$)}

When a gap $2\Delta$ opens in a narrow electronic band, the resulting  trivially gapped state consists of two narrow levels centered about $\mu$. One can model the density of states resulting from spin singlet bound state formation as delta functions so that $D(\varepsilon)=\delta(|\varepsilon\pm sg\mu_{\rm B}B|-\Delta)$, for which we obtain
\begin{equation}\label{heatcapacity2}
\gamma=2N_{\rm A}
\frac{\Delta^2}{k_{\rm B}T^3 }
\frac{{\rm e}^{\Delta/k_{\rm B}T}}{({\rm e}^{\Delta/k_{\rm B}T}+1)^2}
\end{equation}
and
\begin{equation}\label{susceptibility2}
\chi=\mu_0(sg\mu_{\rm B})^2N_{\rm A}
\frac{1}{k_{\rm B}T }
\frac{{\rm e}^{\Delta/k_{\rm B}T}}{({\rm e}^{\Delta/k_{\rm B}T}+1)^2},
\end{equation}
respectively. We locate the maxima in $\gamma$ and $\chi$ by setting $\partial\gamma/\partial T=0$ and $\partial\chi/\partial T=0$, yielding the transcendental equations $3\theta^{\rm max}_\gamma=\tanh(1/2\theta^{\rm max}_\gamma)$ and $\theta^{\rm max}_\chi=\tanh(1/2\theta^{\rm max}_\chi)$, for which we obtain $\theta^{\rm max}_\gamma=k_{\rm B}T_\gamma/\Delta\approx$~0.308296 and $\theta^{\rm max}_\chi=k_{\rm B}T_\chi/\Delta\approx$~0.647918; hence $2\Delta/k_{\rm B}T_\gamma\approx$~6.4873 and $2\Delta/k_{\rm B}T_\chi\approx$~3.0868.

For spin triplet bound state formation, the density of states has the form $D(\varepsilon)=\delta(|\varepsilon|-\sqrt{\Delta^2+(sg\mu_{\rm B}B)^2})$, yielding the same $\gamma$ as given by Equation~(\ref{heatcapacity2}). However, for a spin triplet bound state, $\chi=0$ in the limit $B\rightarrow0$.

\subsubsection{$d$-wave and $s$-wave electronic gaps in a simple tight binding dispersion}

For the $d$-wave gap used to model the peaks in $\gamma$ and $\chi$ in Fig.~3 of the main paper, the electronic density of states is given by
$D(\varepsilon)=\frac{1}{\pi^2}\frac{\partial}{\partial\varepsilon}\int^\pi_0k_y(\varepsilon){\rm d}k_x$,
where we consider a conventional two-dimensional tight-binding dispersion
\[\varepsilon=\sqrt{4t^{2}(\cos k_x+\cos k_y)^2+\frac{\Delta^2}{4}(\cos k_x-\cos k_y)^2}\] with a conventional $d$-wave gap and an assumed in-plane hopping $5t=3\Delta$. For simplicity we neglect higher order hopping terms and set $p=0$ (i.e. half filling). In this case the integrals for $\gamma$ and $\chi$ are performed numerically, and we obtain $\theta^{\rm max}_\gamma=k_{\rm B}T_\gamma/\Delta\approx$~0.311 and $\theta^{\rm max}_\chi=k_{\rm B}T_\chi/\Delta\approx$~0.683; hence $2\Delta/k_{\rm B}T_\gamma\approx$~6.43 and $2\Delta/k_{\rm B}T_\chi\approx$~2.93. In the case of an $s$-wave gap, we simply use \[\varepsilon=\sqrt{4t^{2}(\cos k_x+\cos k_y)^2+\Delta^2}.\]

\subsection{Coherence length}

\begin{figure}[h]
\begin{center}
\includegraphics[width=0.9\linewidth]{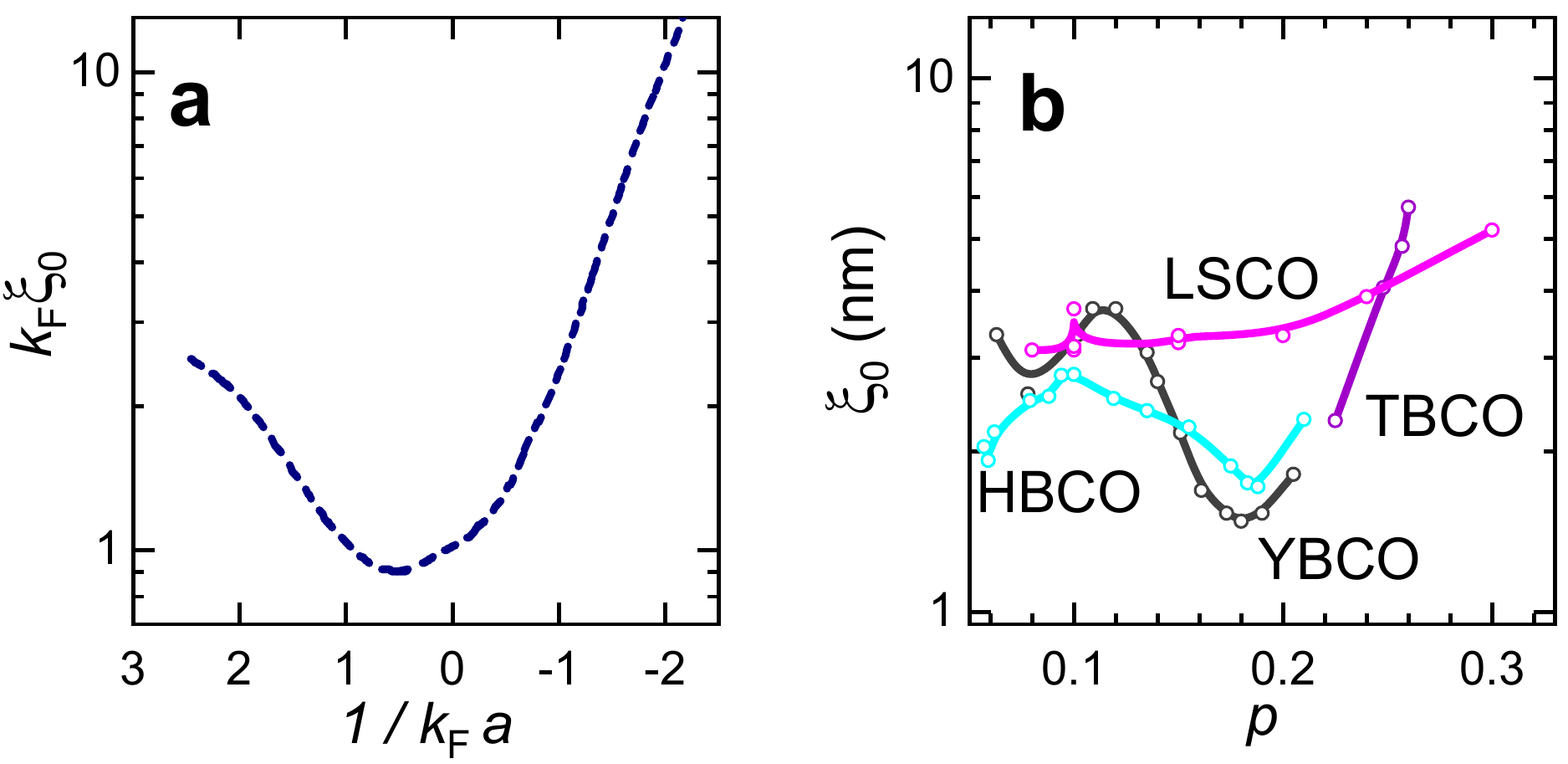}
\textsf{\caption{{\bf Coherence lengths.} {\bf a}, Coherence length $\xi_0$ versus $1/k_{\rm F}a$ for a cold atomic Fermi gas\cite{suzuki1991,grissonnanche2014,chan2020}. {\bf b}, Estimates of the in-plane coherence length as a function of $p$ in various cuprates.  
}
\label{coherence}}
\end{center}
\end{figure}

Figure~\ref{coherence}a shows the coherence length $\xi_0$ estimated for a cold atomic Fermi gas\cite{engelbrecht1997} while Fig.~\ref{coherence}b shows the in-plane coherence length estimated from $H_{\rm c2}$ using $H_{\rm c2}=\Phi_0/2\pi\xi_0^2$, where $H_{\rm c2}$ is itself estimated by extrapolating fits of the vortex melting line as a function of temperature\cite{suzuki1991,grissonnanche2014,chan2020}.

\subsection{The Fermi energy in the cuprates}

One way to infer proximity to a BCS-BEC crossover is to determine the ratio $k_{\rm B}T_{\rm c}/\varepsilon_{\rm F}$, where $T_{\rm F}=k_{\rm F}\varepsilon_{\rm F}$ is the Fermi temperature. At the unitary point, experimental and Quantum Monte Carlo studies have shown that in a three-dimensional system, this ratio reaches a maximum value of $k_{\rm B}T_{\rm c}/\varepsilon_{\rm F}\approx$~0.17 whereas in a ideal two-dimensional system it reaches a maximum value of  $k_{\rm B}T_{\rm c}/\varepsilon_{\rm F}\approx$~0.125~\cite{hazra2019}. Both of these estimates assume a parabolic band approximation. In the BCS and BEC regimes, $k_{\rm B}T_{\rm c}/\varepsilon_{\rm F}$ falls below this value, with the reduction being most severe in the BCS limit.

Assuming a parabolic band approximation, magnetic quantum oscillations measurements of the frequency $F$ and effective mass $m^\ast$ provide an estimate of the Fermi energy via the ratio $\varepsilon_{\rm F}=\hbar eF/m^\ast$. In the overdoped regime, quantum oscillation measurements pertain to the entire Fermi surface. Using average values $F=$~17.8~kT and $m^\ast=$~5.2~$m_{\rm e}$ (where $m_{\rm e}$ is the free electron mass) in TBCO~\cite{rourke2010}, we obtain $\varepsilon_{\rm F}\approx$~400~meV. From this we obtain $k_{\rm B}T_{\rm c}/\varepsilon_{\rm F}\approx$~0.006 (using $T_{\rm c}=$~26~K for the highest transition temperature of samples exhibiting quantum oscillations). This ratio is consistent with the very overdoped cuprate TBCO being deep in the BCS regime. 

Recent studies of the low temperature Sommerfeld coefficient of Eu- and Nd-substituted LSCO~\cite{michon2019} have suggested that the effective mass could increase by as much as a factor of 4 with decreasing hole doping, reaching a maximum value around $p=$~0.23 in that specific material. If a similar 4-fold increase in $m^\ast$ occurs in TBCO, then this combined with its higher maximum  $T_{\rm c}\approx$~96~K would produce a somewhat larger estimate of the ratio $k_{\rm B}T_{\rm c}/\varepsilon_{\rm F}\approx$~0.09 --- a value much closer to that expected in the unitary regime. Realization of the unitary regime is therefore plausible from the perspective of quantum oscillation and heat capacity measurements. 

In the underdoped regime of YBCO, quantum oscillation studies have found much smaller sections of Fermi surface with the effective mass exhibiting a non-monotonic behavior as a function of $p$~\cite{ramshaw2015}. Close to optimal $T_{\rm c}$ at $p=$~0.153 (where $T_{\rm c}\approx$~92~K and $F\approx$~600~T), $m^\ast\approx$~3.6~$m_{\rm e}$. This then falls to $m^\ast\approx$~1.4~$m_{\rm e}$ at $p=$~0.116 (where $T_{\rm c}\approx$~67~K and $F\approx$~560~T) before increasing again to $m^\ast\approx$~4.6~$m_{\rm e}$ at $p=$~0.09 (where $T_{\rm c}\approx$~57~K and $F\approx$~510~T). This yields Fermi energies in the range 13~$\leq\varepsilon_{\rm F}\leq$~46~meV and 0.3~$\leq k_{\rm B}T_{\rm c}/\varepsilon_{\rm F}\leq$~0.6. These values of $k_{\rm B}T_{\rm c}/\varepsilon_{\rm F}$ significantly exceed that expected at the unitary point. Based on Hall effect and other measurements~\cite{leboeuf2007}, however, these small pockets are the product of a broken symmetry phase that reconstructs the Fermi surface. 

In a BEC scenario, we might expect the pairing to onset at temperatures significantly higher than the onset of Fermi surface reconstruction. Provided the orbitally-averaged Fermi velocity $v_{\rm F}$ of the quasiparticles on the small pocket is similar to that of the quasiparticles on the the unreconstructed Fermi surface (which is generally true for translational symmetry breaking) and the parabolic approximation remains valid, we might expect the unreconstructed Fermi energy to be given by $\varepsilon_{\rm F}=\hbar^2k^2_{\rm F}/2=(\frac{\hbar}{2})v_{\rm F}k_{\rm F}$, where $k_{\rm F}$ is the orbitally-averaged Fermi radius. We might therefore expect $\varepsilon_{\rm F}$ to scale as $\varepsilon_{\rm F}\propto k_{\rm F}\propto\sqrt{F}$, where $F\propto(1+p)$ for an unreconstructed Fermi surface consistent with Luttinger's therorem. Assuming such scaling, we obtain unreconstructed Fermi energies in the range 80~$\lesssim\varepsilon_{\rm F}\lesssim$~270~meV, yielding 0.02~$\lesssim k_{\rm B}T_{\rm c}/\varepsilon_{\rm F}\lesssim$~0.07. The upper limit of this $k_{\rm B}T_{\rm c}/\varepsilon_{\rm F}$ estimate is a significant fraction of that ($k_{\rm B}T_{\rm c}/\varepsilon_{\rm F}\approx$~0.125) expected at the unitary point in a two-dimensional system, which is therefore consistent with the pairing having crossed over into the BEC regime in underdoped YBCO.

%
%\section{$d$-wave pairing and `Fermi arcs'}
%
%One departure of the cuprates from a cold atomic gas that appears to have little impact on the overarching similarities in Figs.~1 and~2a of the main paper is the existence of nodes due to $d$-wave pairing symmetry, which have been shown to give rise to gapless regions of the Fermi surface termed `Fermi arcs' at $T>T_{\rm c}$\cite{damascelli2003,norman1998}. It is still debated whether these arcs are intrinsic to the normal state $d$-wave pairing gap\cite{engelbrecht1998}, the consequence of phase fluctuations primarily affecting the $d$-wave nodes\cite{chubukov2007,demello2012}, or the product of a coexisting charge-density wave\cite{hucker2014,blancocanosa2014} or pair-density wave\cite{fradkin2015,agterberg2020} phase. 
%

\subsection{Differences in $p^\ast$ between cuprates with high and low $T_{\rm c}$'s}

Figure~1c of the main paper reveals what appear to be significant differences between cuprates that attain high values of  $T_{\rm c}\sim$~100~K and those that do not. 
From the requirement that $T_\gamma$ and $T_{\rm c}$ intersect in order for superconductivity to become optimally robust, we can now understand why $\delta\gamma(T_{\rm c})$ is peaked at different values of $p^\ast$ in different cuprate families, and not at all in others (e.g. BSLCO)\cite{michon2019,loram1998,wade1994,loram2001,loram1993,wen2009,mirmelstein1995}.  In the case of YBCO, Ca-YBCO, BSCCO and TBCO, the observation of pronounced peaks in $\delta\gamma(T_{\rm c})$ at $p^\ast$ strongly suggests that these systems undergo a BEC to BCS crossover as a function of $p$. 
In the case of LSCO, by contrast, $T_{\rm c}$ appears not to reach a sufficiently high value for it to intersect with $ T_\gamma$ until significantly higher dopings, giving rise to a much broader peak in $\delta\gamma(T_{\rm c})$. By extrapolation, we estimate $p\approx$~0.26 for LSCO in Fig.~1c of the main paper, suggesting that it has a increased inclination to remain within the BEC regime over a wider range of $p$, as has also recently been concluded from measurements of the superfluid density as a function of $p$\cite{bozovic2016}. 
BSLCO and Nd-LSCO appear to lie even deeper in the BEC regime (see Fig.~1a and b of the main paper), on account of their small $T_{\rm c}$ and tiny $\delta\gamma(T_{\rm c})$ values.  
Factors that could contribute (by way of a reduced superfluid density\cite{uemura1989}) in LSCO, BSLCO and Nd-LSCO to an increased tendency for these systems to remain in the BEC limit include their closer proximities to a van Hove singularity in the electronic structure\cite{damascelli2003} and increased levels of disorder\cite{rullieralbenque2008}. For example, systematic studies of disorder have shown this to suppress $T_{\rm c}$ and the superfluid density in a proportionate manner at low $p$, giving rise to a superconductor-to-insulator transition\cite{fukuzumi1996,nachumi1996}.

\subsection{`Fermi arcs'}
It is debated whether these arcs are intrinsic to a normal state $d$-wave pairing gap~\cite{engelbrecht1998}, are the consequence of phase fluctuations primarily affecting the $d$-wave nodes~\cite{chubukov2007,demello2012}, or are the product of a coexisting charge-density wave~\cite{hucker2014,blancocanosa2014} or a pair-density wave~\cite{fradkin2015,agterberg2020} phase. 

\subsection{$T$-dependence of the antinodal gap}

Spectroscopic measurements (including measurements made on the same composition~\cite{ding1996,renner1998}) differ on the degree to which the antinodal gap decreases with increasing temperature. Our finding of a quantitative consistency between Figs.~2c and d of the main paper supports a scenario in which the antinodal gap is largely unsuppressed at $T=T_\gamma$.

\subsection{Lifetime effects on $N_0$ in cold atomic Fermi gas measurements}
For $N_0$, we assume the experimentally quoted conversion between the tuning magnetic field and $1/k_{\rm F}a$~\cite{regal2004,zwielein2004}. Measurements of $N_0$ are affected by lifetime effects. Since the lifetime is also longest at the unitary point, corresponding to $1/k_{\rm F}a=0$, this can cause the measured peak in $N_0$ to become sharper than that in $\delta\gamma(T_{\rm c})$. In Ref.~\cite{regal2004}, the experimental uncertainty in the location of the unitary point in $1/k_{\rm F}a$ is $\approx$~0.2 while in Ref.~\cite{zwielein2004} it is $\approx$~0.03. The lower uncertainty of the latter may account for its greater similarity to $\delta\gamma(T_{\rm c})$. 

\subsection{Alternative choice of a smaller gap for $\Delta$ in the cuprates}
An alternative option is that $\Delta$ corresponds to a smaller secondary gap~\cite{hufner2008} comparable in energy to the ${\bf q}=(\pi,\pi)$ resonance energy seen in neutron scattering measurements. Apart from having a much weaker spectral weight and also having no reported signature in $\gamma$ or $\chi$, this gap yields a near constant gap ratio: $2\Delta/k_{\rm B}T_{\rm c}\approx$~5~\cite{hufner2008}, placing it within the BCS regime that is usually adequately accounted for by Eliashberg theory in other superconductors~\cite{carbotte1990}. Such a constant gap ratio would, however, cause all of the $\delta\gamma(T_{\rm c})$ data points to be stacked vertically at a constant $2\Delta/k_{\rm B}T_{\rm c}$ in Fig.~2a of the main paper. Such behavior is in contravention of the strong dependence of $\delta\gamma(T_{\rm c})$ on $2\Delta/k_{\rm B}T_{\rm c}$ in known boson-mediated (i.e. BCS) superconductors~\cite{carbotte1990}.

\subsection{Evidence for a hidden maximum in the heat capacity of a cold atomic gas}
In the case of heat capacity measurements on a cold atomic Fermi gas~\cite{ku2012}, %
a pseudogap with the same energy as the pairing gap $\Delta$ is expected to produce a maximum in $C$ at $T/T_{\rm F}\approx$~0.176. A maximum at this location cannot be observed owing to its proximity to $T_{\rm c}$ at the unitary point. The $\approx$ 40\% deviation of the normal state $C$ from that of a non interacting Fermi gas in the region of the normal state just above $T_{\rm c}$ is, however, consistent with a pseudogap~\cite{ku2012}. The normal state $\gamma$ (in the absence of a superfluid transition) would need to exhibit a downturn at low temperatures in order to conserve entropy. 

\subsection{Peaks in $\delta\gamma(T_{\rm c})$, $\gamma$ and $m^\ast$ associated with quantum criticality}

The peak in $\delta\gamma(T_{\rm c})$ as a function of $p$ for Ca-YBCO in Fig.~1c of the meain paper was recently attributed to a $T=0$ maximum in the normal state effective mass or $\gamma$ at $p^\ast$~\cite{michon2019}. 
However, in Fig. 3a of the main paper we find it to coincide with a finite temperature maximum in $\gamma$ originating from excitations across $\Delta$. Hence, the maximum in $\gamma$ coincides with $p^\ast$ only at $T=T_{\rm c}$. 

Given the thermodynamic evidence for the continued presence of a normal state gap at $p>p^\ast$ in Fig. 2d of the main paper, 
the peak in $\delta\gamma(T_{\rm c})$ at $T_{\rm c}=T_\gamma$  could not be easily understood were the normal state gap to have a non pairing origin~\cite{norman2014}. For a non pairing gap, we would expect the peak in $\delta\gamma(T_{\rm c})$ to occur only as the normal state gap vanishes, as has been found experimentally at the antiferromagnetic quantum phase transition in heavy fermion superconductors~\cite{park2008} and in thermodynamic simulations of a non-pairing pseudogap state critical point~\cite{leblanc2009,rice2012}.

The reported peaks in $m^\ast$ and $\gamma$ at low $T$ are at temperatures $T\lesssim$~10~K~\cite{ramshaw2015,michon2019}, implying that their contributions to $S_{\rm n}$ near $T_{\rm c}$ are overwhelmed by much larger contributions from excitations across $\Delta$.

\end{document}